\documentclass[%
 aip,
 amsmath,amssymb,
 reprint,%
]{revtex4-1}

\usepackage{graphicx}
\usepackage{dcolumn}
\usepackage{bm}

\usepackage[utf8]{inputenc}
\usepackage[T1]{fontenc}
\usepackage{mathptmx}
\usepackage{etoolbox}
\usepackage{CJKutf8}

\makeatletter
\def\@email#1#2{%
 \endgroup
 \patchcmd{\titleblock@produce}
  {\frontmatter@RRAPformat}
  {\frontmatter@RRAPformat{\produce@RRAP{*#1\href{mailto:#2}{#2}}}\frontmatter@RRAPformat}
  {}{}
}%
\makeatother
\begin{document}

\preprint{AIP/123-QED}

\newcommand{\LANL}{Los Alamos National Laboratory, Los Alamos, New Mexico, 87545, USA}
\newcommand{\UKy}{University of Kentucky, Lexington, Kentucky, 40506, USA}
\newcommand{\NCSU}{North Carolina State University, Raleigh, North Carolina 27695, USA}
\newcommand{\Caltech}{W. K. Kellogg Radiation Laboratory, California Institute of Technology, Pasadena, California 91125, USA}
\newcommand{\TTU}{Tennessee Technological University, Cookeville, Tennessee 38505, USA}
\newcommand{\IU}{Indiana University, Bloomington, Indiana 47405, USA}
\newcommand{\UIUC}{University of Illinois, Champaign, Illinois 61820, USA}
\newcommand{\ETSU}{East Tennessee State University, Johnson City, Tennessee 37614, USA}

\title[Scintillation characteristics of the EJ-299-02H scintillator]{Scintillation characteristics of the EJ-299-02H scintillator}
\author{N. Floyd}
\affiliation{\LANL}
\affiliation{\UKy
}
\email{nfloyd@lanl.gov}
\author{Md. T. Hassan}%
\author{Z. Tang (汤兆文)}
\author{M. Krivo\v{s}}
\affiliation{\LANL}
\author{M. Blatnik}
\affiliation{\LANL}
\affiliation{\Caltech}
\author{C. Cude-Woods}
\affiliation{\LANL}
\affiliation{\NCSU}
\author{S. M. Clayton}
\affiliation{\LANL}
\author{A. T. Holley}
\affiliation{\TTU}
\author{T. M. Ito}
\affiliation{\LANL}
\author{B. A. Johnson}
\affiliation{\IU}
\author{C.-Y. Liu}
\affiliation{\UIUC}
\author{M. Makela}
\author{C. L. Morris}
\affiliation{\LANL}
\author{A. S. C. Navazo}
\affiliation{\LANL}
\author{C. M. O'Shaughnessy}
\affiliation{\LANL}
\author{E. L. Renner}
\affiliation{\LANL}
\author{R. W. Pattie}
\affiliation{\ETSU}

\author{A. R. Young}
\affiliation{\NCSU}
\date{\today}

\begin{abstract}

A study of the dead layer thickness and quenching factor of a plastic scintillator for use in ultracold neutron (UCN) experiments is described. Alpha spectroscopy was used to determine the thickness of a thin surface dead layer, and the relative light outputs from the decay of $^{241}$Am  and Compton scattering of electrons were used to extract the quenching parameter. With these characteristics of the material known, the light yield of the scintillator can be calculated. The ability to make these scintillators deuterated, accompanied by its relatively thin dead layer, make it ideal for use in UCN experiment, where the light yield of decay electrons and alphas from neutron capture are critical for counting events. 

\end{abstract}

\begin{CJK*}{UTF8}{gbsn}
\maketitle
\end{CJK*}

\section{\label{sec:intro}Introduction}

According to the Standard Model of particle physics, the free neutron decays via $n \rightarrow p + e^- + \nu_e$ with a characteristic lifetime $\tau_n$ of approximately $15$~minutes. This lifetime is typically extracted either by measurement of the decay rate  of a cold neutron beam or by counting of surviving ultracold neutrons (UCN) after some amount of storage time. Due to their extremely low kinetic energy, UCN are easily manipulated via magnetic potentials, gravitation, and material barriers. The neutron lifetime has been measured by storing neutrons in material bottles \cite{Serebrov2005, Mampe1993, Arzumanov2015, Pichlmaier2010}, magnetic bottles \cite{ Pattie2018, Ezhov2018, Gonzalez2021},and by measuring the decay rate in beam experiments.\cite{Byrne1996, Nico2005, Yue2013} The most precise neutron lifetime measured by trapping UCN is $877.75 \pm 0.36$~s\cite{Gonzalez2021} and from the decay rate measurements via cold neutrons is $887.7 \pm 2.2$~s.\cite{Yue2013} For the decay rate measurements, absolute measurements of the flux of the neutron beam and decay rate detection efficiency are needed. The difference in measured $\tau_n$ of $\sim4$~standard deviations has been the source of much theoretical and experimental scrutiny\cite{Byrne2019,Serebrov2021}, with explanations such as hidden decay modes and oscillation channels of the neutron\cite{Fornal2018, Berezhiani2009, Berezhiani2019, Broussard2022, Tang2018, Sun2018}. A precise determination of the neutron lifetime is critical due to its role in probing the unitarity of the CKM matrix and as a parameter governing the abundances of light elements from Big Bang Nucleosynthesis.\cite{Tanabashi2018} Several experiments are proposed or underway that may resolve this discrepancy by measuring $\tau_n$ in new ways.\cite{Lawrence2021, Tanabashi2018, Wei2020} In the UCNProBe experiment at Los Alamos National Laboratory (LANL), the neutron beta decay lifetime will be measured using a material trap of deuterated plastic scintillator to count the number of trapped UCN and the number of electrons from beta decay absolutely. In this paper, we report measurements of the dead layer thickness and light yield quenching of this plastic scintillator and discuss the implications for the UCNProBe experiment.

\section{\label{sec:motivation}Motivation}

In the UCNProBe experiment, a storage volume defined by a box of deuterated polystyrene (dPS) plastic scintillator is filled with UCN from the LANL UCN source \cite{Saunders2013, Ito2018}, where neutrons are confined by the Fermi potential of dPS\cite{Tang2021} and their decay events are counted by the collection of scintillation light of decay electrons by photomultiplier tubes (PMT). The measurement of the total number of UCN will be measured in two separate sets of configurations. In the first configuration, neutrons available for decay are counted by capture upon a thin deposited film of boron\cite{Wang2015} on a high Fermi potential scintillator, via $n + ^{10}$B $\rightarrow \alpha + ^{7}$Li. The second configuration will utilize the injection of $^3$He gas into the storage volume for neutron counting via the $n + ^{3}$He $\rightarrow p + T$ capture reaction. The protons and tritons carry 573~keV and 191~keV of kinetic energy, respectively, and thus have an effective range shorter than 10~microns in dPS. In order to understand the systematic error in detection efficiencies for either configuration, several properties of the scintillator have been measured. The first is the fractional decrease in light yield of the scintillation due to quenching for nuclear recoil events. The scintillation light yield for a charged particle can be reasonably expressed with Birks' formula:\cite{Knoll1979}
\begin{equation}
    L = S \int_{0}^{E_i} \frac{\mathrm{d}E}{1 + kB \frac{\mathrm{d}E}{\mathrm{d}x}}
    \label{eq:lightYield}
\end{equation}
where $S$ is the scintillation efficiency of the material which is typically taken as a constant, $E_i$ is the initial energy of the incident particle within the scintillating volume, $\mathrm{d}E/\mathrm{d}x$ is the energy-dependent stopping power of the material for the particle, and $kB$ is a constant known as Birks' coefficient, which characterizes the loss of scintillation light due to quenching effects.\cite{Leo1994} The manufacturer quotes the scintillation efficiency for the Eljen EJ-299-02H dPS scintillator as 8500 photons/MeV. For a toy $kB$ value of 0.12~mm/MeV and 5.48 MeV alpha particle, a 33\% decrease in quenching would result in an almost 50\% gain in light output.

The other critical property of the scintillator is its dead layer, a thin surface region where inelastic scatters by incident particles are not detected as scintillation light. As noted in Ref.~\onlinecite{Roick2018}, a dead layer of a few $\mu$m would manifest as a cutoff in the low energy tail of the beta spectrum. Alpha spectroscopy is customarily used to determine the thickness of a detector dead layer due to the higher corresponding stopping power of alphas leading to larger, more easily detectable energy losses. If the dead layer is assumed to have uniform thickness, then particles at nonzero incidence angle will traverse an effective dead layer thickness $t_{\mathrm{eff}} = t/\cos \theta$, where $\theta$ is the angle of incidence of the particle with respect to the surface normal. In order to account for for multiple scattering and energy straggling,\cite{Leo1994} an energy loss simulation is used to determine the energy deposition in the active volume of the scintillator. The simulated energy loss is then be compared with measured values from experiments to determine the dead layer size.

\section{\label{sec:deadlayer}Measurement Using Alpha Particles}

\begin{figure}
\includegraphics[width = 0.9999\linewidth]{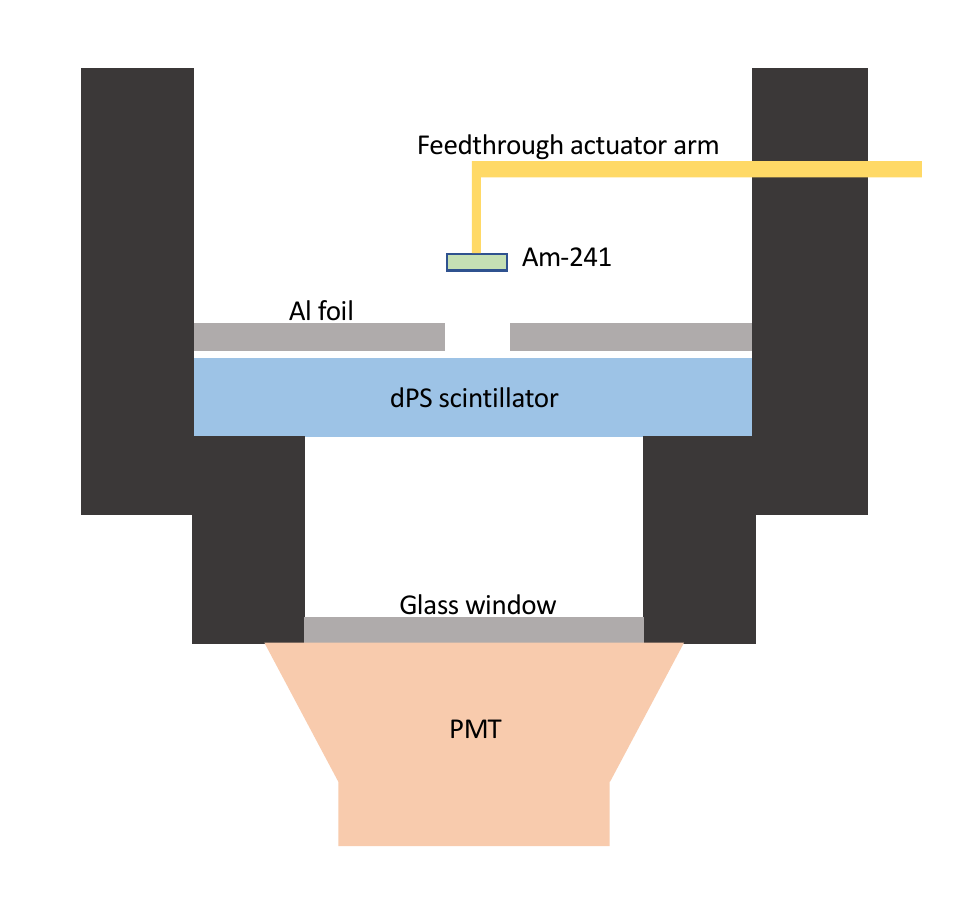}
\caption{\label{fig:setupDeadLayer} Diagram of the dead layer experiment.}
\end{figure}

A measurement of the dead layer thickness of the scintillator was performed by varying the lateral position of a 20.32~mm diameter $^{241}$Am electroplated sample over a 5~mm diameter pinhole in a thin 1100~series aluminum foil barrier over the scintillator. The experimental setup is shown in Figure~\ref{fig:setupDeadLayer}. The alpha source was suspended in a stainless steel holding cell from the arm of a vacuum feedthrough manual actuator, which kept the source at a constant vertical separation of 5.3$\pm$0.1~mm above the pinhole layer and allowed its lateral position to be adjusted without breaking vacuum. The thickness of the aluminum foil barrier was 0.1 mm, which served to attenuate alpha particles outside the pinhole. The underside of the foil was painted black to mitigate diffuse light reflection. The scintillator itself was 13~cm in diameter and 3~mm thick. During the experiments, the vacuum chamber was pumped to below $1.0 \times 10^{-5}$~torr to eliminate systematics from the energy loss of alpha particles in air. Scintillation light was collected by a Hamamatsu~R4017 7.62~mm diameter PMT which was biased at -1700~V. An Ortec~671 shaping amplifier was used, and the output of the shaping amplifier was then recorded using an Amptek~MCA-8000 multichannel analyzer, which served as the data acquisition system for spectrum collection. Measurements with the 2~$\mu$Ci activity alpha source were taken for lateral spacings from 0 to 35~mm in increments of 5~mm, where the pinhole was located at 0 mm. Measurements were also conducted in the opposite direction to verify that the pinhole was correctly aligned to 0 mm.

Studies have shown that oxygen contamination can lead to reduction of scintillation light in liquid scintillators \cite{Seliger1956,Acciarri2010}. If surface oxygen contamination is the cause of the reduction of scintillattion light on the surface, thus producing a dead layer, then a heat treatment may be able to reduce the thickness of this layer. To study this, the vacuum chamber was filled with Ar~gas while the scintillator was inside, heated to 40\textdegree~C and held at temperature for one week, and subsequently pumped down to below $1.0\times10^{-5}$~torr vacuum so that an identical measurement procedure as before could be performed. Afterwards, the process was repeated, but using N$_2$~gas. The results of each experiment are described below.

\subsection{\label{sec:getDataAndError}Data Analysis}

The dead layer thickness was extracted by comparing the mean energy deposition between simulation and experiment. To determine the measured energy deposited in the scintillator, several background effects were accounted for. First, background measurements were performed before and after the alpha source measurements to account for cosmic ray backgrounds. The observed cosmic ray background varied up to 20$\%$ with time, which could be accounted for by comparing the background subtraction of the average cosmic ray background with the subtraction of the average cosmic ray background scaled by up to 20\%. This produced a contribution to the statistical uncertainty, shown in Table~\ref{tab:tableAlphaFit}. For small lateral spacing of the alpha source, the impact of the background subtraction on the mean energy calculation was negligible, while beyond 25~mm, the low count rate, due to a smaller solid angle, was around 0.35\%.

\begin{figure}
\includegraphics[width = 0.9999\linewidth]{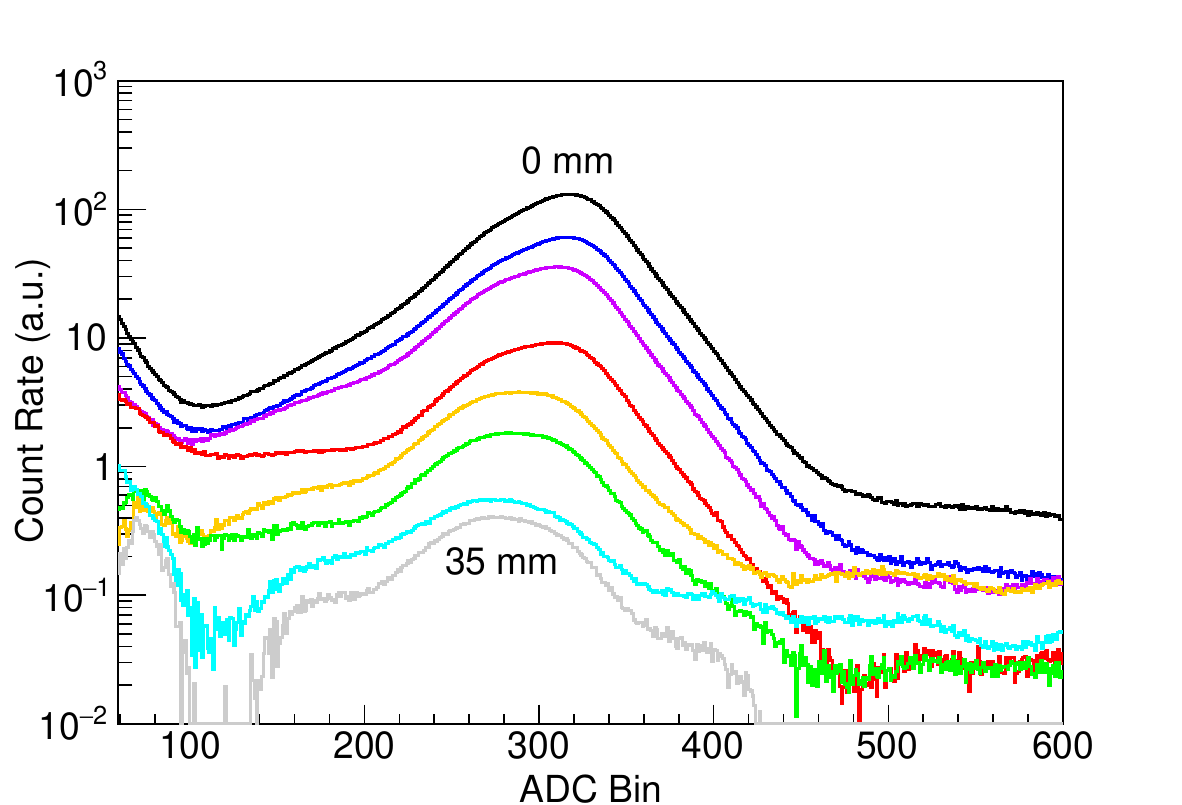}
\caption{\label{fig:backgroundSubtractionFit} Background-subtracted alpha spectra for different lateral positions of the source. Data was not taken beyond 35~mm.}
\end{figure}

\begin{figure}
\includegraphics[width = 0.9999\linewidth]{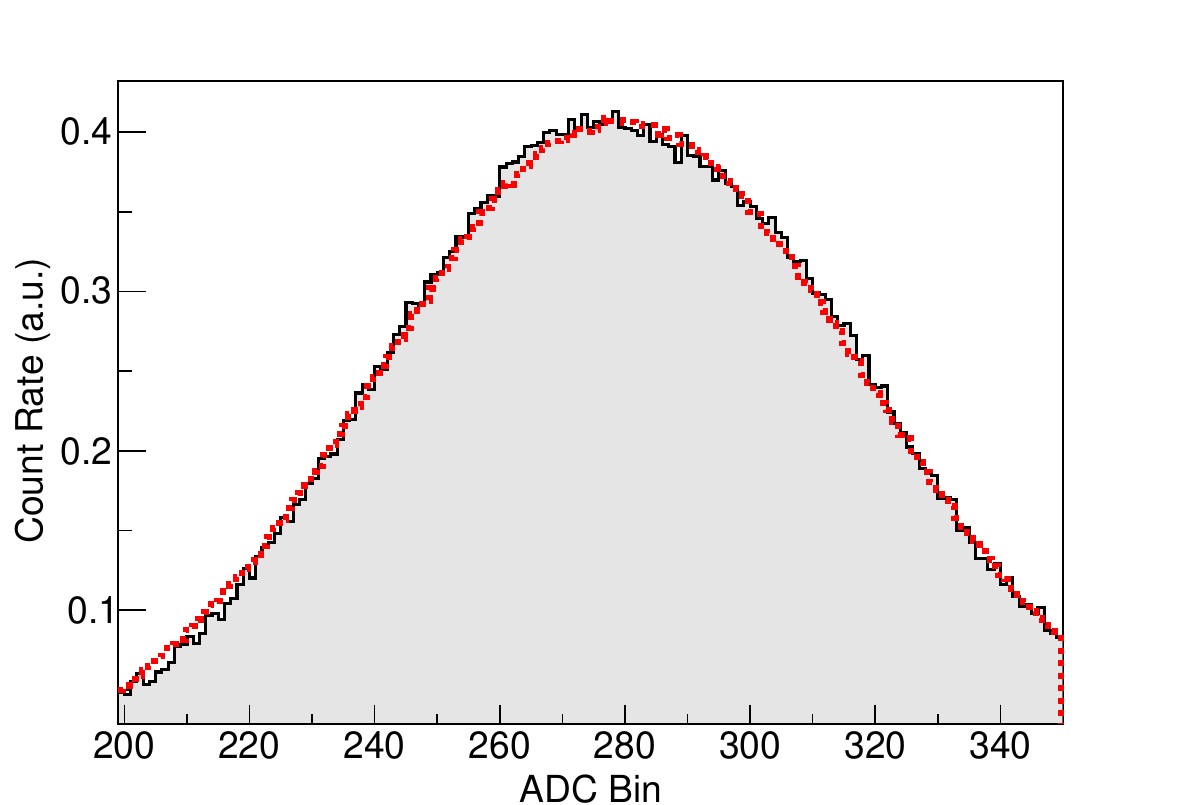}
\caption{\label{fig:blurFit}
Experimentally measured alpha spectrum (solid) compared with GEANT4 simulation histogram blurred to account for detector resolution (dotted).}
\end{figure}

The mean energy deposition of the alpha particles was determined by applying the following fit procedure to the measured spectra: beyond 15~mm lateral position of the source, the spectrum was well-approximated by a single Gaussian; however, this was not the case below 15~mm, as the spectrum was better fit by a double Gaussian. This was likely due to the comparable diameter of the alpha source and pinhole and small vertical separation between the two---despite the underside of the pinhole layer being painted black to mitigate diffuse reflections, there may still have been sufficient scintillation light reflection off of the stainless steel source itself to distort the spectrum. Because the relative amplitude of the higher energy peak decreased with distance and could be neglected beyond 15~mm, it was assumed that the peak represented this light reflection effect. The gamma ray background due to the emission of a 60~keV photon by the decay of $^{241}$Am was sufficiently small in the vicinity of the alpha spectrum that it was neglected. The detector response could be well-matched to GEANT4 simulation, as shown in Figure~\ref{fig:blurFit}.

To account for potential gain drift in the electronics, multiple runs were taken for each lateral spacing, and the weighted average across all runs for a position calculated. The average spectrum fit parameters are shown in Table~\ref{tab:tableAlphaFit}. While the statistical uncertainties due to cosmic ray background subtraction and peak fitting were at the $10^{-3}$ level, the systematic uncertainty associated with the standard deviation of the mean across different runs was the dominant source of uncertainty, at the one percent level. The total uncertainty was the sum in quadrature of the two error contributions.

\begin{table}
\caption{\label{tab:tableAlphaFit}Determined mean peak location for the experimentally measured alpha spectra of $^{241}$Am before heat treatment of the scintillator. Uncertainties due to fitting of the alpha spectrum and cosmic ray subtraction versus gain drift between different runs of the same lateral position are shown separately.}
\begin{ruledtabular}
\begin{tabular}{cccc}
 lateral position (mm) & fitted ADC bin & $\sigma_{\mathrm{statistical}}$ & $\sigma_{\mathrm{systematic}}$ \\
\hline
0 & 302.7 & 0.03 & 3.2 \\
5 & 300.6 & 0.05 & 3.4 \\
10 & 296.9 & 0.03 & 2.3 \\
15 & 292.8 & 0.22 & 4.0 \\
20 & 287.2 & 0.03 & 3.5 \\
25 & 284.2 & 0.05 & 1.1 \\
30 & 275.1 & 0.22 & 1.0 \\
35 & 277.5 & 0.25 & 0.6 \\
\end{tabular}
\end{ruledtabular}
\end{table}

\subsection{\label{sec:simulation}GEANT4 Simulation and Comparison}

\begin{figure}
\includegraphics[width = 0.9999\linewidth]{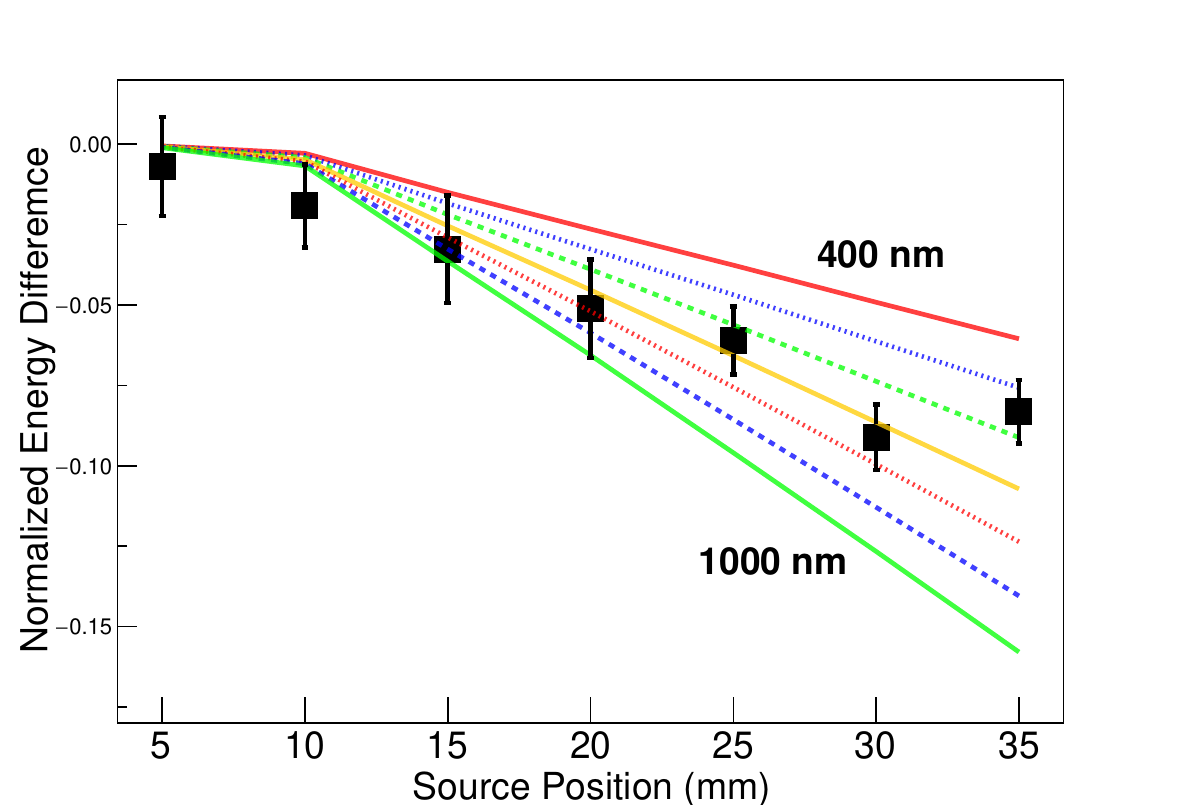}
\caption{\label{fig:deadLayerResults} GEANT4 simulated energy loss curves for different source positions compared with the measured pre-heat treatment data.}
\end{figure}

To determine the dead layer thickness, the energy falloff curves from the experimentally measured data were compared to those from Monte Carlo simulations, where the energy falloff curves refer to the decrease in deposited energy within the scintillator at greater incidence angle due to a larger effective dead layer. Monte Carlo energy spectra were generated using GEANT4\cite{Agostinelli2003, Allison2006, Allison2016} using the \verb=EmStandard_option4= physics list. The dead layer was modeled as a uniform thin surface layer within which no energy deposition was recorded. On the movable source, the alpha particles were generated within a 10~nm thick layer to model an electroplated surface.\cite{Shretter1979} Decay events were generated by random sampling of the position within the source material, momentum direction, and allowed alpha particle energies from the decay of $^{241}$Am. Only alpha energies with probability greater than 0.1\% were considered. The average energy deposition for a particular source position was then determined by calculating the mean of the simulated alpha energy deposition distribution for multiple dead layer thicknesses, ranging from 100~nm to 2000~nm, in increments of 10~nm. A comparison of the measured data and selected dead layer sizes is shown in Figure~\ref{fig:deadLayerResults}.

In order to compare measured and simulated data sets, each were normalized to the respective energy at 0 mm, with the normalized energy difference defined as:
\begin{equation}
    e_i = \frac{E_i - E_o}{E_o}
    \label{eq:normalizedEnergy}
\end{equation}
where $E_o$ and $E_i$ are the measured mean at 0 mm and nonzero position, respectively. The $\chi^2$ similarity was calculated for source positions between 5 and 35 mm. The best fit dead layer thickness was then extracted from the minimum $\chi^2$ value, with corresponding errors set by the $\chi^2_{\mathrm{min}}+1$ limit. The same process was followed for both the argon and nitrogen gas heat treatment data, with the respective dead layer thicknesses shown in Table~\ref{tab:deadLayerResults}. For the pre-heat treatment data, the dead layer thickness was determined to be $630 \pm 50$ nm, in agreement with previously measured values for polystyrene and other plastic scintillators.\cite{Hokin1988, Nefedov2016} 

\begin{table}
\caption{\label{tab:deadLayerResults}Experimentally determined dead layer thicknesses of dPS in response to heat treatment (HT).}
\begin{ruledtabular}
\begin{tabular}{ccc}
Experiment & Dead Layer Thickness (nm)  & $\chi^2_{\nu}$\\
\hline
No HT & $ 630 \pm 50 $ & 0.93 \\
Argon HT & $ 770 \pm 120 $ & 1.61 \\
Nitrogen HT & $ 680 \pm 60 $ & 0.92 \\
\end{tabular}
\end{ruledtabular}
\end{table}

\section{\label{sec:birks}Determination of Quenching Parameter}

The quenching of scintillation light is governed by the value of Birks' coefficient kB in Eq.~(\ref{eq:lightYield}). Its value is extracted from the ratio of measured light outputs for particles with different $\rm{d}E/\rm{d}x$, and was accomplished using alpha particles and Compton scattering electrons from the decay of $^{241}$Am and $^{137}$Cs, respectively. The light output of the two sources was similar such that the two spectra could not be easily distinguished, thus measurements with each source were performed separately. The alpha source was mounted in the same source holder apparatus as was done for the dead layer measurement. The Compton scattering source was placed directly outside the chamber. The walls of the stainless steel chamber were sufficiently thick to attenuate any decay electrons so that only the Compton scattering of decay gammas could be studied. The decay of $^{137}$Cs produces a characteristic 662~keV gamma ray, which Compton scatters within a material up to a maximum energy $E_{\mathrm{CE}} = 478$ keV, referred to as the Compton edge, defined as:\cite{Knoll1979}
\begin{equation}
    E_{\mathrm{CE}} = E_{\gamma} \Bigg[ \frac{2E_{\gamma}}{m_{\mathrm{e}} c^2 + 2 E_{\gamma}} \Bigg]
    \label{eq:comptonEdge}
\end{equation}
where $E_{\gamma}$ is the incident gamma energy and $m_{\mathrm{e}}$ is the electron mass. The photopeak was not observed due to the small thickness of the material and low gamma attenuation of polystyrene. With the Compton edge energy known and the average alpha energy deposition set by the previous dead layer measurement, the value of Birks' coefficient was estimated numerically from the ratio of alpha and electron light yields:
\begin{equation}
    \frac{D_{\beta}}{D_{\alpha}} = \frac{\int_0^{E_{\beta}}\mathrm{d}E_{\beta}[1 + kB (\frac{\mathrm{d}E}{\mathrm{d}x})_{\beta}]^{-1}}{\int_0^{E_{\alpha}}\mathrm{d}E_{\alpha}[1 + kB (\frac{\mathrm{d}E}{\mathrm{d}x})_{\alpha}]^{-1}} = \frac{L_{\beta}(kB)}{L_{\alpha}(kB)}
    \label{eq:birksRatio}
\end{equation}
where $D_e$ and $D_{\alpha}$ are the measured channel numbers of the Compton edge and average alpha energy, respectively, and $L_i(kB)$ is the light yield for a particle for some particular quenching factor as defined in Eq.~(\ref{eq:lightYield}). In order to determine the position of the Compton edge, the procedure outlined in Ref.~\onlinecite{Ashrafi2011} was followed. The energy spectrum of the 662~keV Compton scattering was simulated using GEANT4 and subsequently broadened to account for the energy resolution of the detector. The energy resolution was assumed to scale as $aE^{-1/2}$, where $a$ was a fitting parameter. The spectrum was also stretched horizontally to account for the energy calibration of the detector. Above 100~keV, the light output of a scintillator in response to electrons is approximately linear, so that the detector response or ADC bin $D(E)$ can be expressed as:
\begin{equation}
    D(E) = b (E - E_o) \approx b E
    \label{eq:channelNumber}
\end{equation}
where $E$ is the deposited energy within the scintillator, and $E_o$ is a small offset typically taken to be negligible.\cite{Ashrafi2011}

The best shape agreement between the GEANT4 blurred Compton spectrum and measured Compton spectrum was determined by the minimum $\chi^2$ similarity, corresponding to values of $a$ and $b$ of 0.136 and 1125, respectively, so that the Compton edge of $^{137}$Cs corresponded to bin 538$\pm$30. The mean alpha energy at zero lateral spacing was already known from the dead layer measurement to be $5.341\pm0.010$ MeV. The uncertainty in alpha energy was governed by the error in the dead layer measurement---the difference in mean energy evaluated between the determined dead layer thickness and the $\mathrm{t}_{\mathrm{dead}} \pm \sigma_{\mathrm{dead}}$, where $\sigma_{\mathrm{dead}}$ is the determined dead layer uncertainty. The contribution of the dead layer to the uncertainty of the electron energy in Eq.~(\ref{eq:birksRatio}) is negligible. The measured alpha peak location can be determined as before to be bin 517.6$\pm$1.2. Thus there are two uncertainties in the evaluation of Birks' coefficient: the ability to fit the two spectra and the theoretical uncertainty in incident energy from the dead layer. Inserting the measured positions and expected energies into Eq.~(\ref{eq:birksRatio}) leads to a value of Birks' coefficient of $0.0801 \pm 0.0039$~mm/MeV, in agreement with previously measured values for polystyrene.\cite{Kelleter2020, Tretyak2010}

\begin{figure}
\includegraphics[width = 0.9999\linewidth]{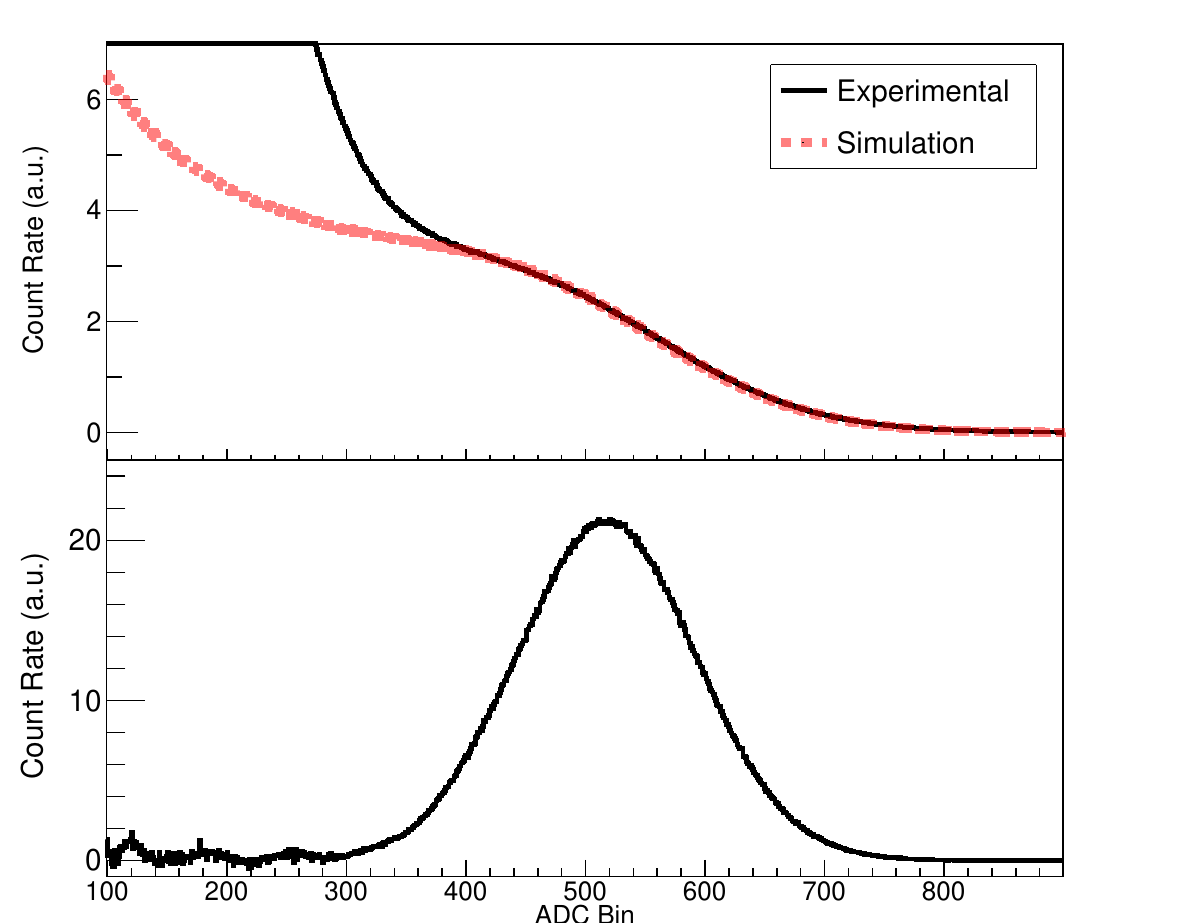}
\caption{\label{fig:comptonBirks} Energy spectra used to determine the quenching of dPS. Top: Measured Compton spectrum of $^{137}$Cs in dPS (black) and blurred GEANT4 spectrum (red). Bottom: measured alpha spectrum of $^{241}$Am in dPS.}
\end{figure}

\section{\label{sec:experiment}Impact on detection efficiencies}

With dead layer size and quenching parameter measured, the light yield of the scintillator in response to different particles can be calculated. For both protons and tritons from the neutron capture reaction $^3He(n, p)T$ and electrons from neutron beta decay, the expected light yields for normal incidence on the scintillator are shown in Table \ref{tab:tableLightYield}. Particles at glancing incident angles encounter an effective dead layer thickness comparable to their range and thus have a low probability for detection. Using simulation, the decay of UCN and the counting of neutrons with $^3$He can be modeled to determine the systematic error in counting efficiencies. It is assumed that the cubic storage volume is sufficiently small that the spatial density of UCN is approximately uniform. Using GEANT4, the storage volume is modeled with a uniform surface dead layer. Incident electrons are generated with uniformly distributed momenta, and proton/triton pairs are generated with uniform but back-to-back trajectories. Their corresponding light yields can then be calculated by inserting their energy after traversing the dead layer into Eq.~(\ref{eq:lightYield}). For electrons and protons, stopping power was calculated using the energy loss obtained from the NIST ESTAR and PSTAR tables, respectively, while for tritons, the SRIM code\cite{Ziegler2010} was used to calculate stopping power.

\begin{table}
\caption{\label{tab:tableLightYield}Light yields of different particles assuming normal incidence upon the scintillator and zero dead layer thickness.}
\begin{ruledtabular}
\begin{tabular}{ccccc}
 &$E$ (keV) & Range in dPS ($\mu$m) & Light Yield & Expected PEs \\
\hline
$p$ & 573 & 9.9 & 856 & 158 \\
$T$ & 191 & 2.8 & 242 & 45 \\
$e^-$ & 782\footnotemark[1] & 3,160 & 6,480 & 1,202 \\
$e^-$ & 300\footnotemark[2] & 830 & 2,450 & 454 \\
$e^-$ & 50 & 43.5 & 382 & 71 \\
$e^-$ & 6\footnotemark[3] & 2.1 & 41 & 8 \\
\end{tabular}
\end{ruledtabular}
\footnotetext[1]{Beta endpoint energy.}
\footnotetext[2]{Most probable beta energy.}
\footnotetext[3]{Estimated lower limit of detection.}
\end{table}

\begin{figure}
\includegraphics[width = 0.9999\linewidth]{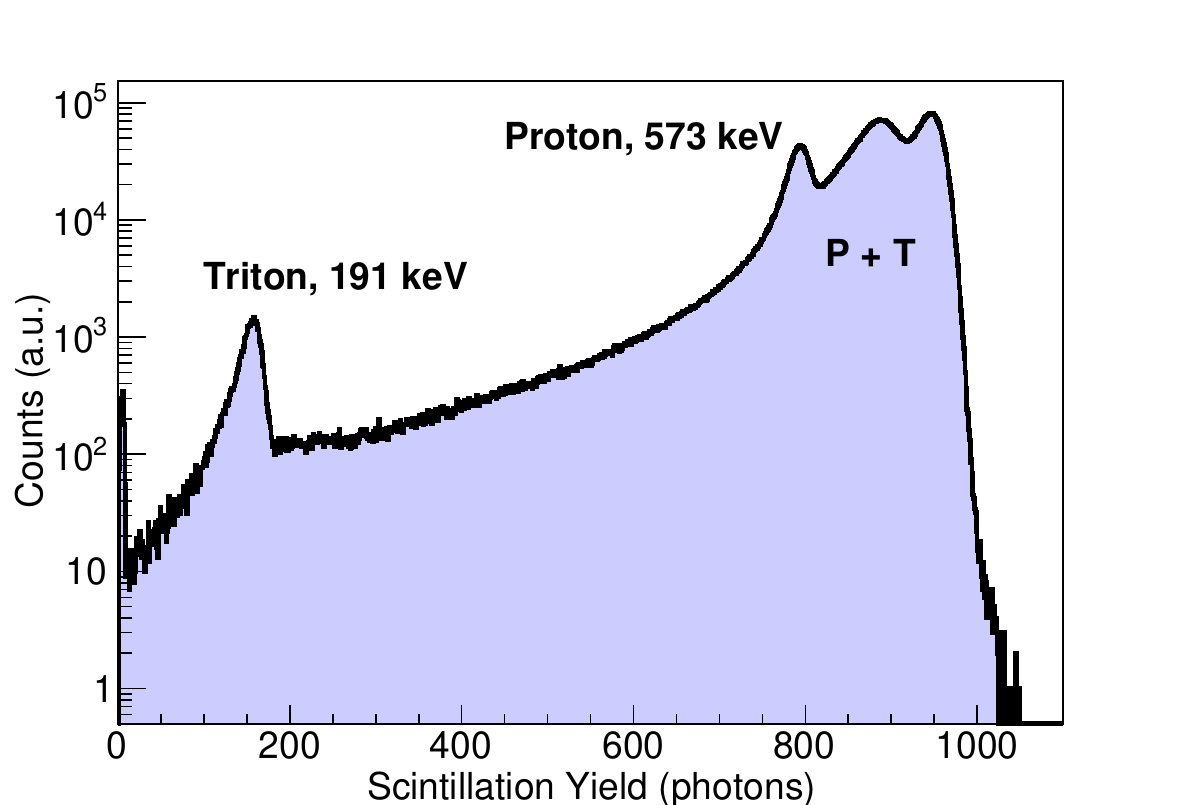}
\caption{\label{fig:3He_1D} Expected light output from $^{3}He$ neutron counting, taking the surface dead layer thickness into account. A small fraction of particles incident at a large angle with respect to the surface normal traverse an effective dead layer thickness greater than the mean free path within polystyrene and lose all kinetic energy without being detected. The peak at 150~photons corresponds to triton-only events, and similarly the peak at 775 indicates proton-only events. The bulk of scintillation events correspond to energy deposition by both the proton and triton.}
\end{figure}

Detection efficiency can be estimated by assuming a conservative threshold of 8~photoelectrons~(PE) across all PMTs and applying that limit to the expected light yields. In the final experiment, four ET Enterprises 9305~PMTs will be used to collect scintillation light, and the $\sim 720~\mathrm{cm}^3$ housing chamber will be coated in 97\%~reflective teflon, so that $\sim 80$\% of the total light can be collected. The detector quantum efficiency can be determined by averaging the PMT quantum efficiency across the emission spectrum of the scintillator, giving a value of 23.2\%. Based on the above assumptions, the respective detection efficiencies for different dead layer sizes are calculated and shown in Table \ref{tab:tableEfficiency}. For the measured dead layer thickness and quenching coefficient, the expected detection efficiency is 99.97\%, and the minimum detectable electron energy is 6~keV. 

\begin{table}
\caption{\label{tab:tableEfficiency}Fraction of $^{3}$He neutron counting events above the detection threshold for different surface dead layer thicknesses.}
\begin{ruledtabular}
\begin{tabular}{ccc}
 $\mathrm{t}_{\mathrm{dead}}$ (nm) & Average PE Yield & Fraction Detected ($\%$) \\
\hline
0 & 203 & 100 \\
200 & 189 & 99.999 \\
500 & 169 & 99.99 \\
700 & 157 & 99.96 \\
1000 & 140 & 99.8 \\
1200 & 131 & 99.3 \\
1500 & 119 & 98.8 \\
1700 & 112 & 97.5 \\
2000 & 101 & 95.7 \\
\end{tabular}
\end{ruledtabular}
\end{table}

\section{\label{sec:conclusion}Conclusion}

We have measured the dead layer thickness of the EJ-299-02H polystyrene scintillator to be $630\pm50$~nm by use of alpha spectroscopy and simulation.  Heat treatment was performed on the scintillator and the dead layer was remeasured, but results were within uncertainty in agreement with the non-HT value, so there was no noticeable change in dead layer thickness. We have also measured the quenching of the scintillator to be $0.0801\pm0.0039$~mm/MeV by comparing the light yields of the same alpha source and a Compton scattering source. Knowledge of these parameters is critical for estimating the detection efficiencies of the UCNProBe neutron lifetime measurement at LANL. For the measured dead layer size and quenching coefficient, the loss of neutron counting efficiency should be below 0.1\%. Additionally, these properties are relevant to beta decay correlation experiments,\cite{Severijns2006, Saul2020} particularly measurements of the $\beta$ asymmetry parameter $A$,\cite{Plaster2012, Konrad2012, Roick2019} where understanding of backscattering of electrons is important. If there are variations of the dead layer size across the detector, the non-uniformity will be accounted for in the final experiment by {\it in situ} measuring the dead layer at different positions on each scintillator with point sources and monitored over time. 

\section*{\label{sec:acknowledgements}Acknowledgements}

This work was supported by the Los Alamos National
Laboratory LDRD Program (Project No. 20190048 ER), National
Science Foundation (Grant No. PHY1914133), and U.S. Department of Energy, Office of Nuclear Physics (Grant No. DE-FG02-97ER41042).

\section*{\label{sec:data}Data availability}

The data that support the findings of this study are available from the corresponding author upon reasonable request.

\section*{\label{sec:references}References}

\nocite{}
\bibliography{DeadLayerPaper_bib}

\begin{thebibliography}{45}%
\makeatletter
\providecommand \@ifxundefined [1]{%
 \@ifx{#1\undefined}
}%
\providecommand \@ifnum [1]{%
 \ifnum #1\expandafter \@firstoftwo
 \else \expandafter \@secondoftwo
 \fi
}%
\providecommand \@ifx [1]{%
 \ifx #1\expandafter \@firstoftwo
 \else \expandafter \@secondoftwo
 \fi
}%
\providecommand \natexlab [1]{#1}%
\providecommand \enquote  [1]{``#1''}%
\providecommand \bibnamefont  [1]{#1}%
\providecommand \bibfnamefont [1]{#1}%
\providecommand \citenamefont [1]{#1}%
\providecommand \href@noop [0]{\@secondoftwo}%
\providecommand \href [0]{\begingroup \@sanitize@url \@href}%
\providecommand \@href[1]{\@@startlink{#1}\@@href}%
\providecommand \@@href[1]{\endgroup#1\@@endlink}%
\providecommand \@sanitize@url [0]{\catcode `\\12\catcode `\$12\catcode
  `\&12\catcode `\#12\catcode `\^12\catcode `\_12\catcode `\%12\relax}%
\providecommand \@@startlink[1]{}%
\providecommand \@@endlink[0]{}%
\providecommand \url  [0]{\begingroup\@sanitize@url \@url }%
\providecommand \@url [1]{\endgroup\@href {#1}{\urlprefix }}%
\providecommand \urlprefix  [0]{URL }%
\providecommand \Eprint [0]{\href }%
\providecommand \doibase [0]{http://dx.doi.org/}%
\providecommand \selectlanguage [0]{\@gobble}%
\providecommand \bibinfo  [0]{\@secondoftwo}%
\providecommand \bibfield  [0]{\@secondoftwo}%
\providecommand \translation [1]{[#1]}%
\providecommand \BibitemOpen [0]{}%
\providecommand \bibitemStop [0]{}%
\providecommand \bibitemNoStop [0]{.\EOS\space}%
\providecommand \EOS [0]{\spacefactor3000\relax}%
\providecommand \BibitemShut  [1]{\csname bibitem#1\endcsname}%
\let\auto@bib@innerbib\@empty
\bibitem [{\citenamefont {Serebrov}\ \emph {et~al.}(2005)\citenamefont
  {Serebrov}, \citenamefont {Varlamov}, \citenamefont {Kharitonov},
  \citenamefont {Fomin}, \citenamefont {Pokotilovski}, \citenamefont
  {Geltenbort}, \citenamefont {Butterworth}, \citenamefont {Krasnoschekova},
  \citenamefont {Lasakov}, \citenamefont {Tal'daev}, \citenamefont
  {Vassiljev},\ and\ \citenamefont {Zherebtsov}}]{Serebrov2005}%
  \BibitemOpen
  \bibfield  {author} {\bibinfo {author} {\bibfnamefont {A.}~\bibnamefont
  {Serebrov}}, \bibinfo {author} {\bibfnamefont {V.}~\bibnamefont {Varlamov}},
  \bibinfo {author} {\bibfnamefont {A.}~\bibnamefont {Kharitonov}}, \bibinfo
  {author} {\bibfnamefont {A.}~\bibnamefont {Fomin}}, \bibinfo {author}
  {\bibfnamefont {Y.}~\bibnamefont {Pokotilovski}}, \bibinfo {author}
  {\bibfnamefont {P.}~\bibnamefont {Geltenbort}}, \bibinfo {author}
  {\bibfnamefont {J.}~\bibnamefont {Butterworth}}, \bibinfo {author}
  {\bibfnamefont {I.}~\bibnamefont {Krasnoschekova}}, \bibinfo {author}
  {\bibfnamefont {M.}~\bibnamefont {Lasakov}}, \bibinfo {author} {\bibfnamefont
  {R.}~\bibnamefont {Tal'daev}}, \bibinfo {author} {\bibfnamefont
  {A.}~\bibnamefont {Vassiljev}}, \ and\ \bibinfo {author} {\bibfnamefont
  {O.}~\bibnamefont {Zherebtsov}},\ }\bibfield  {title} {\enquote {\bibinfo
  {title} {Measurement of the neutron lifetime using a gravitational trap and a
  low-temperature fomblin coating},}\ }\href {\doibase
  10.1016/J.PHYSLETB.2004.11.013} {\bibfield  {journal} {\bibinfo  {journal}
  {Physics Letters B}\ }\textbf {\bibinfo {volume} {605}},\ \bibinfo {pages}
  {72--78} (\bibinfo {year} {2005})}\BibitemShut {NoStop}%
\bibitem [{\citenamefont {Mampe}\ \emph {et~al.}(1993)\citenamefont {Mampe},
  \citenamefont {Bondarenko}, \citenamefont {Morozov}, \citenamefont {Panin},\
  and\ \citenamefont {Fomin}}]{Mampe1993}%
  \BibitemOpen
  \bibfield  {author} {\bibinfo {author} {\bibfnamefont {W.}~\bibnamefont
  {Mampe}}, \bibinfo {author} {\bibfnamefont {L.}~\bibnamefont {Bondarenko}},
  \bibinfo {author} {\bibfnamefont {V.}~\bibnamefont {Morozov}}, \bibinfo
  {author} {\bibfnamefont {Y.}~\bibnamefont {Panin}}, \ and\ \bibinfo {author}
  {\bibfnamefont {A.}~\bibnamefont {Fomin}},\ }\bibfield  {title} {\enquote
  {\bibinfo {title} {Measuring neutron lifetime by storing ultracold neutrons
  and detecting inelastically scattered neutrons},}\ }\href@noop {} {\bibfield
  {journal} {\bibinfo  {journal} {JETP Lett.}\ }\textbf {\bibinfo {volume}
  {57}},\ \bibinfo {pages} {82--87} (\bibinfo {year} {1993})}\BibitemShut
  {NoStop}%
\bibitem [{\citenamefont {Arzumanov}\ \emph {et~al.}(2015)\citenamefont
  {Arzumanov}, \citenamefont {Bondarenko}, \citenamefont {Chernyavsky},
  \citenamefont {Geltenbort}, \citenamefont {Morozov}, \citenamefont
  {Nesvizhevsky}, \citenamefont {Panin},\ and\ \citenamefont
  {Strepetov}}]{Arzumanov2015}%
  \BibitemOpen
  \bibfield  {author} {\bibinfo {author} {\bibfnamefont {S.}~\bibnamefont
  {Arzumanov}}, \bibinfo {author} {\bibfnamefont {L.}~\bibnamefont
  {Bondarenko}}, \bibinfo {author} {\bibfnamefont {S.}~\bibnamefont
  {Chernyavsky}}, \bibinfo {author} {\bibfnamefont {P.}~\bibnamefont
  {Geltenbort}}, \bibinfo {author} {\bibfnamefont {V.}~\bibnamefont {Morozov}},
  \bibinfo {author} {\bibfnamefont {V.}~\bibnamefont {Nesvizhevsky}}, \bibinfo
  {author} {\bibfnamefont {Y.}~\bibnamefont {Panin}}, \ and\ \bibinfo {author}
  {\bibfnamefont {A.}~\bibnamefont {Strepetov}},\ }\bibfield  {title} {\enquote
  {\bibinfo {title} {A measurement of the neutron lifetime using the method of
  storage of ultracold neutrons and detection of inelastically up-scattered
  neutrons},}\ }\href {\doibase 10.1016/J.PHYSLETB.2015.04.021} {\bibfield
  {journal} {\bibinfo  {journal} {Physics Letters B}\ }\textbf {\bibinfo
  {volume} {745}},\ \bibinfo {pages} {79--89} (\bibinfo {year}
  {2015})}\BibitemShut {NoStop}%
\bibitem [{\citenamefont {Pichlmaier}\ \emph {et~al.}(2010)\citenamefont
  {Pichlmaier}, \citenamefont {Varlamov}, \citenamefont {Schreckenbach},\ and\
  \citenamefont {Geltenbort}}]{Pichlmaier2010}%
  \BibitemOpen
  \bibfield  {author} {\bibinfo {author} {\bibfnamefont {A.}~\bibnamefont
  {Pichlmaier}}, \bibinfo {author} {\bibfnamefont {V.}~\bibnamefont
  {Varlamov}}, \bibinfo {author} {\bibfnamefont {K.}~\bibnamefont
  {Schreckenbach}}, \ and\ \bibinfo {author} {\bibfnamefont {P.}~\bibnamefont
  {Geltenbort}},\ }\bibfield  {title} {\enquote {\bibinfo {title} {Neutron
  lifetime measurement with the ucn trap-in-trap mambo ii},}\ }\href {\doibase
  10.1016/J.PHYSLETB.2010.08.032} {\bibfield  {journal} {\bibinfo  {journal}
  {Physics Letters B}\ }\textbf {\bibinfo {volume} {693}},\ \bibinfo {pages}
  {221--226} (\bibinfo {year} {2010})}\BibitemShut {NoStop}%
\bibitem [{\citenamefont {Pattie}\ \emph {et~al.}(2018)\citenamefont {Pattie},
  \citenamefont {Callahan}, \citenamefont {Cude-Woods}, \citenamefont {Adamek},
  \citenamefont {Broussard}, \citenamefont {Clayton}, \citenamefont {Currie},
  \citenamefont {Dees}, \citenamefont {Ding}, \citenamefont {Engel},
  \citenamefont {Fellers}, \citenamefont {Fox}, \citenamefont {Geltenbort},
  \citenamefont {Hickerson}, \citenamefont {Hoffbauer}, \citenamefont {Holley},
  \citenamefont {Komives}, \citenamefont {Liu}, \citenamefont {MacDonald},
  \citenamefont {Makela}, \citenamefont {Morris}, \citenamefont {Ortiz},
  \citenamefont {Ramsey}, \citenamefont {Salvat}, \citenamefont {Saunders},
  \citenamefont {Seestrom}, \citenamefont {Sharapov}, \citenamefont {Sjue},
  \citenamefont {Tang}, \citenamefont {Vanderwerp}, \citenamefont {Vogelaar},
  \citenamefont {Walstrom}, \citenamefont {Wang}, \citenamefont {Wei},
  \citenamefont {Weaver}, \citenamefont {Wexler}, \citenamefont {Womack},
  \citenamefont {Young},\ and\ \citenamefont {Zeck}}]{Pattie2018}%
  \BibitemOpen
  \bibfield  {author} {\bibinfo {author} {\bibfnamefont {R.~W.}\ \bibnamefont
  {Pattie}}, \bibinfo {author} {\bibfnamefont {N.~B.}\ \bibnamefont
  {Callahan}}, \bibinfo {author} {\bibfnamefont {C.}~\bibnamefont
  {Cude-Woods}}, \bibinfo {author} {\bibfnamefont {E.~R.}\ \bibnamefont
  {Adamek}}, \bibinfo {author} {\bibfnamefont {L.~J.}\ \bibnamefont
  {Broussard}}, \bibinfo {author} {\bibfnamefont {S.~M.}\ \bibnamefont
  {Clayton}}, \bibinfo {author} {\bibfnamefont {S.~A.}\ \bibnamefont {Currie}},
  \bibinfo {author} {\bibfnamefont {E.~B.}\ \bibnamefont {Dees}}, \bibinfo
  {author} {\bibfnamefont {X.}~\bibnamefont {Ding}}, \bibinfo {author}
  {\bibfnamefont {E.~M.}\ \bibnamefont {Engel}}, \bibinfo {author}
  {\bibfnamefont {D.~E.}\ \bibnamefont {Fellers}}, \bibinfo {author}
  {\bibfnamefont {W.}~\bibnamefont {Fox}}, \bibinfo {author} {\bibfnamefont
  {P.}~\bibnamefont {Geltenbort}}, \bibinfo {author} {\bibfnamefont {K.~P.}\
  \bibnamefont {Hickerson}}, \bibinfo {author} {\bibfnamefont {M.~A.}\
  \bibnamefont {Hoffbauer}}, \bibinfo {author} {\bibfnamefont {A.~T.}\
  \bibnamefont {Holley}}, \bibinfo {author} {\bibfnamefont {A.}~\bibnamefont
  {Komives}}, \bibinfo {author} {\bibfnamefont {C.-Y.}\ \bibnamefont {Liu}},
  \bibinfo {author} {\bibfnamefont {S.~W.~T.}\ \bibnamefont {MacDonald}},
  \bibinfo {author} {\bibfnamefont {M.}~\bibnamefont {Makela}}, \bibinfo
  {author} {\bibfnamefont {C.~L.}\ \bibnamefont {Morris}}, \bibinfo {author}
  {\bibfnamefont {J.~D.}\ \bibnamefont {Ortiz}}, \bibinfo {author}
  {\bibfnamefont {J.}~\bibnamefont {Ramsey}}, \bibinfo {author} {\bibfnamefont
  {D.~J.}\ \bibnamefont {Salvat}}, \bibinfo {author} {\bibfnamefont
  {A.}~\bibnamefont {Saunders}}, \bibinfo {author} {\bibfnamefont {S.~J.}\
  \bibnamefont {Seestrom}}, \bibinfo {author} {\bibfnamefont {E.~I.}\
  \bibnamefont {Sharapov}}, \bibinfo {author} {\bibfnamefont {S.~K.}\
  \bibnamefont {Sjue}}, \bibinfo {author} {\bibfnamefont {Z.}~\bibnamefont
  {Tang}}, \bibinfo {author} {\bibfnamefont {J.}~\bibnamefont {Vanderwerp}},
  \bibinfo {author} {\bibfnamefont {B.}~\bibnamefont {Vogelaar}}, \bibinfo
  {author} {\bibfnamefont {P.~L.}\ \bibnamefont {Walstrom}}, \bibinfo {author}
  {\bibfnamefont {Z.}~\bibnamefont {Wang}}, \bibinfo {author} {\bibfnamefont
  {W.}~\bibnamefont {Wei}}, \bibinfo {author} {\bibfnamefont {H.~L.}\
  \bibnamefont {Weaver}}, \bibinfo {author} {\bibfnamefont {J.~W.}\
  \bibnamefont {Wexler}}, \bibinfo {author} {\bibfnamefont {T.~L.}\
  \bibnamefont {Womack}}, \bibinfo {author} {\bibfnamefont {A.~R.}\
  \bibnamefont {Young}}, \ and\ \bibinfo {author} {\bibfnamefont {B.~A.}\
  \bibnamefont {Zeck}},\ }\bibfield  {title} {\enquote {\bibinfo {title}
  {Measurement of the neutron lifetime using a magneto-gravitational trap and
  in situ detection.}}\ }\href {\doibase 10.1126/science.aan8895} {\bibfield
  {journal} {\bibinfo  {journal} {Science (New York, N.Y.)}\ }\textbf {\bibinfo
  {volume} {360}},\ \bibinfo {pages} {627--632} (\bibinfo {year}
  {2018})}\BibitemShut {NoStop}%
\bibitem [{\citenamefont {Ezhov}\ \emph {et~al.}(2018)\citenamefont {Ezhov},
  \citenamefont {Andreev}, \citenamefont {Ban}, \citenamefont {Bazarov},
  \citenamefont {Geltenbort}, \citenamefont {Glushkov}, \citenamefont
  {Knyazkov}, \citenamefont {Kovrizhnykh}, \citenamefont {Krygin},
  \citenamefont {Naviliat-Cuncic},\ and\ \citenamefont {Ryabov}}]{Ezhov2018}%
  \BibitemOpen
  \bibfield  {author} {\bibinfo {author} {\bibfnamefont {V.~F.}\ \bibnamefont
  {Ezhov}}, \bibinfo {author} {\bibfnamefont {A.~Z.}\ \bibnamefont {Andreev}},
  \bibinfo {author} {\bibfnamefont {G.}~\bibnamefont {Ban}}, \bibinfo {author}
  {\bibfnamefont {B.~A.}\ \bibnamefont {Bazarov}}, \bibinfo {author}
  {\bibfnamefont {P.}~\bibnamefont {Geltenbort}}, \bibinfo {author}
  {\bibfnamefont {A.~G.}\ \bibnamefont {Glushkov}}, \bibinfo {author}
  {\bibfnamefont {V.~A.}\ \bibnamefont {Knyazkov}}, \bibinfo {author}
  {\bibfnamefont {N.~A.}\ \bibnamefont {Kovrizhnykh}}, \bibinfo {author}
  {\bibfnamefont {G.~B.}\ \bibnamefont {Krygin}}, \bibinfo {author}
  {\bibfnamefont {O.}~\bibnamefont {Naviliat-Cuncic}}, \ and\ \bibinfo {author}
  {\bibfnamefont {V.~L.}\ \bibnamefont {Ryabov}},\ }\bibfield  {title}
  {\enquote {\bibinfo {title} {Measurement of the neutron lifetime with
  ultracold neutrons stored in a magneto-gravitational trap},}\ }\href
  {\doibase 10.1134/S0021364018110024} {\bibfield  {journal} {\bibinfo
  {journal} {JETP Letters}\ }\textbf {\bibinfo {volume} {107}},\ \bibinfo
  {pages} {671--675} (\bibinfo {year} {2018})}\BibitemShut {NoStop}%
\bibitem [{\citenamefont {Gonzalez}\ \emph {et~al.}(2021)\citenamefont
  {Gonzalez}, \citenamefont {Fries}, \citenamefont {Cude-Woods}, \citenamefont
  {Bailey}, \citenamefont {Blatnik}, \citenamefont {Broussard}, \citenamefont
  {Callahan}, \citenamefont {Choi}, \citenamefont {Clayton}, \citenamefont
  {Currie}, \citenamefont {Dawid}, \citenamefont {Dees}, \citenamefont
  {Filippone}, \citenamefont {Fox}, \citenamefont {Geltenbort}, \citenamefont
  {George}, \citenamefont {Hayen}, \citenamefont {Hickerson}, \citenamefont
  {Hoffbauer}, \citenamefont {Hoffman}, \citenamefont {Holley}, \citenamefont
  {Ito}, \citenamefont {Komives}, \citenamefont {Liu}, \citenamefont {Makela},
  \citenamefont {Morris}, \citenamefont {Musedinovic}, \citenamefont
  {O’Shaughnessy}, \citenamefont {Pattie}, \citenamefont {Ramsey},
  \citenamefont {Salvat}, \citenamefont {Saunders}, \citenamefont {Sharapov},
  \citenamefont {Slutsky}, \citenamefont {Su}, \citenamefont {Sun},
  \citenamefont {Swank}, \citenamefont {Tang}, \citenamefont {Uhrich},
  \citenamefont {Vanderwerp}, \citenamefont {Walstrom}, \citenamefont {Wang},
  \citenamefont {Wei},\ and\ \citenamefont {Young}}]{Gonzalez2021}%
  \BibitemOpen
  \bibfield  {author} {\bibinfo {author} {\bibfnamefont {F.~M.}\ \bibnamefont
  {Gonzalez}}, \bibinfo {author} {\bibfnamefont {E.~M.}\ \bibnamefont {Fries}},
  \bibinfo {author} {\bibfnamefont {C.}~\bibnamefont {Cude-Woods}}, \bibinfo
  {author} {\bibfnamefont {T.}~\bibnamefont {Bailey}}, \bibinfo {author}
  {\bibfnamefont {M.}~\bibnamefont {Blatnik}}, \bibinfo {author} {\bibfnamefont
  {L.~J.}\ \bibnamefont {Broussard}}, \bibinfo {author} {\bibfnamefont {N.~B.}\
  \bibnamefont {Callahan}}, \bibinfo {author} {\bibfnamefont {J.~H.}\
  \bibnamefont {Choi}}, \bibinfo {author} {\bibfnamefont {S.~M.}\ \bibnamefont
  {Clayton}}, \bibinfo {author} {\bibfnamefont {S.~A.}\ \bibnamefont {Currie}},
  \bibinfo {author} {\bibfnamefont {M.}~\bibnamefont {Dawid}}, \bibinfo
  {author} {\bibfnamefont {E.~B.}\ \bibnamefont {Dees}}, \bibinfo {author}
  {\bibfnamefont {B.~W.}\ \bibnamefont {Filippone}}, \bibinfo {author}
  {\bibfnamefont {W.}~\bibnamefont {Fox}}, \bibinfo {author} {\bibfnamefont
  {P.}~\bibnamefont {Geltenbort}}, \bibinfo {author} {\bibfnamefont
  {E.}~\bibnamefont {George}}, \bibinfo {author} {\bibfnamefont
  {L.}~\bibnamefont {Hayen}}, \bibinfo {author} {\bibfnamefont {K.~P.}\
  \bibnamefont {Hickerson}}, \bibinfo {author} {\bibfnamefont {M.~A.}\
  \bibnamefont {Hoffbauer}}, \bibinfo {author} {\bibfnamefont {K.}~\bibnamefont
  {Hoffman}}, \bibinfo {author} {\bibfnamefont {A.~T.}\ \bibnamefont {Holley}},
  \bibinfo {author} {\bibfnamefont {T.~M.}\ \bibnamefont {Ito}}, \bibinfo
  {author} {\bibfnamefont {A.}~\bibnamefont {Komives}}, \bibinfo {author}
  {\bibfnamefont {C.~Y.}\ \bibnamefont {Liu}}, \bibinfo {author} {\bibfnamefont
  {M.}~\bibnamefont {Makela}}, \bibinfo {author} {\bibfnamefont {C.~L.}\
  \bibnamefont {Morris}}, \bibinfo {author} {\bibfnamefont {R.}~\bibnamefont
  {Musedinovic}}, \bibinfo {author} {\bibfnamefont {C.}~\bibnamefont
  {O’Shaughnessy}}, \bibinfo {author} {\bibfnamefont {R.~W.}\ \bibnamefont
  {Pattie}}, \bibinfo {author} {\bibfnamefont {J.}~\bibnamefont {Ramsey}},
  \bibinfo {author} {\bibfnamefont {D.~J.}\ \bibnamefont {Salvat}}, \bibinfo
  {author} {\bibfnamefont {A.}~\bibnamefont {Saunders}}, \bibinfo {author}
  {\bibfnamefont {E.~I.}\ \bibnamefont {Sharapov}}, \bibinfo {author}
  {\bibfnamefont {S.}~\bibnamefont {Slutsky}}, \bibinfo {author} {\bibfnamefont
  {V.}~\bibnamefont {Su}}, \bibinfo {author} {\bibfnamefont {X.}~\bibnamefont
  {Sun}}, \bibinfo {author} {\bibfnamefont {C.}~\bibnamefont {Swank}}, \bibinfo
  {author} {\bibfnamefont {Z.}~\bibnamefont {Tang}}, \bibinfo {author}
  {\bibfnamefont {W.}~\bibnamefont {Uhrich}}, \bibinfo {author} {\bibfnamefont
  {J.}~\bibnamefont {Vanderwerp}}, \bibinfo {author} {\bibfnamefont
  {P.}~\bibnamefont {Walstrom}}, \bibinfo {author} {\bibfnamefont
  {Z.}~\bibnamefont {Wang}}, \bibinfo {author} {\bibfnamefont {W.}~\bibnamefont
  {Wei}}, \ and\ \bibinfo {author} {\bibfnamefont {A.~R.}\ \bibnamefont
  {Young}},\ }\bibfield  {title} {\enquote {\bibinfo {title} {Improved neutron
  lifetime measurement with},}\ }\href {\doibase
  10.1103/PhysRevLett.127.162501} {\bibfield  {journal} {\bibinfo  {journal}
  {Physical Review Letters}\ }\textbf {\bibinfo {volume} {127}} (\bibinfo
  {year} {2021}),\ 10.1103/PhysRevLett.127.162501}\BibitemShut {NoStop}%
\bibitem [{\citenamefont {Byrne}\ \emph {et~al.}(1996)\citenamefont {Byrne},
  \citenamefont {Dawber}, \citenamefont {Habeck}, \citenamefont {Smidt},
  \citenamefont {Spain},\ and\ \citenamefont {Williams}}]{Byrne1996}%
  \BibitemOpen
  \bibfield  {author} {\bibinfo {author} {\bibfnamefont {J.}~\bibnamefont
  {Byrne}}, \bibinfo {author} {\bibfnamefont {P.~G.}\ \bibnamefont {Dawber}},
  \bibinfo {author} {\bibfnamefont {C.~G.}\ \bibnamefont {Habeck}}, \bibinfo
  {author} {\bibfnamefont {S.~J.}\ \bibnamefont {Smidt}}, \bibinfo {author}
  {\bibfnamefont {J.~A.}\ \bibnamefont {Spain}}, \ and\ \bibinfo {author}
  {\bibfnamefont {A.~P.}\ \bibnamefont {Williams}},\ }\bibfield  {title}
  {\enquote {\bibinfo {title} {A revised value for the neutron lifetime
  measured using a penning trap},}\ }\href {\doibase 10.1209/epl/i1996-00319-x}
  {\bibfield  {journal} {\bibinfo  {journal} {Europhysics Letters (EPL)}\
  }\textbf {\bibinfo {volume} {33}},\ \bibinfo {pages} {187--192} (\bibinfo
  {year} {1996})}\BibitemShut {NoStop}%
\bibitem [{\citenamefont {Nico}\ \emph {et~al.}(2005)\citenamefont {Nico},
  \citenamefont {Dewey}, \citenamefont {Gilliam}, \citenamefont {Wietfeldt},
  \citenamefont {Fei}, \citenamefont {Snow}, \citenamefont {Greene},
  \citenamefont {Pauwels}, \citenamefont {Eykens}, \citenamefont {Lamberty},
  \citenamefont {Gestel},\ and\ \citenamefont {Scott}}]{Nico2005}%
  \BibitemOpen
  \bibfield  {author} {\bibinfo {author} {\bibfnamefont {J.~S.}\ \bibnamefont
  {Nico}}, \bibinfo {author} {\bibfnamefont {M.~S.}\ \bibnamefont {Dewey}},
  \bibinfo {author} {\bibfnamefont {D.~M.}\ \bibnamefont {Gilliam}}, \bibinfo
  {author} {\bibfnamefont {F.~E.}\ \bibnamefont {Wietfeldt}}, \bibinfo {author}
  {\bibfnamefont {X.}~\bibnamefont {Fei}}, \bibinfo {author} {\bibfnamefont
  {W.~M.}\ \bibnamefont {Snow}}, \bibinfo {author} {\bibfnamefont {G.~L.}\
  \bibnamefont {Greene}}, \bibinfo {author} {\bibfnamefont {J.}~\bibnamefont
  {Pauwels}}, \bibinfo {author} {\bibfnamefont {R.}~\bibnamefont {Eykens}},
  \bibinfo {author} {\bibfnamefont {A.}~\bibnamefont {Lamberty}}, \bibinfo
  {author} {\bibfnamefont {J.~V.}\ \bibnamefont {Gestel}}, \ and\ \bibinfo
  {author} {\bibfnamefont {R.~D.}\ \bibnamefont {Scott}},\ }\bibfield  {title}
  {\enquote {\bibinfo {title} {Measurement of the neutron lifetime by counting
  trapped protons in a cold neutron beam},}\ }\href {\doibase
  10.1103/PhysRevC.71.055502} {\bibfield  {journal} {\bibinfo  {journal}
  {Physical Review C}\ }\textbf {\bibinfo {volume} {71}},\ \bibinfo {pages}
  {055502} (\bibinfo {year} {2005})}\BibitemShut {NoStop}%
\bibitem [{\citenamefont {Yue}\ \emph {et~al.}(2013)\citenamefont {Yue},
  \citenamefont {Dewey}, \citenamefont {Gilliam}, \citenamefont {Greene},
  \citenamefont {Laptev}, \citenamefont {Nico}, \citenamefont {Snow},\ and\
  \citenamefont {Wietfeldt}}]{Yue2013}%
  \BibitemOpen
  \bibfield  {author} {\bibinfo {author} {\bibfnamefont {A.~T.}\ \bibnamefont
  {Yue}}, \bibinfo {author} {\bibfnamefont {M.~S.}\ \bibnamefont {Dewey}},
  \bibinfo {author} {\bibfnamefont {D.~M.}\ \bibnamefont {Gilliam}}, \bibinfo
  {author} {\bibfnamefont {G.~L.}\ \bibnamefont {Greene}}, \bibinfo {author}
  {\bibfnamefont {A.~B.}\ \bibnamefont {Laptev}}, \bibinfo {author}
  {\bibfnamefont {J.~S.}\ \bibnamefont {Nico}}, \bibinfo {author}
  {\bibfnamefont {W.~M.}\ \bibnamefont {Snow}}, \ and\ \bibinfo {author}
  {\bibfnamefont {F.~E.}\ \bibnamefont {Wietfeldt}},\ }\bibfield  {title}
  {\enquote {\bibinfo {title} {Improved determination of the neutron
  lifetime},}\ }\href {\doibase 10.1103/PhysRevLett.111.222501} {\bibfield
  {journal} {\bibinfo  {journal} {Physical Review Letters}\ }\textbf {\bibinfo
  {volume} {111}},\ \bibinfo {pages} {222501} (\bibinfo {year}
  {2013})}\BibitemShut {NoStop}%
\bibitem [{\citenamefont {Byrne}\ and\ \citenamefont
  {Worcester}(2019)}]{Byrne2019}%
  \BibitemOpen
  \bibfield  {author} {\bibinfo {author} {\bibfnamefont {J.}~\bibnamefont
  {Byrne}}\ and\ \bibinfo {author} {\bibfnamefont {D.~L.}\ \bibnamefont
  {Worcester}},\ }\bibfield  {title} {\enquote {\bibinfo {title} {The neutron
  lifetime anomaly and charge exchange collisions of trapped protons*},}\
  }\href {\doibase 10.1088/1361-6471/AB256B} {\bibfield  {journal} {\bibinfo
  {journal} {Journal of Physics G: Nuclear and Particle Physics}\ }\textbf
  {\bibinfo {volume} {46}},\ \bibinfo {pages} {085001} (\bibinfo {year}
  {2019})}\BibitemShut {NoStop}%
\bibitem [{\citenamefont {Serebrov}\ \emph {et~al.}(2021)\citenamefont
  {Serebrov}, \citenamefont {Chaikovskii}, \citenamefont {Klyushnikov},
  \citenamefont {Zherebtsov},\ and\ \citenamefont {Chechkin}}]{Serebrov2021}%
  \BibitemOpen
  \bibfield  {author} {\bibinfo {author} {\bibfnamefont {A.~P.}\ \bibnamefont
  {Serebrov}}, \bibinfo {author} {\bibfnamefont {M.~E.}\ \bibnamefont
  {Chaikovskii}}, \bibinfo {author} {\bibfnamefont {G.~N.}\ \bibnamefont
  {Klyushnikov}}, \bibinfo {author} {\bibfnamefont {O.~M.}\ \bibnamefont
  {Zherebtsov}}, \ and\ \bibinfo {author} {\bibfnamefont {A.~V.}\ \bibnamefont
  {Chechkin}},\ }\bibfield  {title} {\enquote {\bibinfo {title} {Search for
  explanation of the neutron lifetime anomaly},}\ }\href {\doibase
  10.1103/PhysRevD.103.074010} {\bibfield  {journal} {\bibinfo  {journal}
  {Physical Review D}\ }\textbf {\bibinfo {volume} {103}},\ \bibinfo {pages}
  {074010} (\bibinfo {year} {2021})}\BibitemShut {NoStop}%
\bibitem [{\citenamefont {Fornal}\ and\ \citenamefont
  {Grinstein}(2018)}]{Fornal2018}%
  \BibitemOpen
  \bibfield  {author} {\bibinfo {author} {\bibfnamefont {B.}~\bibnamefont
  {Fornal}}\ and\ \bibinfo {author} {\bibfnamefont {B.}~\bibnamefont
  {Grinstein}},\ }\bibfield  {title} {\enquote {\bibinfo {title} {Dark matter
  interpretation of the neutron decay anomaly},}\ }\href {\doibase
  10.1103/PhysRevLett.120.191801} {\bibfield  {journal} {\bibinfo  {journal}
  {Physical Review Letters}\ }\textbf {\bibinfo {volume} {120}},\ \bibinfo
  {pages} {191801} (\bibinfo {year} {2018})}\BibitemShut {NoStop}%
\bibitem [{\citenamefont {Berezhiani}(2009)}]{Berezhiani2009}%
  \BibitemOpen
  \bibfield  {author} {\bibinfo {author} {\bibfnamefont {Z.}~\bibnamefont
  {Berezhiani}},\ }\bibfield  {title} {\enquote {\bibinfo {title} {More about
  neutron–mirror neutron oscillation},}\ }\href {\doibase
  10.1140/epjc/s10052-009-1165-1} {\bibfield  {journal} {\bibinfo  {journal}
  {The European Physical Journal C}\ }\textbf {\bibinfo {volume} {64}},\
  \bibinfo {pages} {421--431} (\bibinfo {year} {2009})}\BibitemShut {NoStop}%
\bibitem [{\citenamefont {Berezhiani}(2019)}]{Berezhiani2019}%
  \BibitemOpen
  \bibfield  {author} {\bibinfo {author} {\bibfnamefont {Z.}~\bibnamefont
  {Berezhiani}},\ }\bibfield  {title} {\enquote {\bibinfo {title} {Neutron
  lifetime puzzle and neutron–mirror neutron oscillation},}\ }\href {\doibase
  10.1140/epjc/s10052-019-6995-x} {\bibfield  {journal} {\bibinfo  {journal}
  {The European Physical Journal C}\ }\textbf {\bibinfo {volume} {79}},\
  \bibinfo {pages} {484} (\bibinfo {year} {2019})}\BibitemShut {NoStop}%
\bibitem [{\citenamefont {Broussard}\ \emph {et~al.}(2022)\citenamefont
  {Broussard}, \citenamefont {Barrow}, \citenamefont {Debeer-Schmitt},
  \citenamefont {Dennis}, \citenamefont {Fitzsimmons}, \citenamefont {Frost},
  \citenamefont {Gilbert}, \citenamefont {Gonzalez}, \citenamefont {Heilbronn},
  \citenamefont {Iverson}, \citenamefont {Johnston}, \citenamefont {Kamyshkov},
  \citenamefont {Kline}, \citenamefont {Lewiz}, \citenamefont {Matteson},
  \citenamefont {Ternullo}, \citenamefont {Varriano},\ and\ \citenamefont
  {Vavra}}]{Broussard2022}%
  \BibitemOpen
  \bibfield  {author} {\bibinfo {author} {\bibfnamefont {L.~J.}\ \bibnamefont
  {Broussard}}, \bibinfo {author} {\bibfnamefont {J.~L.}\ \bibnamefont
  {Barrow}}, \bibinfo {author} {\bibfnamefont {L.}~\bibnamefont
  {Debeer-Schmitt}}, \bibinfo {author} {\bibfnamefont {T.}~\bibnamefont
  {Dennis}}, \bibinfo {author} {\bibfnamefont {M.~R.}\ \bibnamefont
  {Fitzsimmons}}, \bibinfo {author} {\bibfnamefont {M.~J.}\ \bibnamefont
  {Frost}}, \bibinfo {author} {\bibfnamefont {C.~E.}\ \bibnamefont {Gilbert}},
  \bibinfo {author} {\bibfnamefont {F.~M.}\ \bibnamefont {Gonzalez}}, \bibinfo
  {author} {\bibfnamefont {L.}~\bibnamefont {Heilbronn}}, \bibinfo {author}
  {\bibfnamefont {E.~B.}\ \bibnamefont {Iverson}}, \bibinfo {author}
  {\bibfnamefont {A.}~\bibnamefont {Johnston}}, \bibinfo {author}
  {\bibfnamefont {Y.}~\bibnamefont {Kamyshkov}}, \bibinfo {author}
  {\bibfnamefont {M.}~\bibnamefont {Kline}}, \bibinfo {author} {\bibfnamefont
  {P.}~\bibnamefont {Lewiz}}, \bibinfo {author} {\bibfnamefont
  {C.}~\bibnamefont {Matteson}}, \bibinfo {author} {\bibfnamefont
  {J.}~\bibnamefont {Ternullo}}, \bibinfo {author} {\bibfnamefont
  {L.}~\bibnamefont {Varriano}}, \ and\ \bibinfo {author} {\bibfnamefont
  {S.}~\bibnamefont {Vavra}},\ }\bibfield  {title} {\enquote {\bibinfo {title}
  {Experimental search for neutron to mirror neutron oscillations as an
  explanation of the neutron lifetime anomaly},}\ }\href {\doibase
  10.1103/PhysRevLett.128.212503} {\bibfield  {journal} {\bibinfo  {journal}
  {Physical Review Letters}\ }\textbf {\bibinfo {volume} {128}} (\bibinfo
  {year} {2022}),\ 10.1103/PhysRevLett.128.212503}\BibitemShut {NoStop}%
\bibitem [{\citenamefont {Tang}\ \emph {et~al.}(2018)\citenamefont {Tang},
  \citenamefont {Blatnik}, \citenamefont {Broussard}, \citenamefont {Choi},
  \citenamefont {Clayton}, \citenamefont {Cude-Woods}, \citenamefont {Currie},
  \citenamefont {Fellers}, \citenamefont {Fries}, \citenamefont {Geltenbort},
  \citenamefont {Gonzalez}, \citenamefont {Hickerson}, \citenamefont {Ito},
  \citenamefont {Liu}, \citenamefont {Macdonald}, \citenamefont {Makela},
  \citenamefont {Morris}, \citenamefont {O'Shaughnessy}, \citenamefont
  {Pattie}, \citenamefont {Plaster}, \citenamefont {Salvat}, \citenamefont
  {Saunders}, \citenamefont {Wang}, \citenamefont {Young},\ and\ \citenamefont
  {Zeck}}]{Tang2018}%
  \BibitemOpen
  \bibfield  {author} {\bibinfo {author} {\bibfnamefont {Z.}~\bibnamefont
  {Tang}}, \bibinfo {author} {\bibfnamefont {M.}~\bibnamefont {Blatnik}},
  \bibinfo {author} {\bibfnamefont {L.~J.}\ \bibnamefont {Broussard}}, \bibinfo
  {author} {\bibfnamefont {J.~H.}\ \bibnamefont {Choi}}, \bibinfo {author}
  {\bibfnamefont {S.~M.}\ \bibnamefont {Clayton}}, \bibinfo {author}
  {\bibfnamefont {C.}~\bibnamefont {Cude-Woods}}, \bibinfo {author}
  {\bibfnamefont {S.}~\bibnamefont {Currie}}, \bibinfo {author} {\bibfnamefont
  {D.~E.}\ \bibnamefont {Fellers}}, \bibinfo {author} {\bibfnamefont {E.~M.}\
  \bibnamefont {Fries}}, \bibinfo {author} {\bibfnamefont {P.}~\bibnamefont
  {Geltenbort}}, \bibinfo {author} {\bibfnamefont {F.}~\bibnamefont
  {Gonzalez}}, \bibinfo {author} {\bibfnamefont {K.~P.}\ \bibnamefont
  {Hickerson}}, \bibinfo {author} {\bibfnamefont {T.~M.}\ \bibnamefont {Ito}},
  \bibinfo {author} {\bibfnamefont {C.~Y.}\ \bibnamefont {Liu}}, \bibinfo
  {author} {\bibfnamefont {S.~W.}\ \bibnamefont {Macdonald}}, \bibinfo {author}
  {\bibfnamefont {M.}~\bibnamefont {Makela}}, \bibinfo {author} {\bibfnamefont
  {C.~L.}\ \bibnamefont {Morris}}, \bibinfo {author} {\bibfnamefont {C.~M.}\
  \bibnamefont {O'Shaughnessy}}, \bibinfo {author} {\bibfnamefont {R.~W.}\
  \bibnamefont {Pattie}}, \bibinfo {author} {\bibfnamefont {B.}~\bibnamefont
  {Plaster}}, \bibinfo {author} {\bibfnamefont {D.~J.}\ \bibnamefont {Salvat}},
  \bibinfo {author} {\bibfnamefont {A.}~\bibnamefont {Saunders}}, \bibinfo
  {author} {\bibfnamefont {Z.}~\bibnamefont {Wang}}, \bibinfo {author}
  {\bibfnamefont {A.~R.}\ \bibnamefont {Young}}, \ and\ \bibinfo {author}
  {\bibfnamefont {B.~A.}\ \bibnamefont {Zeck}},\ }\bibfield  {title} {\enquote
  {\bibinfo {title} {Search for the neutron decay n→x+gamma, where x is a
  dark matter particle},}\ }\href {\doibase 10.1103/PhysRevLett.121.022505}
  {\bibfield  {journal} {\bibinfo  {journal} {Physical Review Letters}\
  }\textbf {\bibinfo {volume} {121}} (\bibinfo {year} {2018}),\
  10.1103/PhysRevLett.121.022505}\BibitemShut {NoStop}%
\bibitem [{\citenamefont {Sun}\ \emph {et~al.}(2018)\citenamefont {Sun},
  \citenamefont {Adamek}, \citenamefont {Allgeier}, \citenamefont {Blatnik},
  \citenamefont {Bowles}, \citenamefont {Broussard}, \citenamefont {Brown},
  \citenamefont {Carr}, \citenamefont {Clayton}, \citenamefont {Cude-Woods},
  \citenamefont {Currie}, \citenamefont {Dees}, \citenamefont {Ding},
  \citenamefont {Filippone}, \citenamefont {García}, \citenamefont
  {Geltenbort}, \citenamefont {Hasan}, \citenamefont {Hickerson}, \citenamefont
  {Hoagland}, \citenamefont {Hong}, \citenamefont {Hogan}, \citenamefont
  {Holley}, \citenamefont {Ito}, \citenamefont {Knecht}, \citenamefont {Liu},
  \citenamefont {Liu}, \citenamefont {Makela}, \citenamefont {Mammei},
  \citenamefont {Martin}, \citenamefont {Melconian}, \citenamefont
  {Mendenhall}, \citenamefont {Moore}, \citenamefont {Morris}, \citenamefont
  {Nepal}, \citenamefont {Nouri}, \citenamefont {Pattie}, \citenamefont
  {Galván}, \citenamefont {Phillips}, \citenamefont {Picker}, \citenamefont
  {Pitt}, \citenamefont {Plaster}, \citenamefont {Ramsey}, \citenamefont
  {Rios}, \citenamefont {Salvat}, \citenamefont {Saunders}, \citenamefont
  {Sondheim}, \citenamefont {Sjue}, \citenamefont {Slutsky}, \citenamefont
  {Swank}, \citenamefont {Swift}, \citenamefont {Tatar}, \citenamefont
  {Vogelaar}, \citenamefont {Vorndick}, \citenamefont {Wang}, \citenamefont
  {Wei}, \citenamefont {Wexler}, \citenamefont {Womack}, \citenamefont {Wrede},
  \citenamefont {Young},\ and\ \citenamefont {Zeck}}]{Sun2018}%
  \BibitemOpen
  \bibfield  {author} {\bibinfo {author} {\bibfnamefont {X.}~\bibnamefont
  {Sun}}, \bibinfo {author} {\bibfnamefont {E.}~\bibnamefont {Adamek}},
  \bibinfo {author} {\bibfnamefont {B.}~\bibnamefont {Allgeier}}, \bibinfo
  {author} {\bibfnamefont {M.}~\bibnamefont {Blatnik}}, \bibinfo {author}
  {\bibfnamefont {T.~J.}\ \bibnamefont {Bowles}}, \bibinfo {author}
  {\bibfnamefont {L.~J.}\ \bibnamefont {Broussard}}, \bibinfo {author}
  {\bibfnamefont {M.~A.}\ \bibnamefont {Brown}}, \bibinfo {author}
  {\bibfnamefont {R.}~\bibnamefont {Carr}}, \bibinfo {author} {\bibfnamefont
  {S.}~\bibnamefont {Clayton}}, \bibinfo {author} {\bibfnamefont
  {C.}~\bibnamefont {Cude-Woods}}, \bibinfo {author} {\bibfnamefont
  {S.}~\bibnamefont {Currie}}, \bibinfo {author} {\bibfnamefont {E.~B.}\
  \bibnamefont {Dees}}, \bibinfo {author} {\bibfnamefont {X.}~\bibnamefont
  {Ding}}, \bibinfo {author} {\bibfnamefont {B.~W.}\ \bibnamefont {Filippone}},
  \bibinfo {author} {\bibfnamefont {A.}~\bibnamefont {García}}, \bibinfo
  {author} {\bibfnamefont {P.}~\bibnamefont {Geltenbort}}, \bibinfo {author}
  {\bibfnamefont {S.}~\bibnamefont {Hasan}}, \bibinfo {author} {\bibfnamefont
  {K.~P.}\ \bibnamefont {Hickerson}}, \bibinfo {author} {\bibfnamefont
  {J.}~\bibnamefont {Hoagland}}, \bibinfo {author} {\bibfnamefont
  {R.}~\bibnamefont {Hong}}, \bibinfo {author} {\bibfnamefont {G.~E.}\
  \bibnamefont {Hogan}}, \bibinfo {author} {\bibfnamefont {A.~T.}\ \bibnamefont
  {Holley}}, \bibinfo {author} {\bibfnamefont {T.~M.}\ \bibnamefont {Ito}},
  \bibinfo {author} {\bibfnamefont {A.}~\bibnamefont {Knecht}}, \bibinfo
  {author} {\bibfnamefont {C.~Y.}\ \bibnamefont {Liu}}, \bibinfo {author}
  {\bibfnamefont {J.}~\bibnamefont {Liu}}, \bibinfo {author} {\bibfnamefont
  {M.}~\bibnamefont {Makela}}, \bibinfo {author} {\bibfnamefont
  {R.}~\bibnamefont {Mammei}}, \bibinfo {author} {\bibfnamefont {J.~W.}\
  \bibnamefont {Martin}}, \bibinfo {author} {\bibfnamefont {D.}~\bibnamefont
  {Melconian}}, \bibinfo {author} {\bibfnamefont {M.~P.}\ \bibnamefont
  {Mendenhall}}, \bibinfo {author} {\bibfnamefont {S.~D.}\ \bibnamefont
  {Moore}}, \bibinfo {author} {\bibfnamefont {C.~L.}\ \bibnamefont {Morris}},
  \bibinfo {author} {\bibfnamefont {S.}~\bibnamefont {Nepal}}, \bibinfo
  {author} {\bibfnamefont {N.}~\bibnamefont {Nouri}}, \bibinfo {author}
  {\bibfnamefont {R.~W.}\ \bibnamefont {Pattie}}, \bibinfo {author}
  {\bibfnamefont {A.~P.}\ \bibnamefont {Galván}}, \bibinfo {author}
  {\bibfnamefont {D.~G.}\ \bibnamefont {Phillips}}, \bibinfo {author}
  {\bibfnamefont {R.}~\bibnamefont {Picker}}, \bibinfo {author} {\bibfnamefont
  {M.~L.}\ \bibnamefont {Pitt}}, \bibinfo {author} {\bibfnamefont
  {B.}~\bibnamefont {Plaster}}, \bibinfo {author} {\bibfnamefont {J.~C.}\
  \bibnamefont {Ramsey}}, \bibinfo {author} {\bibfnamefont {R.}~\bibnamefont
  {Rios}}, \bibinfo {author} {\bibfnamefont {D.~J.}\ \bibnamefont {Salvat}},
  \bibinfo {author} {\bibfnamefont {A.}~\bibnamefont {Saunders}}, \bibinfo
  {author} {\bibfnamefont {W.}~\bibnamefont {Sondheim}}, \bibinfo {author}
  {\bibfnamefont {S.}~\bibnamefont {Sjue}}, \bibinfo {author} {\bibfnamefont
  {S.}~\bibnamefont {Slutsky}}, \bibinfo {author} {\bibfnamefont
  {C.}~\bibnamefont {Swank}}, \bibinfo {author} {\bibfnamefont
  {G.}~\bibnamefont {Swift}}, \bibinfo {author} {\bibfnamefont
  {E.}~\bibnamefont {Tatar}}, \bibinfo {author} {\bibfnamefont {R.~B.}\
  \bibnamefont {Vogelaar}}, \bibinfo {author} {\bibfnamefont {B.}~\bibnamefont
  {Vorndick}}, \bibinfo {author} {\bibfnamefont {Z.}~\bibnamefont {Wang}},
  \bibinfo {author} {\bibfnamefont {W.}~\bibnamefont {Wei}}, \bibinfo {author}
  {\bibfnamefont {J.}~\bibnamefont {Wexler}}, \bibinfo {author} {\bibfnamefont
  {T.}~\bibnamefont {Womack}}, \bibinfo {author} {\bibfnamefont
  {C.}~\bibnamefont {Wrede}}, \bibinfo {author} {\bibfnamefont {A.~R.}\
  \bibnamefont {Young}}, \ and\ \bibinfo {author} {\bibfnamefont {B.~A.}\
  \bibnamefont {Zeck}},\ }\bibfield  {title} {\enquote {\bibinfo {title}
  {Search for dark matter decay of the free neutron from the ucna experiment: N
  $\rightarrow$ $\chi$ +$e^+e^-$},}\ }\href {\doibase
  10.1103/PhysRevC.97.052501} {\bibfield  {journal} {\bibinfo  {journal}
  {Physical Review C}\ }\textbf {\bibinfo {volume} {97}},\ \bibinfo {pages}
  {052501} (\bibinfo {year} {2018})}\BibitemShut {NoStop}%
\bibitem [{\citenamefont {Tanabashi}\ \emph {et~al.}(2018)\citenamefont
  {Tanabashi}, \citenamefont {Hagiwara}, \citenamefont {Hikasa}, \citenamefont
  {Nakamura}, \citenamefont {Sumino}, \citenamefont {Takahashi}, \citenamefont
  {Tanaka}, \citenamefont {Agashe}, \citenamefont {Aielli}, \citenamefont
  {Amsler}, \citenamefont {Antonelli}, \citenamefont {Asner}, \citenamefont
  {Baer}, \citenamefont {Banerjee}, \citenamefont {Barnett}, \citenamefont
  {Basaglia}, \citenamefont {Bauer}, \citenamefont {Beatty}, \citenamefont
  {Belousov}, \citenamefont {Beringer}, \citenamefont {Bethke}, \citenamefont
  {Bettini}, \citenamefont {Bichsel}, \citenamefont {Biebel}, \citenamefont
  {Black}, \citenamefont {Blucher}, \citenamefont {Buchmuller}, \citenamefont
  {Burkert}, \citenamefont {Bychkov}, \citenamefont {Cahn}, \citenamefont
  {Carena}, \citenamefont {Ceccucci}, \citenamefont {Cerri}, \citenamefont
  {Chakraborty}, \citenamefont {Chen}, \citenamefont {Chivukula}, \citenamefont
  {Cowan}, \citenamefont {Dahl}, \citenamefont {D'Ambrosio}, \citenamefont
  {Damour}, \citenamefont {Florian}, \citenamefont {Gouvêa}, \citenamefont
  {Degrand}, \citenamefont {Jong}, \citenamefont {Dissertori}, \citenamefont
  {Dobrescu}, \citenamefont {D'Onofrio}, \citenamefont {Doser}, \citenamefont
  {Drees}, \citenamefont {Dreiner}, \citenamefont {Dwyer}, \citenamefont
  {Eerola}, \citenamefont {Eidelman}, \citenamefont {Ellis}, \citenamefont
  {Erler}, \citenamefont {Ezhela}, \citenamefont {Fetscher}, \citenamefont
  {Fields}, \citenamefont {Firestone}, \citenamefont {Foster}, \citenamefont
  {Freitas}, \citenamefont {Gallagher}, \citenamefont {Garren}, \citenamefont
  {Gerber}, \citenamefont {Gerbier}, \citenamefont {Gershon}, \citenamefont
  {Gershtein}, \citenamefont {Gherghetta}, \citenamefont {Godizov},
  \citenamefont {Goodman}, \citenamefont {Grab}, \citenamefont {Gritsan},
  \citenamefont {Grojean}, \citenamefont {Groom}, \citenamefont {Grünewald},
  \citenamefont {Gurtu}, \citenamefont {Gutsche}, \citenamefont {Haber},
  \citenamefont {Hanhart}, \citenamefont {Hashimoto}, \citenamefont {Hayato},
  \citenamefont {Hayes}, \citenamefont {Hebecker}, \citenamefont {Heinemeyer},
  \citenamefont {Heltsley}, \citenamefont {Hernández-Rey}, \citenamefont
  {Hisano}, \citenamefont {Höcker}, \citenamefont {Holder}, \citenamefont
  {Holtkamp}, \citenamefont {Hyodo}, \citenamefont {Irwin}, \citenamefont
  {Johnson}, \citenamefont {Kado}, \citenamefont {Karliner}, \citenamefont
  {Katz}, \citenamefont {Klein}, \citenamefont {Klempt}, \citenamefont
  {Kowalewski}, \citenamefont {Krauss}, \citenamefont {Kreps}, \citenamefont
  {Krusche}, \citenamefont {Kuyanov}, \citenamefont {Kwon}, \citenamefont
  {Lahav}, \citenamefont {Laiho}, \citenamefont {Lesgourgues}, \citenamefont
  {Liddle}, \citenamefont {Ligeti}, \citenamefont {Lin}, \citenamefont
  {Lippmann}, \citenamefont {Liss}, \citenamefont {Littenberg}, \citenamefont
  {Lugovsky}, \citenamefont {Lugovsky}, \citenamefont {Lusiani}, \citenamefont
  {Makida}, \citenamefont {Maltoni}, \citenamefont {Mannel}, \citenamefont
  {Manohar}, \citenamefont {Marciano}, \citenamefont {Martin}, \citenamefont
  {Masoni}, \citenamefont {Matthews}, \citenamefont {Meißner}, \citenamefont
  {Milstead}, \citenamefont {Mitchell}, \citenamefont {Mönig}, \citenamefont
  {Molaro}, \citenamefont {Moortgat}, \citenamefont {Moskovic}, \citenamefont
  {Murayama}, \citenamefont {Narain}, \citenamefont {Nason}, \citenamefont
  {Navas}, \citenamefont {Neubert}, \citenamefont {Nevski}, \citenamefont
  {Nir}, \citenamefont {Olive}, \citenamefont {Griso}, \citenamefont {Parsons},
  \citenamefont {Patrignani}, \citenamefont {Peacock}, \citenamefont
  {Pennington}, \citenamefont {Petcov}, \citenamefont {Petrov}, \citenamefont
  {Pianori}, \citenamefont {Piepke}, \citenamefont {Pomarol}, \citenamefont
  {Quadt}, \citenamefont {Rademacker}, \citenamefont {Raffelt}, \citenamefont
  {Ratcliff}, \citenamefont {Richardson}, \citenamefont {Ringwald},
  \citenamefont {Roesler}, \citenamefont {Rolli}, \citenamefont {Romaniouk},
  \citenamefont {Rosenberg}, \citenamefont {Rosner}, \citenamefont {Rybka},
  \citenamefont {Ryutin}, \citenamefont {Sachrajda}, \citenamefont {Sakai},
  \citenamefont {Salam}, \citenamefont {Sarkar}, \citenamefont {Sauli},
  \citenamefont {Schneider}, \citenamefont {Scholberg}, \citenamefont
  {Schwartz}, \citenamefont {Scott}, \citenamefont {Sharma}, \citenamefont
  {Sharpe}, \citenamefont {Shutt}, \citenamefont {Silari}, \citenamefont
  {Sjöstrand}, \citenamefont {Skands}, \citenamefont {Skwarnicki},
  \citenamefont {Smith}, \citenamefont {Smoot}, \citenamefont {Spanier},
  \citenamefont {Spieler}, \citenamefont {Spiering}, \citenamefont {Stahl},
  \citenamefont {Stone}, \citenamefont {Sumiyoshi}, \citenamefont {Syphers},
  \citenamefont {Terashi}, \citenamefont {Terning}, \citenamefont {Thoma},
  \citenamefont {Thorne}, \citenamefont {Tiator}, \citenamefont {Titov},
  \citenamefont {Tkachenko}, \citenamefont {Törnqvist}, \citenamefont {Tovey},
  \citenamefont {Valencia}, \citenamefont {Water}, \citenamefont {Varelas},
  \citenamefont {Venanzoni}, \citenamefont {Verde}, \citenamefont {Vincter},
  \citenamefont {Vogel}, \citenamefont {Vogt}, \citenamefont {Wakely},
  \citenamefont {Walkowiak}, \citenamefont {Walter}, \citenamefont {Wands},
  \citenamefont {Ward}, \citenamefont {Wascko}, \citenamefont {Weiglein},
  \citenamefont {Weinberg}, \citenamefont {Weinberg}, \citenamefont {White},
  \citenamefont {Wiencke}, \citenamefont {Willocq}, \citenamefont {Wohl},
  \citenamefont {Womersley}, \citenamefont {Woody}, \citenamefont {Workman},
  \citenamefont {Yao}, \citenamefont {Zeller}, \citenamefont {Zenin},
  \citenamefont {Zhu}, \citenamefont {Zhu}, \citenamefont {Zimmermann},
  \citenamefont {Zyla}, \citenamefont {Anderson}, \citenamefont {Fuller},
  \citenamefont {Lugovsky},\ and\ \citenamefont {Schaffner}}]{Tanabashi2018}%
  \BibitemOpen
  \bibfield  {author} {\bibinfo {author} {\bibfnamefont {M.}~\bibnamefont
  {Tanabashi}}, \bibinfo {author} {\bibfnamefont {K.}~\bibnamefont {Hagiwara}},
  \bibinfo {author} {\bibfnamefont {K.}~\bibnamefont {Hikasa}}, \bibinfo
  {author} {\bibfnamefont {K.}~\bibnamefont {Nakamura}}, \bibinfo {author}
  {\bibfnamefont {Y.}~\bibnamefont {Sumino}}, \bibinfo {author} {\bibfnamefont
  {F.}~\bibnamefont {Takahashi}}, \bibinfo {author} {\bibfnamefont
  {J.}~\bibnamefont {Tanaka}}, \bibinfo {author} {\bibfnamefont
  {K.}~\bibnamefont {Agashe}}, \bibinfo {author} {\bibfnamefont
  {G.}~\bibnamefont {Aielli}}, \bibinfo {author} {\bibfnamefont
  {C.}~\bibnamefont {Amsler}}, \bibinfo {author} {\bibfnamefont
  {M.}~\bibnamefont {Antonelli}}, \bibinfo {author} {\bibfnamefont {D.~M.}\
  \bibnamefont {Asner}}, \bibinfo {author} {\bibfnamefont {H.}~\bibnamefont
  {Baer}}, \bibinfo {author} {\bibfnamefont {S.}~\bibnamefont {Banerjee}},
  \bibinfo {author} {\bibfnamefont {R.~M.}\ \bibnamefont {Barnett}}, \bibinfo
  {author} {\bibfnamefont {T.}~\bibnamefont {Basaglia}}, \bibinfo {author}
  {\bibfnamefont {C.~W.}\ \bibnamefont {Bauer}}, \bibinfo {author}
  {\bibfnamefont {J.~J.}\ \bibnamefont {Beatty}}, \bibinfo {author}
  {\bibfnamefont {V.~I.}\ \bibnamefont {Belousov}}, \bibinfo {author}
  {\bibfnamefont {J.}~\bibnamefont {Beringer}}, \bibinfo {author}
  {\bibfnamefont {S.}~\bibnamefont {Bethke}}, \bibinfo {author} {\bibfnamefont
  {A.}~\bibnamefont {Bettini}}, \bibinfo {author} {\bibfnamefont
  {H.}~\bibnamefont {Bichsel}}, \bibinfo {author} {\bibfnamefont
  {O.}~\bibnamefont {Biebel}}, \bibinfo {author} {\bibfnamefont {K.~M.}\
  \bibnamefont {Black}}, \bibinfo {author} {\bibfnamefont {E.}~\bibnamefont
  {Blucher}}, \bibinfo {author} {\bibfnamefont {O.}~\bibnamefont {Buchmuller}},
  \bibinfo {author} {\bibfnamefont {V.}~\bibnamefont {Burkert}}, \bibinfo
  {author} {\bibfnamefont {M.~A.}\ \bibnamefont {Bychkov}}, \bibinfo {author}
  {\bibfnamefont {R.~N.}\ \bibnamefont {Cahn}}, \bibinfo {author}
  {\bibfnamefont {M.}~\bibnamefont {Carena}}, \bibinfo {author} {\bibfnamefont
  {A.}~\bibnamefont {Ceccucci}}, \bibinfo {author} {\bibfnamefont
  {A.}~\bibnamefont {Cerri}}, \bibinfo {author} {\bibfnamefont
  {D.}~\bibnamefont {Chakraborty}}, \bibinfo {author} {\bibfnamefont {M.~C.}\
  \bibnamefont {Chen}}, \bibinfo {author} {\bibfnamefont {R.~S.}\ \bibnamefont
  {Chivukula}}, \bibinfo {author} {\bibfnamefont {G.}~\bibnamefont {Cowan}},
  \bibinfo {author} {\bibfnamefont {O.}~\bibnamefont {Dahl}}, \bibinfo {author}
  {\bibfnamefont {G.}~\bibnamefont {D'Ambrosio}}, \bibinfo {author}
  {\bibfnamefont {T.}~\bibnamefont {Damour}}, \bibinfo {author} {\bibfnamefont
  {D.~D.}\ \bibnamefont {Florian}}, \bibinfo {author} {\bibfnamefont {A.~D.}\
  \bibnamefont {Gouvêa}}, \bibinfo {author} {\bibfnamefont {T.}~\bibnamefont
  {Degrand}}, \bibinfo {author} {\bibfnamefont {P.~D.}\ \bibnamefont {Jong}},
  \bibinfo {author} {\bibfnamefont {G.}~\bibnamefont {Dissertori}}, \bibinfo
  {author} {\bibfnamefont {B.~A.}\ \bibnamefont {Dobrescu}}, \bibinfo {author}
  {\bibfnamefont {M.}~\bibnamefont {D'Onofrio}}, \bibinfo {author}
  {\bibfnamefont {M.}~\bibnamefont {Doser}}, \bibinfo {author} {\bibfnamefont
  {M.}~\bibnamefont {Drees}}, \bibinfo {author} {\bibfnamefont {H.~K.}\
  \bibnamefont {Dreiner}}, \bibinfo {author} {\bibfnamefont {D.~A.}\
  \bibnamefont {Dwyer}}, \bibinfo {author} {\bibfnamefont {P.}~\bibnamefont
  {Eerola}}, \bibinfo {author} {\bibfnamefont {S.}~\bibnamefont {Eidelman}},
  \bibinfo {author} {\bibfnamefont {J.}~\bibnamefont {Ellis}}, \bibinfo
  {author} {\bibfnamefont {J.}~\bibnamefont {Erler}}, \bibinfo {author}
  {\bibfnamefont {V.~V.}\ \bibnamefont {Ezhela}}, \bibinfo {author}
  {\bibfnamefont {W.}~\bibnamefont {Fetscher}}, \bibinfo {author}
  {\bibfnamefont {B.~D.}\ \bibnamefont {Fields}}, \bibinfo {author}
  {\bibfnamefont {R.}~\bibnamefont {Firestone}}, \bibinfo {author}
  {\bibfnamefont {B.}~\bibnamefont {Foster}}, \bibinfo {author} {\bibfnamefont
  {A.}~\bibnamefont {Freitas}}, \bibinfo {author} {\bibfnamefont
  {H.}~\bibnamefont {Gallagher}}, \bibinfo {author} {\bibfnamefont
  {L.}~\bibnamefont {Garren}}, \bibinfo {author} {\bibfnamefont {H.~J.}\
  \bibnamefont {Gerber}}, \bibinfo {author} {\bibfnamefont {G.}~\bibnamefont
  {Gerbier}}, \bibinfo {author} {\bibfnamefont {T.}~\bibnamefont {Gershon}},
  \bibinfo {author} {\bibfnamefont {Y.}~\bibnamefont {Gershtein}}, \bibinfo
  {author} {\bibfnamefont {T.}~\bibnamefont {Gherghetta}}, \bibinfo {author}
  {\bibfnamefont {A.~A.}\ \bibnamefont {Godizov}}, \bibinfo {author}
  {\bibfnamefont {M.}~\bibnamefont {Goodman}}, \bibinfo {author} {\bibfnamefont
  {C.}~\bibnamefont {Grab}}, \bibinfo {author} {\bibfnamefont {A.~V.}\
  \bibnamefont {Gritsan}}, \bibinfo {author} {\bibfnamefont {C.}~\bibnamefont
  {Grojean}}, \bibinfo {author} {\bibfnamefont {D.~E.}\ \bibnamefont {Groom}},
  \bibinfo {author} {\bibfnamefont {M.}~\bibnamefont {Grünewald}}, \bibinfo
  {author} {\bibfnamefont {A.}~\bibnamefont {Gurtu}}, \bibinfo {author}
  {\bibfnamefont {T.}~\bibnamefont {Gutsche}}, \bibinfo {author} {\bibfnamefont
  {H.~E.}\ \bibnamefont {Haber}}, \bibinfo {author} {\bibfnamefont
  {C.}~\bibnamefont {Hanhart}}, \bibinfo {author} {\bibfnamefont
  {S.}~\bibnamefont {Hashimoto}}, \bibinfo {author} {\bibfnamefont
  {Y.}~\bibnamefont {Hayato}}, \bibinfo {author} {\bibfnamefont {K.~G.}\
  \bibnamefont {Hayes}}, \bibinfo {author} {\bibfnamefont {A.}~\bibnamefont
  {Hebecker}}, \bibinfo {author} {\bibfnamefont {S.}~\bibnamefont
  {Heinemeyer}}, \bibinfo {author} {\bibfnamefont {B.}~\bibnamefont
  {Heltsley}}, \bibinfo {author} {\bibfnamefont {J.~J.}\ \bibnamefont
  {Hernández-Rey}}, \bibinfo {author} {\bibfnamefont {J.}~\bibnamefont
  {Hisano}}, \bibinfo {author} {\bibfnamefont {A.}~\bibnamefont {Höcker}},
  \bibinfo {author} {\bibfnamefont {J.}~\bibnamefont {Holder}}, \bibinfo
  {author} {\bibfnamefont {A.}~\bibnamefont {Holtkamp}}, \bibinfo {author}
  {\bibfnamefont {T.}~\bibnamefont {Hyodo}}, \bibinfo {author} {\bibfnamefont
  {K.~D.}\ \bibnamefont {Irwin}}, \bibinfo {author} {\bibfnamefont {K.~F.}\
  \bibnamefont {Johnson}}, \bibinfo {author} {\bibfnamefont {M.}~\bibnamefont
  {Kado}}, \bibinfo {author} {\bibfnamefont {M.}~\bibnamefont {Karliner}},
  \bibinfo {author} {\bibfnamefont {U.~F.}\ \bibnamefont {Katz}}, \bibinfo
  {author} {\bibfnamefont {S.~R.}\ \bibnamefont {Klein}}, \bibinfo {author}
  {\bibfnamefont {E.}~\bibnamefont {Klempt}}, \bibinfo {author} {\bibfnamefont
  {R.~V.}\ \bibnamefont {Kowalewski}}, \bibinfo {author} {\bibfnamefont
  {F.}~\bibnamefont {Krauss}}, \bibinfo {author} {\bibfnamefont
  {M.}~\bibnamefont {Kreps}}, \bibinfo {author} {\bibfnamefont
  {B.}~\bibnamefont {Krusche}}, \bibinfo {author} {\bibfnamefont {Y.~V.}\
  \bibnamefont {Kuyanov}}, \bibinfo {author} {\bibfnamefont {Y.}~\bibnamefont
  {Kwon}}, \bibinfo {author} {\bibfnamefont {O.}~\bibnamefont {Lahav}},
  \bibinfo {author} {\bibfnamefont {J.}~\bibnamefont {Laiho}}, \bibinfo
  {author} {\bibfnamefont {J.}~\bibnamefont {Lesgourgues}}, \bibinfo {author}
  {\bibfnamefont {A.}~\bibnamefont {Liddle}}, \bibinfo {author} {\bibfnamefont
  {Z.}~\bibnamefont {Ligeti}}, \bibinfo {author} {\bibfnamefont {C.~J.}\
  \bibnamefont {Lin}}, \bibinfo {author} {\bibfnamefont {C.}~\bibnamefont
  {Lippmann}}, \bibinfo {author} {\bibfnamefont {T.~M.}\ \bibnamefont {Liss}},
  \bibinfo {author} {\bibfnamefont {L.}~\bibnamefont {Littenberg}}, \bibinfo
  {author} {\bibfnamefont {K.~S.}\ \bibnamefont {Lugovsky}}, \bibinfo {author}
  {\bibfnamefont {S.~B.}\ \bibnamefont {Lugovsky}}, \bibinfo {author}
  {\bibfnamefont {A.}~\bibnamefont {Lusiani}}, \bibinfo {author} {\bibfnamefont
  {Y.}~\bibnamefont {Makida}}, \bibinfo {author} {\bibfnamefont
  {F.}~\bibnamefont {Maltoni}}, \bibinfo {author} {\bibfnamefont
  {T.}~\bibnamefont {Mannel}}, \bibinfo {author} {\bibfnamefont {A.~V.}\
  \bibnamefont {Manohar}}, \bibinfo {author} {\bibfnamefont {W.~J.}\
  \bibnamefont {Marciano}}, \bibinfo {author} {\bibfnamefont {A.~D.}\
  \bibnamefont {Martin}}, \bibinfo {author} {\bibfnamefont {A.}~\bibnamefont
  {Masoni}}, \bibinfo {author} {\bibfnamefont {J.}~\bibnamefont {Matthews}},
  \bibinfo {author} {\bibfnamefont {U.~G.}\ \bibnamefont {Meißner}}, \bibinfo
  {author} {\bibfnamefont {D.}~\bibnamefont {Milstead}}, \bibinfo {author}
  {\bibfnamefont {R.~E.}\ \bibnamefont {Mitchell}}, \bibinfo {author}
  {\bibfnamefont {K.}~\bibnamefont {Mönig}}, \bibinfo {author} {\bibfnamefont
  {P.}~\bibnamefont {Molaro}}, \bibinfo {author} {\bibfnamefont
  {F.}~\bibnamefont {Moortgat}}, \bibinfo {author} {\bibfnamefont
  {M.}~\bibnamefont {Moskovic}}, \bibinfo {author} {\bibfnamefont
  {H.}~\bibnamefont {Murayama}}, \bibinfo {author} {\bibfnamefont
  {M.}~\bibnamefont {Narain}}, \bibinfo {author} {\bibfnamefont
  {P.}~\bibnamefont {Nason}}, \bibinfo {author} {\bibfnamefont
  {S.}~\bibnamefont {Navas}}, \bibinfo {author} {\bibfnamefont
  {M.}~\bibnamefont {Neubert}}, \bibinfo {author} {\bibfnamefont
  {P.}~\bibnamefont {Nevski}}, \bibinfo {author} {\bibfnamefont
  {Y.}~\bibnamefont {Nir}}, \bibinfo {author} {\bibfnamefont {K.~A.}\
  \bibnamefont {Olive}}, \bibinfo {author} {\bibfnamefont {S.~P.}\ \bibnamefont
  {Griso}}, \bibinfo {author} {\bibfnamefont {J.}~\bibnamefont {Parsons}},
  \bibinfo {author} {\bibfnamefont {C.}~\bibnamefont {Patrignani}}, \bibinfo
  {author} {\bibfnamefont {J.~A.}\ \bibnamefont {Peacock}}, \bibinfo {author}
  {\bibfnamefont {M.}~\bibnamefont {Pennington}}, \bibinfo {author}
  {\bibfnamefont {S.~T.}\ \bibnamefont {Petcov}}, \bibinfo {author}
  {\bibfnamefont {V.~A.}\ \bibnamefont {Petrov}}, \bibinfo {author}
  {\bibfnamefont {E.}~\bibnamefont {Pianori}}, \bibinfo {author} {\bibfnamefont
  {A.}~\bibnamefont {Piepke}}, \bibinfo {author} {\bibfnamefont
  {A.}~\bibnamefont {Pomarol}}, \bibinfo {author} {\bibfnamefont
  {A.}~\bibnamefont {Quadt}}, \bibinfo {author} {\bibfnamefont
  {J.}~\bibnamefont {Rademacker}}, \bibinfo {author} {\bibfnamefont
  {G.}~\bibnamefont {Raffelt}}, \bibinfo {author} {\bibfnamefont {B.~N.}\
  \bibnamefont {Ratcliff}}, \bibinfo {author} {\bibfnamefont {P.}~\bibnamefont
  {Richardson}}, \bibinfo {author} {\bibfnamefont {A.}~\bibnamefont
  {Ringwald}}, \bibinfo {author} {\bibfnamefont {S.}~\bibnamefont {Roesler}},
  \bibinfo {author} {\bibfnamefont {S.}~\bibnamefont {Rolli}}, \bibinfo
  {author} {\bibfnamefont {A.}~\bibnamefont {Romaniouk}}, \bibinfo {author}
  {\bibfnamefont {L.~J.}\ \bibnamefont {Rosenberg}}, \bibinfo {author}
  {\bibfnamefont {J.~L.}\ \bibnamefont {Rosner}}, \bibinfo {author}
  {\bibfnamefont {G.}~\bibnamefont {Rybka}}, \bibinfo {author} {\bibfnamefont
  {R.~A.}\ \bibnamefont {Ryutin}}, \bibinfo {author} {\bibfnamefont {C.~T.}\
  \bibnamefont {Sachrajda}}, \bibinfo {author} {\bibfnamefont {Y.}~\bibnamefont
  {Sakai}}, \bibinfo {author} {\bibfnamefont {G.~P.}\ \bibnamefont {Salam}},
  \bibinfo {author} {\bibfnamefont {S.}~\bibnamefont {Sarkar}}, \bibinfo
  {author} {\bibfnamefont {F.}~\bibnamefont {Sauli}}, \bibinfo {author}
  {\bibfnamefont {O.}~\bibnamefont {Schneider}}, \bibinfo {author}
  {\bibfnamefont {K.}~\bibnamefont {Scholberg}}, \bibinfo {author}
  {\bibfnamefont {A.~J.}\ \bibnamefont {Schwartz}}, \bibinfo {author}
  {\bibfnamefont {D.}~\bibnamefont {Scott}}, \bibinfo {author} {\bibfnamefont
  {V.}~\bibnamefont {Sharma}}, \bibinfo {author} {\bibfnamefont {S.~R.}\
  \bibnamefont {Sharpe}}, \bibinfo {author} {\bibfnamefont {T.}~\bibnamefont
  {Shutt}}, \bibinfo {author} {\bibfnamefont {M.}~\bibnamefont {Silari}},
  \bibinfo {author} {\bibfnamefont {T.}~\bibnamefont {Sjöstrand}}, \bibinfo
  {author} {\bibfnamefont {P.}~\bibnamefont {Skands}}, \bibinfo {author}
  {\bibfnamefont {T.}~\bibnamefont {Skwarnicki}}, \bibinfo {author}
  {\bibfnamefont {J.~G.}\ \bibnamefont {Smith}}, \bibinfo {author}
  {\bibfnamefont {G.~F.}\ \bibnamefont {Smoot}}, \bibinfo {author}
  {\bibfnamefont {S.}~\bibnamefont {Spanier}}, \bibinfo {author} {\bibfnamefont
  {H.}~\bibnamefont {Spieler}}, \bibinfo {author} {\bibfnamefont
  {C.}~\bibnamefont {Spiering}}, \bibinfo {author} {\bibfnamefont
  {A.}~\bibnamefont {Stahl}}, \bibinfo {author} {\bibfnamefont {S.~L.}\
  \bibnamefont {Stone}}, \bibinfo {author} {\bibfnamefont {T.}~\bibnamefont
  {Sumiyoshi}}, \bibinfo {author} {\bibfnamefont {M.~J.}\ \bibnamefont
  {Syphers}}, \bibinfo {author} {\bibfnamefont {K.}~\bibnamefont {Terashi}},
  \bibinfo {author} {\bibfnamefont {J.}~\bibnamefont {Terning}}, \bibinfo
  {author} {\bibfnamefont {U.}~\bibnamefont {Thoma}}, \bibinfo {author}
  {\bibfnamefont {R.~S.}\ \bibnamefont {Thorne}}, \bibinfo {author}
  {\bibfnamefont {L.}~\bibnamefont {Tiator}}, \bibinfo {author} {\bibfnamefont
  {M.}~\bibnamefont {Titov}}, \bibinfo {author} {\bibfnamefont {N.~P.}\
  \bibnamefont {Tkachenko}}, \bibinfo {author} {\bibfnamefont {N.~A.}\
  \bibnamefont {Törnqvist}}, \bibinfo {author} {\bibfnamefont {D.~R.}\
  \bibnamefont {Tovey}}, \bibinfo {author} {\bibfnamefont {G.}~\bibnamefont
  {Valencia}}, \bibinfo {author} {\bibfnamefont {R.~V.~D.}\ \bibnamefont
  {Water}}, \bibinfo {author} {\bibfnamefont {N.}~\bibnamefont {Varelas}},
  \bibinfo {author} {\bibfnamefont {G.}~\bibnamefont {Venanzoni}}, \bibinfo
  {author} {\bibfnamefont {L.}~\bibnamefont {Verde}}, \bibinfo {author}
  {\bibfnamefont {M.~G.}\ \bibnamefont {Vincter}}, \bibinfo {author}
  {\bibfnamefont {P.}~\bibnamefont {Vogel}}, \bibinfo {author} {\bibfnamefont
  {A.}~\bibnamefont {Vogt}}, \bibinfo {author} {\bibfnamefont {S.~P.}\
  \bibnamefont {Wakely}}, \bibinfo {author} {\bibfnamefont {W.}~\bibnamefont
  {Walkowiak}}, \bibinfo {author} {\bibfnamefont {C.~W.}\ \bibnamefont
  {Walter}}, \bibinfo {author} {\bibfnamefont {D.}~\bibnamefont {Wands}},
  \bibinfo {author} {\bibfnamefont {D.~R.}\ \bibnamefont {Ward}}, \bibinfo
  {author} {\bibfnamefont {M.~O.}\ \bibnamefont {Wascko}}, \bibinfo {author}
  {\bibfnamefont {G.}~\bibnamefont {Weiglein}}, \bibinfo {author}
  {\bibfnamefont {D.~H.}\ \bibnamefont {Weinberg}}, \bibinfo {author}
  {\bibfnamefont {E.~J.}\ \bibnamefont {Weinberg}}, \bibinfo {author}
  {\bibfnamefont {M.}~\bibnamefont {White}}, \bibinfo {author} {\bibfnamefont
  {L.~R.}\ \bibnamefont {Wiencke}}, \bibinfo {author} {\bibfnamefont
  {S.}~\bibnamefont {Willocq}}, \bibinfo {author} {\bibfnamefont {C.~G.}\
  \bibnamefont {Wohl}}, \bibinfo {author} {\bibfnamefont {J.}~\bibnamefont
  {Womersley}}, \bibinfo {author} {\bibfnamefont {C.~L.}\ \bibnamefont
  {Woody}}, \bibinfo {author} {\bibfnamefont {R.~L.}\ \bibnamefont {Workman}},
  \bibinfo {author} {\bibfnamefont {W.~M.}\ \bibnamefont {Yao}}, \bibinfo
  {author} {\bibfnamefont {G.~P.}\ \bibnamefont {Zeller}}, \bibinfo {author}
  {\bibfnamefont {O.~V.}\ \bibnamefont {Zenin}}, \bibinfo {author}
  {\bibfnamefont {R.~Y.}\ \bibnamefont {Zhu}}, \bibinfo {author} {\bibfnamefont
  {S.~L.}\ \bibnamefont {Zhu}}, \bibinfo {author} {\bibfnamefont
  {F.}~\bibnamefont {Zimmermann}}, \bibinfo {author} {\bibfnamefont {P.~A.}\
  \bibnamefont {Zyla}}, \bibinfo {author} {\bibfnamefont {J.}~\bibnamefont
  {Anderson}}, \bibinfo {author} {\bibfnamefont {L.}~\bibnamefont {Fuller}},
  \bibinfo {author} {\bibfnamefont {V.~S.}\ \bibnamefont {Lugovsky}}, \ and\
  \bibinfo {author} {\bibfnamefont {P.}~\bibnamefont {Schaffner}},\ }\href
  {\doibase 10.1103/PhysRevD.98.030001} {\enquote {\bibinfo {title} {Review of
  particle physics},}\ } (\bibinfo {year} {2018})\BibitemShut {NoStop}%
\bibitem [{\citenamefont {Lawrence}, \citenamefont {Wilson},\ and\
  \citenamefont {Peplowski}(2021)}]{Lawrence2021}%
  \BibitemOpen
  \bibfield  {author} {\bibinfo {author} {\bibfnamefont {D.~J.}\ \bibnamefont
  {Lawrence}}, \bibinfo {author} {\bibfnamefont {J.~T.}\ \bibnamefont
  {Wilson}}, \ and\ \bibinfo {author} {\bibfnamefont {P.~N.}\ \bibnamefont
  {Peplowski}},\ }\bibfield  {title} {\enquote {\bibinfo {title} {Space-based
  measurements of neutron lifetime: Approaches to resolving the neutron
  lifetime anomaly},}\ }\href {\doibase 10.1016/j.nima.2020.164919} {\bibfield
  {journal} {\bibinfo  {journal} {Nuclear Instruments and Methods in Physics
  Research, Section A: Accelerators, Spectrometers, Detectors and Associated
  Equipment}\ }\textbf {\bibinfo {volume} {988}} (\bibinfo {year} {2021}),\
  10.1016/j.nima.2020.164919}\BibitemShut {NoStop}%
\bibitem [{\citenamefont {Wei}(2020)}]{Wei2020}%
  \BibitemOpen
  \bibfield  {author} {\bibinfo {author} {\bibfnamefont {W.}~\bibnamefont
  {Wei}},\ }\bibfield  {title} {\enquote {\bibinfo {title} {A new neutron
  lifetime experiment with cold neutron beam decay in superfluid helium-4},}\
  }\href {\doibase 10.1088/1361-6471/abacdb} {\bibfield  {journal} {\bibinfo
  {journal} {Journal of Physics G: Nuclear and Particle Physics}\ }\textbf
  {\bibinfo {volume} {47}} (\bibinfo {year} {2020}),\
  10.1088/1361-6471/abacdb}\BibitemShut {NoStop}%
\bibitem [{\citenamefont {Saunders}\ \emph {et~al.}(2013)\citenamefont
  {Saunders}, \citenamefont {Makela}, \citenamefont {Bagdasarova},
  \citenamefont {Back}, \citenamefont {Boissevain}, \citenamefont {Broussard},
  \citenamefont {Bowles}, \citenamefont {Carr}, \citenamefont {Currie},
  \citenamefont {Filippone}, \citenamefont {García}, \citenamefont
  {Geltenbort}, \citenamefont {Hickerson}, \citenamefont {Hill}, \citenamefont
  {Hoagland}, \citenamefont {Hoedl}, \citenamefont {Holley}, \citenamefont
  {Hogan}, \citenamefont {Ito}, \citenamefont {Lamoreaux}, \citenamefont {Liu},
  \citenamefont {Liu}, \citenamefont {Mammei}, \citenamefont {Martin},
  \citenamefont {Melconian}, \citenamefont {Mendenhall}, \citenamefont
  {Morris}, \citenamefont {Mortensen}, \citenamefont {Pattie}, \citenamefont
  {Pitt}, \citenamefont {Plaster}, \citenamefont {Ramsey}, \citenamefont
  {Rios}, \citenamefont {Sallaska}, \citenamefont {Seestrom}, \citenamefont
  {Sharapov}, \citenamefont {Sjue}, \citenamefont {Sondheim}, \citenamefont
  {Teasdale}, \citenamefont {Young}, \citenamefont {Vorndick}, \citenamefont
  {Vogelaar}, \citenamefont {Wang},\ and\ \citenamefont {Xu}}]{Saunders2013}%
  \BibitemOpen
  \bibfield  {author} {\bibinfo {author} {\bibfnamefont {A.}~\bibnamefont
  {Saunders}}, \bibinfo {author} {\bibfnamefont {M.}~\bibnamefont {Makela}},
  \bibinfo {author} {\bibfnamefont {Y.}~\bibnamefont {Bagdasarova}}, \bibinfo
  {author} {\bibfnamefont {H.~O.}\ \bibnamefont {Back}}, \bibinfo {author}
  {\bibfnamefont {J.}~\bibnamefont {Boissevain}}, \bibinfo {author}
  {\bibfnamefont {L.~J.}\ \bibnamefont {Broussard}}, \bibinfo {author}
  {\bibfnamefont {T.~J.}\ \bibnamefont {Bowles}}, \bibinfo {author}
  {\bibfnamefont {R.}~\bibnamefont {Carr}}, \bibinfo {author} {\bibfnamefont
  {S.~A.}\ \bibnamefont {Currie}}, \bibinfo {author} {\bibfnamefont
  {B.}~\bibnamefont {Filippone}}, \bibinfo {author} {\bibfnamefont
  {A.}~\bibnamefont {García}}, \bibinfo {author} {\bibfnamefont
  {P.}~\bibnamefont {Geltenbort}}, \bibinfo {author} {\bibfnamefont {K.~P.}\
  \bibnamefont {Hickerson}}, \bibinfo {author} {\bibfnamefont {R.~E.}\
  \bibnamefont {Hill}}, \bibinfo {author} {\bibfnamefont {J.}~\bibnamefont
  {Hoagland}}, \bibinfo {author} {\bibfnamefont {S.}~\bibnamefont {Hoedl}},
  \bibinfo {author} {\bibfnamefont {A.~T.}\ \bibnamefont {Holley}}, \bibinfo
  {author} {\bibfnamefont {G.}~\bibnamefont {Hogan}}, \bibinfo {author}
  {\bibfnamefont {T.~M.}\ \bibnamefont {Ito}}, \bibinfo {author} {\bibfnamefont
  {S.}~\bibnamefont {Lamoreaux}}, \bibinfo {author} {\bibfnamefont {C.~Y.}\
  \bibnamefont {Liu}}, \bibinfo {author} {\bibfnamefont {J.}~\bibnamefont
  {Liu}}, \bibinfo {author} {\bibfnamefont {R.~R.}\ \bibnamefont {Mammei}},
  \bibinfo {author} {\bibfnamefont {J.}~\bibnamefont {Martin}}, \bibinfo
  {author} {\bibfnamefont {D.}~\bibnamefont {Melconian}}, \bibinfo {author}
  {\bibfnamefont {M.~P.}\ \bibnamefont {Mendenhall}}, \bibinfo {author}
  {\bibfnamefont {C.~L.}\ \bibnamefont {Morris}}, \bibinfo {author}
  {\bibfnamefont {R.~N.}\ \bibnamefont {Mortensen}}, \bibinfo {author}
  {\bibfnamefont {R.~W.}\ \bibnamefont {Pattie}}, \bibinfo {author}
  {\bibfnamefont {M.}~\bibnamefont {Pitt}}, \bibinfo {author} {\bibfnamefont
  {B.}~\bibnamefont {Plaster}}, \bibinfo {author} {\bibfnamefont
  {J.}~\bibnamefont {Ramsey}}, \bibinfo {author} {\bibfnamefont
  {R.}~\bibnamefont {Rios}}, \bibinfo {author} {\bibfnamefont {A.}~\bibnamefont
  {Sallaska}}, \bibinfo {author} {\bibfnamefont {S.~J.}\ \bibnamefont
  {Seestrom}}, \bibinfo {author} {\bibfnamefont {E.~I.}\ \bibnamefont
  {Sharapov}}, \bibinfo {author} {\bibfnamefont {S.}~\bibnamefont {Sjue}},
  \bibinfo {author} {\bibfnamefont {W.~E.}\ \bibnamefont {Sondheim}}, \bibinfo
  {author} {\bibfnamefont {W.}~\bibnamefont {Teasdale}}, \bibinfo {author}
  {\bibfnamefont {A.~R.}\ \bibnamefont {Young}}, \bibinfo {author}
  {\bibfnamefont {B.}~\bibnamefont {Vorndick}}, \bibinfo {author}
  {\bibfnamefont {R.~B.}\ \bibnamefont {Vogelaar}}, \bibinfo {author}
  {\bibfnamefont {Z.}~\bibnamefont {Wang}}, \ and\ \bibinfo {author}
  {\bibfnamefont {Y.}~\bibnamefont {Xu}},\ }\bibfield  {title} {\enquote
  {\bibinfo {title} {Performance of the los alamos national laboratory
  spallation-driven solid-deuterium ultra-cold neutron source},}\ }\href
  {\doibase 10.1063/1.4770063} {\bibfield  {journal} {\bibinfo  {journal}
  {Review of Scientific Instruments}\ }\textbf {\bibinfo {volume} {84}}
  (\bibinfo {year} {2013}),\ 10.1063/1.4770063}\BibitemShut {NoStop}%
\bibitem [{\citenamefont {Ito}\ \emph {et~al.}(2018)\citenamefont {Ito},
  \citenamefont {Adamek}, \citenamefont {Callahan}, \citenamefont {Choi},
  \citenamefont {Clayton}, \citenamefont {Cude-Woods}, \citenamefont {Currie},
  \citenamefont {Ding}, \citenamefont {Fellers}, \citenamefont {Geltenbort},
  \citenamefont {Lamoreaux}, \citenamefont {Liu}, \citenamefont {Macdonald},
  \citenamefont {Makela}, \citenamefont {Morris}, \citenamefont {Pattie},
  \citenamefont {Ramsey}, \citenamefont {Salvat}, \citenamefont {Saunders},
  \citenamefont {Sharapov}, \citenamefont {Sjue}, \citenamefont {Sprow},
  \citenamefont {Tang}, \citenamefont {Weaver}, \citenamefont {Wei},\ and\
  \citenamefont {Young}}]{Ito2018}%
  \BibitemOpen
  \bibfield  {author} {\bibinfo {author} {\bibfnamefont {T.~M.}\ \bibnamefont
  {Ito}}, \bibinfo {author} {\bibfnamefont {E.~R.}\ \bibnamefont {Adamek}},
  \bibinfo {author} {\bibfnamefont {N.~B.}\ \bibnamefont {Callahan}}, \bibinfo
  {author} {\bibfnamefont {J.~H.}\ \bibnamefont {Choi}}, \bibinfo {author}
  {\bibfnamefont {S.~M.}\ \bibnamefont {Clayton}}, \bibinfo {author}
  {\bibfnamefont {C.}~\bibnamefont {Cude-Woods}}, \bibinfo {author}
  {\bibfnamefont {S.}~\bibnamefont {Currie}}, \bibinfo {author} {\bibfnamefont
  {X.}~\bibnamefont {Ding}}, \bibinfo {author} {\bibfnamefont {D.~E.}\
  \bibnamefont {Fellers}}, \bibinfo {author} {\bibfnamefont {P.}~\bibnamefont
  {Geltenbort}}, \bibinfo {author} {\bibfnamefont {S.~K.}\ \bibnamefont
  {Lamoreaux}}, \bibinfo {author} {\bibfnamefont {C.~Y.}\ \bibnamefont {Liu}},
  \bibinfo {author} {\bibfnamefont {S.}~\bibnamefont {Macdonald}}, \bibinfo
  {author} {\bibfnamefont {M.}~\bibnamefont {Makela}}, \bibinfo {author}
  {\bibfnamefont {C.~L.}\ \bibnamefont {Morris}}, \bibinfo {author}
  {\bibfnamefont {R.~W.}\ \bibnamefont {Pattie}}, \bibinfo {author}
  {\bibfnamefont {J.~C.}\ \bibnamefont {Ramsey}}, \bibinfo {author}
  {\bibfnamefont {D.~J.}\ \bibnamefont {Salvat}}, \bibinfo {author}
  {\bibfnamefont {A.}~\bibnamefont {Saunders}}, \bibinfo {author}
  {\bibfnamefont {E.~I.}\ \bibnamefont {Sharapov}}, \bibinfo {author}
  {\bibfnamefont {S.}~\bibnamefont {Sjue}}, \bibinfo {author} {\bibfnamefont
  {A.~P.}\ \bibnamefont {Sprow}}, \bibinfo {author} {\bibfnamefont
  {Z.}~\bibnamefont {Tang}}, \bibinfo {author} {\bibfnamefont {H.~L.}\
  \bibnamefont {Weaver}}, \bibinfo {author} {\bibfnamefont {W.}~\bibnamefont
  {Wei}}, \ and\ \bibinfo {author} {\bibfnamefont {A.~R.}\ \bibnamefont
  {Young}},\ }\bibfield  {title} {\enquote {\bibinfo {title} {Performance of
  the upgraded ultracold neutron source at los alamos national laboratory and
  its implication for a possible neutron electric dipole moment experiment},}\
  }\href {\doibase 10.1103/PhysRevC.97.012501} {\bibfield  {journal} {\bibinfo
  {journal} {Physical Review C}\ }\textbf {\bibinfo {volume} {97}} (\bibinfo
  {year} {2018}),\ 10.1103/PhysRevC.97.012501}\BibitemShut {NoStop}%
\bibitem [{\citenamefont {Tang}\ \emph {et~al.}(2021)\citenamefont {Tang},
  \citenamefont {Watkins}, \citenamefont {Clayton}, \citenamefont {Currie},
  \citenamefont {Fellers}, \citenamefont {Hassan}, \citenamefont {Hooks},
  \citenamefont {Ito}, \citenamefont {Lawrence}, \citenamefont {MacDonald},
  \citenamefont {Makela}, \citenamefont {Morris}, \citenamefont {Neukirch},
  \citenamefont {Saunders}, \citenamefont {O'Shaughnessy}, \citenamefont
  {Cude-Woods}, \citenamefont {Choi}, \citenamefont {Young}, \citenamefont
  {Zeck}, \citenamefont {Gonzalez}, \citenamefont {Liu}, \citenamefont {Floyd},
  \citenamefont {Hickerson}, \citenamefont {Holley}, \citenamefont {Johnson},
  \citenamefont {Lambert},\ and\ \citenamefont {Pattie}}]{Tang2021}%
  \BibitemOpen
  \bibfield  {author} {\bibinfo {author} {\bibfnamefont {Z.}~\bibnamefont
  {Tang}}, \bibinfo {author} {\bibfnamefont {E.~B.}\ \bibnamefont {Watkins}},
  \bibinfo {author} {\bibfnamefont {S.~M.}\ \bibnamefont {Clayton}}, \bibinfo
  {author} {\bibfnamefont {S.~A.}\ \bibnamefont {Currie}}, \bibinfo {author}
  {\bibfnamefont {D.~E.}\ \bibnamefont {Fellers}}, \bibinfo {author}
  {\bibfnamefont {M.~T.}\ \bibnamefont {Hassan}}, \bibinfo {author}
  {\bibfnamefont {D.~E.}\ \bibnamefont {Hooks}}, \bibinfo {author}
  {\bibfnamefont {T.~M.}\ \bibnamefont {Ito}}, \bibinfo {author} {\bibfnamefont
  {S.~K.}\ \bibnamefont {Lawrence}}, \bibinfo {author} {\bibfnamefont {S.~W.}\
  \bibnamefont {MacDonald}}, \bibinfo {author} {\bibfnamefont {M.}~\bibnamefont
  {Makela}}, \bibinfo {author} {\bibfnamefont {C.~L.}\ \bibnamefont {Morris}},
  \bibinfo {author} {\bibfnamefont {L.~P.}\ \bibnamefont {Neukirch}}, \bibinfo
  {author} {\bibfnamefont {A.}~\bibnamefont {Saunders}}, \bibinfo {author}
  {\bibfnamefont {C.~M.}\ \bibnamefont {O'Shaughnessy}}, \bibinfo {author}
  {\bibfnamefont {C.}~\bibnamefont {Cude-Woods}}, \bibinfo {author}
  {\bibfnamefont {J.~H.}\ \bibnamefont {Choi}}, \bibinfo {author}
  {\bibfnamefont {A.~R.}\ \bibnamefont {Young}}, \bibinfo {author}
  {\bibfnamefont {B.~A.}\ \bibnamefont {Zeck}}, \bibinfo {author}
  {\bibfnamefont {F.}~\bibnamefont {Gonzalez}}, \bibinfo {author}
  {\bibfnamefont {C.~Y.}\ \bibnamefont {Liu}}, \bibinfo {author} {\bibfnamefont
  {N.~C.}\ \bibnamefont {Floyd}}, \bibinfo {author} {\bibfnamefont {K.~P.}\
  \bibnamefont {Hickerson}}, \bibinfo {author} {\bibfnamefont {A.~T.}\
  \bibnamefont {Holley}}, \bibinfo {author} {\bibfnamefont {B.~A.}\
  \bibnamefont {Johnson}}, \bibinfo {author} {\bibfnamefont {J.~C.}\
  \bibnamefont {Lambert}}, \ and\ \bibinfo {author} {\bibfnamefont {R.~W.}\
  \bibnamefont {Pattie}},\ }\bibfield  {title} {\enquote {\bibinfo {title}
  {Ultracold neutron properties of the eljen-299-02d deuterated
  scintillator},}\ }\href {\doibase 10.1063/5.0030972} {\bibfield  {journal}
  {\bibinfo  {journal} {Review of Scientific Instruments}\ }\textbf {\bibinfo
  {volume} {92}} (\bibinfo {year} {2021}),\ 10.1063/5.0030972}\BibitemShut
  {NoStop}%
\bibitem [{\citenamefont {Wang}\ \emph {et~al.}(2015)\citenamefont {Wang},
  \citenamefont {Hoffbauer}, \citenamefont {Morris}, \citenamefont {Callahan},
  \citenamefont {Adamek}, \citenamefont {Bacon}, \citenamefont {Blatnik},
  \citenamefont {Brandt}, \citenamefont {Broussard}, \citenamefont {Clayton},
  \citenamefont {Cude-Woods}, \citenamefont {Currie}, \citenamefont {Dees},
  \citenamefont {Ding}, \citenamefont {Gao}, \citenamefont {Gray},
  \citenamefont {Hickerson}, \citenamefont {Holley}, \citenamefont {Ito},
  \citenamefont {Liu}, \citenamefont {Makela}, \citenamefont {Ramsey},
  \citenamefont {Pattie}, \citenamefont {Salvat}, \citenamefont {Saunders},
  \citenamefont {Schmidt}, \citenamefont {Schulze}, \citenamefont {Seestrom},
  \citenamefont {Sharapov}, \citenamefont {Sprow}, \citenamefont {Tang},
  \citenamefont {Wei}, \citenamefont {Wexler}, \citenamefont {Womack},
  \citenamefont {Young},\ and\ \citenamefont {Zeck}}]{Wang2015}%
  \BibitemOpen
  \bibfield  {author} {\bibinfo {author} {\bibfnamefont {Z.}~\bibnamefont
  {Wang}}, \bibinfo {author} {\bibfnamefont {M.~A.}\ \bibnamefont {Hoffbauer}},
  \bibinfo {author} {\bibfnamefont {C.~L.}\ \bibnamefont {Morris}}, \bibinfo
  {author} {\bibfnamefont {N.~B.}\ \bibnamefont {Callahan}}, \bibinfo {author}
  {\bibfnamefont {E.~R.}\ \bibnamefont {Adamek}}, \bibinfo {author}
  {\bibfnamefont {J.~D.}\ \bibnamefont {Bacon}}, \bibinfo {author}
  {\bibfnamefont {M.}~\bibnamefont {Blatnik}}, \bibinfo {author} {\bibfnamefont
  {A.~E.}\ \bibnamefont {Brandt}}, \bibinfo {author} {\bibfnamefont {L.~J.}\
  \bibnamefont {Broussard}}, \bibinfo {author} {\bibfnamefont {S.~M.}\
  \bibnamefont {Clayton}}, \bibinfo {author} {\bibfnamefont {C.}~\bibnamefont
  {Cude-Woods}}, \bibinfo {author} {\bibfnamefont {S.}~\bibnamefont {Currie}},
  \bibinfo {author} {\bibfnamefont {E.~B.}\ \bibnamefont {Dees}}, \bibinfo
  {author} {\bibfnamefont {X.}~\bibnamefont {Ding}}, \bibinfo {author}
  {\bibfnamefont {J.}~\bibnamefont {Gao}}, \bibinfo {author} {\bibfnamefont
  {F.~E.}\ \bibnamefont {Gray}}, \bibinfo {author} {\bibfnamefont {K.~P.}\
  \bibnamefont {Hickerson}}, \bibinfo {author} {\bibfnamefont {A.~T.}\
  \bibnamefont {Holley}}, \bibinfo {author} {\bibfnamefont {T.~M.}\
  \bibnamefont {Ito}}, \bibinfo {author} {\bibfnamefont {C.~Y.}\ \bibnamefont
  {Liu}}, \bibinfo {author} {\bibfnamefont {M.}~\bibnamefont {Makela}},
  \bibinfo {author} {\bibfnamefont {J.~C.}\ \bibnamefont {Ramsey}}, \bibinfo
  {author} {\bibfnamefont {R.~W.}\ \bibnamefont {Pattie}}, \bibinfo {author}
  {\bibfnamefont {D.~J.}\ \bibnamefont {Salvat}}, \bibinfo {author}
  {\bibfnamefont {A.}~\bibnamefont {Saunders}}, \bibinfo {author}
  {\bibfnamefont {D.~W.}\ \bibnamefont {Schmidt}}, \bibinfo {author}
  {\bibfnamefont {R.~K.}\ \bibnamefont {Schulze}}, \bibinfo {author}
  {\bibfnamefont {S.~J.}\ \bibnamefont {Seestrom}}, \bibinfo {author}
  {\bibfnamefont {E.~I.}\ \bibnamefont {Sharapov}}, \bibinfo {author}
  {\bibfnamefont {A.}~\bibnamefont {Sprow}}, \bibinfo {author} {\bibfnamefont
  {Z.}~\bibnamefont {Tang}}, \bibinfo {author} {\bibfnamefont {W.}~\bibnamefont
  {Wei}}, \bibinfo {author} {\bibfnamefont {J.}~\bibnamefont {Wexler}},
  \bibinfo {author} {\bibfnamefont {T.~L.}\ \bibnamefont {Womack}}, \bibinfo
  {author} {\bibfnamefont {A.~R.}\ \bibnamefont {Young}}, \ and\ \bibinfo
  {author} {\bibfnamefont {B.~A.}\ \bibnamefont {Zeck}},\ }\bibfield  {title}
  {\enquote {\bibinfo {title} {A multilayer surface detector for ultracold
  neutrons},}\ }\href {\doibase 10.1016/j.nima.2015.07.010} {\bibfield
  {journal} {\bibinfo  {journal} {Nuclear Instruments and Methods in Physics
  Research, Section A: Accelerators, Spectrometers, Detectors and Associated
  Equipment}\ }\textbf {\bibinfo {volume} {798}},\ \bibinfo {pages} {30--35}
  (\bibinfo {year} {2015})}\BibitemShut {NoStop}%
\bibitem [{\citenamefont {Knoll}(1979)}]{Knoll1979}%
  \BibitemOpen
  \bibfield  {author} {\bibinfo {author} {\bibfnamefont {G.}~\bibnamefont
  {Knoll}},\ }\href@noop {} {\emph {\bibinfo {title} {Radiation Detection and
  Measurement}}},\ \bibinfo {edition} {4th}\ ed.\ (\bibinfo  {publisher}
  {Wiley},\ \bibinfo {year} {1979})\BibitemShut {NoStop}%
\bibitem [{\citenamefont {Leo}(1994)}]{Leo1994}%
  \BibitemOpen
  \bibfield  {author} {\bibinfo {author} {\bibfnamefont {W.~R.}\ \bibnamefont
  {Leo}},\ }\href {\doibase 10.1007/978-3-642-57920-2} {\emph {\bibinfo {title}
  {Techniques for Nuclear and Particle Physics Experiments}}}\ (\bibinfo
  {publisher} {Springer Berlin Heidelberg},\ \bibinfo {year}
  {1994})\BibitemShut {NoStop}%
\bibitem [{\citenamefont {Roick}\ \emph {et~al.}(2018)\citenamefont {Roick},
  \citenamefont {Dubbers}, \citenamefont {Märkisch}, \citenamefont {Saul},\
  and\ \citenamefont {Schmidt}}]{Roick2018}%
  \BibitemOpen
  \bibfield  {author} {\bibinfo {author} {\bibfnamefont {C.}~\bibnamefont
  {Roick}}, \bibinfo {author} {\bibfnamefont {D.}~\bibnamefont {Dubbers}},
  \bibinfo {author} {\bibfnamefont {B.}~\bibnamefont {Märkisch}}, \bibinfo
  {author} {\bibfnamefont {H.}~\bibnamefont {Saul}}, \ and\ \bibinfo {author}
  {\bibfnamefont {U.}~\bibnamefont {Schmidt}},\ }\bibfield  {title} {\enquote
  {\bibinfo {title} {Electron time-of-flight: A new tool in beta -decay
  spectroscopy},}\ }\href {\doibase 10.1103/PhysRevC.97.035502} {\bibfield
  {journal} {\bibinfo  {journal} {Physical Review C}\ }\textbf {\bibinfo
  {volume} {97}} (\bibinfo {year} {2018}),\
  10.1103/PhysRevC.97.035502}\BibitemShut {NoStop}%
\bibitem [{\citenamefont {Seliger}, \citenamefont {Ziegler},\ and\
  \citenamefont {Jaffe}(1956)}]{Seliger1956}%
  \BibitemOpen
  \bibfield  {author} {\bibinfo {author} {\bibfnamefont {H.~H.}\ \bibnamefont
  {Seliger}}, \bibinfo {author} {\bibfnamefont {C.~A.}\ \bibnamefont
  {Ziegler}}, \ and\ \bibinfo {author} {\bibfnamefont {I.}~\bibnamefont
  {Jaffe}},\ }\bibfield  {title} {\enquote {\bibinfo {title} {Role of oxygen in
  the quenching of liquid scintillators},}\ }\href {\doibase
  10.1103/PhysRev.101.998} {\bibfield  {journal} {\bibinfo  {journal} {Physical
  Review}\ }\textbf {\bibinfo {volume} {101}},\ \bibinfo {pages} {998}
  (\bibinfo {year} {1956})}\BibitemShut {NoStop}%
\bibitem [{\citenamefont {Acciarri}\ \emph {et~al.}(2010)\citenamefont
  {Acciarri}, \citenamefont {Antonello}, \citenamefont {Baibussinov},
  \citenamefont {Baldo-Ceolin}, \citenamefont {Benetti}, \citenamefont
  {Calaprice}, \citenamefont {Calligarich}, \citenamefont {Cambiaghi},
  \citenamefont {Canci}, \citenamefont {Carbonara}, \citenamefont {Cavanna},
  \citenamefont {Centro}, \citenamefont {Cocco}, \citenamefont {Pompeo},
  \citenamefont {Fiorillo}, \citenamefont {Galbiati}, \citenamefont {Gallo},
  \citenamefont {Grandi}, \citenamefont {Meng}, \citenamefont {Modena},
  \citenamefont {Montanari}, \citenamefont {Palamara}, \citenamefont {Pandola},
  \citenamefont {Mortari}, \citenamefont {Pietropaolo}, \citenamefont
  {Raselli}, \citenamefont {Roncadelli}, \citenamefont {Rossella},
  \citenamefont {Rubbia}, \citenamefont {Segreto}, \citenamefont {Szelc},
  \citenamefont {Tortorici}, \citenamefont {Ventura},\ and\ \citenamefont
  {Vignoli}}]{Acciarri2010}%
  \BibitemOpen
  \bibfield  {author} {\bibinfo {author} {\bibfnamefont {R.}~\bibnamefont
  {Acciarri}}, \bibinfo {author} {\bibfnamefont {M.}~\bibnamefont {Antonello}},
  \bibinfo {author} {\bibfnamefont {B.}~\bibnamefont {Baibussinov}}, \bibinfo
  {author} {\bibfnamefont {M.}~\bibnamefont {Baldo-Ceolin}}, \bibinfo {author}
  {\bibfnamefont {P.}~\bibnamefont {Benetti}}, \bibinfo {author} {\bibfnamefont
  {F.}~\bibnamefont {Calaprice}}, \bibinfo {author} {\bibfnamefont
  {E.}~\bibnamefont {Calligarich}}, \bibinfo {author} {\bibfnamefont
  {M.}~\bibnamefont {Cambiaghi}}, \bibinfo {author} {\bibfnamefont
  {N.}~\bibnamefont {Canci}}, \bibinfo {author} {\bibfnamefont
  {F.}~\bibnamefont {Carbonara}}, \bibinfo {author} {\bibfnamefont
  {F.}~\bibnamefont {Cavanna}}, \bibinfo {author} {\bibfnamefont
  {S.}~\bibnamefont {Centro}}, \bibinfo {author} {\bibfnamefont {A.~G.}\
  \bibnamefont {Cocco}}, \bibinfo {author} {\bibfnamefont {F.~D.}\ \bibnamefont
  {Pompeo}}, \bibinfo {author} {\bibfnamefont {G.}~\bibnamefont {Fiorillo}},
  \bibinfo {author} {\bibfnamefont {C.}~\bibnamefont {Galbiati}}, \bibinfo
  {author} {\bibfnamefont {V.}~\bibnamefont {Gallo}}, \bibinfo {author}
  {\bibfnamefont {L.}~\bibnamefont {Grandi}}, \bibinfo {author} {\bibfnamefont
  {G.}~\bibnamefont {Meng}}, \bibinfo {author} {\bibfnamefont {I.}~\bibnamefont
  {Modena}}, \bibinfo {author} {\bibfnamefont {C.}~\bibnamefont {Montanari}},
  \bibinfo {author} {\bibfnamefont {O.}~\bibnamefont {Palamara}}, \bibinfo
  {author} {\bibfnamefont {L.}~\bibnamefont {Pandola}}, \bibinfo {author}
  {\bibfnamefont {G.~B.~P.}\ \bibnamefont {Mortari}}, \bibinfo {author}
  {\bibfnamefont {F.}~\bibnamefont {Pietropaolo}}, \bibinfo {author}
  {\bibfnamefont {G.~L.}\ \bibnamefont {Raselli}}, \bibinfo {author}
  {\bibfnamefont {M.}~\bibnamefont {Roncadelli}}, \bibinfo {author}
  {\bibfnamefont {M.}~\bibnamefont {Rossella}}, \bibinfo {author}
  {\bibfnamefont {C.}~\bibnamefont {Rubbia}}, \bibinfo {author} {\bibfnamefont
  {E.}~\bibnamefont {Segreto}}, \bibinfo {author} {\bibfnamefont
  {M.}~\bibnamefont {Szelc}}, \bibinfo {author} {\bibfnamefont
  {F.}~\bibnamefont {Tortorici}}, \bibinfo {author} {\bibfnamefont
  {S.}~\bibnamefont {Ventura}}, \ and\ \bibinfo {author} {\bibfnamefont
  {C.}~\bibnamefont {Vignoli}},\ }\bibfield  {title} {\enquote {\bibinfo
  {title} {Oxygen contamination in liquid argon: combined effects on ionization
  electron charge and scintillation light},}\ }\href {\doibase
  10.1088/1748-0221/5/05/P05003} {\bibfield  {journal} {\bibinfo  {journal}
  {Journal of Instrumentation}\ }\textbf {\bibinfo {volume} {5}},\ \bibinfo
  {pages} {P05003} (\bibinfo {year} {2010})}\BibitemShut {NoStop}%
\bibitem [{\citenamefont {Agostinelli}\ \emph {et~al.}(2003)\citenamefont
  {Agostinelli}, \citenamefont {Allison}, \citenamefont {Amako}, \citenamefont
  {Apostolakis}, \citenamefont {Araujo}, \citenamefont {Arce}, \citenamefont
  {Asai}, \citenamefont {Axen}, \citenamefont {Banerjee}, \citenamefont
  {Barrand}, \citenamefont {Behner}, \citenamefont {Bellagamba}, \citenamefont
  {Boudreau}, \citenamefont {Broglia}, \citenamefont {Brunengo}, \citenamefont
  {Burkhardt}, \citenamefont {Chauvie}, \citenamefont {Chuma}, \citenamefont
  {Chytracek}, \citenamefont {Cooperman}, \citenamefont {Cosmo}, \citenamefont
  {Degtyarenko}, \citenamefont {Dell'Acqua}, \citenamefont {Depaola},
  \citenamefont {Dietrich}, \citenamefont {Enami}, \citenamefont {Feliciello},
  \citenamefont {Ferguson}, \citenamefont {Fesefeldt}, \citenamefont {Folger},
  \citenamefont {Foppiano}, \citenamefont {Forti}, \citenamefont {Garelli},
  \citenamefont {Giani}, \citenamefont {Giannitrapani}, \citenamefont {Gibin},
  \citenamefont {Cadenas}, \citenamefont {Gonzalez}, \citenamefont {Abril},
  \citenamefont {Greeniaus}, \citenamefont {Greiner}, \citenamefont {Grichine},
  \citenamefont {Grossheim}, \citenamefont {Guatelli}, \citenamefont
  {Gumplinger}, \citenamefont {Hamatsu}, \citenamefont {Hashimoto},
  \citenamefont {Hasui}, \citenamefont {Heikkinen}, \citenamefont {Howard},
  \citenamefont {Ivanchenko}, \citenamefont {Johnson}, \citenamefont {Jones},
  \citenamefont {Kallenbach}, \citenamefont {Kanaya}, \citenamefont {Kawabata},
  \citenamefont {Kawabata}, \citenamefont {Kawaguti}, \citenamefont {Kelner},
  \citenamefont {Kent}, \citenamefont {Kimura}, \citenamefont {Kodama},
  \citenamefont {Kokoulin}, \citenamefont {Kossov}, \citenamefont {Kurashige},
  \citenamefont {Lamanna}, \citenamefont {Lampen}, \citenamefont {Lara},
  \citenamefont {Lefebure}, \citenamefont {Lei}, \citenamefont {Liendl},
  \citenamefont {Lockman}, \citenamefont {Longo}, \citenamefont {Magni},
  \citenamefont {Maire}, \citenamefont {Medernach}, \citenamefont {Minamimoto},
  \citenamefont {de~Freitas}, \citenamefont {Morita}, \citenamefont {Murakami},
  \citenamefont {Nagamatu}, \citenamefont {Nartallo}, \citenamefont {Nieminen},
  \citenamefont {Nishimura}, \citenamefont {Ohtsubo}, \citenamefont {Okamura},
  \citenamefont {O'Neale}, \citenamefont {Oohata}, \citenamefont {Paech},
  \citenamefont {Perl}, \citenamefont {Pfeiffer}, \citenamefont {Pia},
  \citenamefont {Ranjard}, \citenamefont {Rybin}, \citenamefont {Sadilov},
  \citenamefont {di~Salvo}, \citenamefont {Santin}, \citenamefont {Sasaki},
  \citenamefont {Savvas}, \citenamefont {Sawada}, \citenamefont {Scherer},
  \citenamefont {Sei}, \citenamefont {Sirotenko}, \citenamefont {Smith},
  \citenamefont {Starkov}, \citenamefont {Stoecker}, \citenamefont {Sulkimo},
  \citenamefont {Takahata}, \citenamefont {Tanaka}, \citenamefont {Tcherniaev},
  \citenamefont {Tehrani}, \citenamefont {Tropeano}, \citenamefont {Truscott},
  \citenamefont {Uno}, \citenamefont {Urban}, \citenamefont {Urban},
  \citenamefont {Verderi}, \citenamefont {Walkden}, \citenamefont {Wander},
  \citenamefont {Weber}, \citenamefont {Wellisch}, \citenamefont {Wenaus},
  \citenamefont {Williams}, \citenamefont {Wright}, \citenamefont {Yamada},
  \citenamefont {Yoshida},\ and\ \citenamefont {Zschiesche}}]{Agostinelli2003}%
  \BibitemOpen
  \bibfield  {author} {\bibinfo {author} {\bibfnamefont {S.}~\bibnamefont
  {Agostinelli}}, \bibinfo {author} {\bibfnamefont {J.}~\bibnamefont
  {Allison}}, \bibinfo {author} {\bibfnamefont {K.}~\bibnamefont {Amako}},
  \bibinfo {author} {\bibfnamefont {J.}~\bibnamefont {Apostolakis}}, \bibinfo
  {author} {\bibfnamefont {H.}~\bibnamefont {Araujo}}, \bibinfo {author}
  {\bibfnamefont {P.}~\bibnamefont {Arce}}, \bibinfo {author} {\bibfnamefont
  {M.}~\bibnamefont {Asai}}, \bibinfo {author} {\bibfnamefont {D.}~\bibnamefont
  {Axen}}, \bibinfo {author} {\bibfnamefont {S.}~\bibnamefont {Banerjee}},
  \bibinfo {author} {\bibfnamefont {G.}~\bibnamefont {Barrand}}, \bibinfo
  {author} {\bibfnamefont {F.}~\bibnamefont {Behner}}, \bibinfo {author}
  {\bibfnamefont {L.}~\bibnamefont {Bellagamba}}, \bibinfo {author}
  {\bibfnamefont {J.}~\bibnamefont {Boudreau}}, \bibinfo {author}
  {\bibfnamefont {L.}~\bibnamefont {Broglia}}, \bibinfo {author} {\bibfnamefont
  {A.}~\bibnamefont {Brunengo}}, \bibinfo {author} {\bibfnamefont
  {H.}~\bibnamefont {Burkhardt}}, \bibinfo {author} {\bibfnamefont
  {S.}~\bibnamefont {Chauvie}}, \bibinfo {author} {\bibfnamefont
  {J.}~\bibnamefont {Chuma}}, \bibinfo {author} {\bibfnamefont
  {R.}~\bibnamefont {Chytracek}}, \bibinfo {author} {\bibfnamefont
  {G.}~\bibnamefont {Cooperman}}, \bibinfo {author} {\bibfnamefont
  {G.}~\bibnamefont {Cosmo}}, \bibinfo {author} {\bibfnamefont
  {P.}~\bibnamefont {Degtyarenko}}, \bibinfo {author} {\bibfnamefont
  {A.}~\bibnamefont {Dell'Acqua}}, \bibinfo {author} {\bibfnamefont
  {G.}~\bibnamefont {Depaola}}, \bibinfo {author} {\bibfnamefont
  {D.}~\bibnamefont {Dietrich}}, \bibinfo {author} {\bibfnamefont
  {R.}~\bibnamefont {Enami}}, \bibinfo {author} {\bibfnamefont
  {A.}~\bibnamefont {Feliciello}}, \bibinfo {author} {\bibfnamefont
  {C.}~\bibnamefont {Ferguson}}, \bibinfo {author} {\bibfnamefont
  {H.}~\bibnamefont {Fesefeldt}}, \bibinfo {author} {\bibfnamefont
  {G.}~\bibnamefont {Folger}}, \bibinfo {author} {\bibfnamefont
  {F.}~\bibnamefont {Foppiano}}, \bibinfo {author} {\bibfnamefont
  {A.}~\bibnamefont {Forti}}, \bibinfo {author} {\bibfnamefont
  {S.}~\bibnamefont {Garelli}}, \bibinfo {author} {\bibfnamefont
  {S.}~\bibnamefont {Giani}}, \bibinfo {author} {\bibfnamefont
  {R.}~\bibnamefont {Giannitrapani}}, \bibinfo {author} {\bibfnamefont
  {D.}~\bibnamefont {Gibin}}, \bibinfo {author} {\bibfnamefont {J.~J.~G.}\
  \bibnamefont {Cadenas}}, \bibinfo {author} {\bibfnamefont {I.}~\bibnamefont
  {Gonzalez}}, \bibinfo {author} {\bibfnamefont {G.~G.}\ \bibnamefont {Abril}},
  \bibinfo {author} {\bibfnamefont {G.}~\bibnamefont {Greeniaus}}, \bibinfo
  {author} {\bibfnamefont {W.}~\bibnamefont {Greiner}}, \bibinfo {author}
  {\bibfnamefont {V.}~\bibnamefont {Grichine}}, \bibinfo {author}
  {\bibfnamefont {A.}~\bibnamefont {Grossheim}}, \bibinfo {author}
  {\bibfnamefont {S.}~\bibnamefont {Guatelli}}, \bibinfo {author}
  {\bibfnamefont {P.}~\bibnamefont {Gumplinger}}, \bibinfo {author}
  {\bibfnamefont {R.}~\bibnamefont {Hamatsu}}, \bibinfo {author} {\bibfnamefont
  {K.}~\bibnamefont {Hashimoto}}, \bibinfo {author} {\bibfnamefont
  {H.}~\bibnamefont {Hasui}}, \bibinfo {author} {\bibfnamefont
  {A.}~\bibnamefont {Heikkinen}}, \bibinfo {author} {\bibfnamefont
  {A.}~\bibnamefont {Howard}}, \bibinfo {author} {\bibfnamefont
  {V.}~\bibnamefont {Ivanchenko}}, \bibinfo {author} {\bibfnamefont
  {A.}~\bibnamefont {Johnson}}, \bibinfo {author} {\bibfnamefont {F.~W.}\
  \bibnamefont {Jones}}, \bibinfo {author} {\bibfnamefont {J.}~\bibnamefont
  {Kallenbach}}, \bibinfo {author} {\bibfnamefont {N.}~\bibnamefont {Kanaya}},
  \bibinfo {author} {\bibfnamefont {M.}~\bibnamefont {Kawabata}}, \bibinfo
  {author} {\bibfnamefont {Y.}~\bibnamefont {Kawabata}}, \bibinfo {author}
  {\bibfnamefont {M.}~\bibnamefont {Kawaguti}}, \bibinfo {author}
  {\bibfnamefont {S.}~\bibnamefont {Kelner}}, \bibinfo {author} {\bibfnamefont
  {P.}~\bibnamefont {Kent}}, \bibinfo {author} {\bibfnamefont {A.}~\bibnamefont
  {Kimura}}, \bibinfo {author} {\bibfnamefont {T.}~\bibnamefont {Kodama}},
  \bibinfo {author} {\bibfnamefont {R.}~\bibnamefont {Kokoulin}}, \bibinfo
  {author} {\bibfnamefont {M.}~\bibnamefont {Kossov}}, \bibinfo {author}
  {\bibfnamefont {H.}~\bibnamefont {Kurashige}}, \bibinfo {author}
  {\bibfnamefont {E.}~\bibnamefont {Lamanna}}, \bibinfo {author} {\bibfnamefont
  {T.}~\bibnamefont {Lampen}}, \bibinfo {author} {\bibfnamefont
  {V.}~\bibnamefont {Lara}}, \bibinfo {author} {\bibfnamefont {V.}~\bibnamefont
  {Lefebure}}, \bibinfo {author} {\bibfnamefont {F.}~\bibnamefont {Lei}},
  \bibinfo {author} {\bibfnamefont {M.}~\bibnamefont {Liendl}}, \bibinfo
  {author} {\bibfnamefont {W.}~\bibnamefont {Lockman}}, \bibinfo {author}
  {\bibfnamefont {F.}~\bibnamefont {Longo}}, \bibinfo {author} {\bibfnamefont
  {S.}~\bibnamefont {Magni}}, \bibinfo {author} {\bibfnamefont
  {M.}~\bibnamefont {Maire}}, \bibinfo {author} {\bibfnamefont
  {E.}~\bibnamefont {Medernach}}, \bibinfo {author} {\bibfnamefont
  {K.}~\bibnamefont {Minamimoto}}, \bibinfo {author} {\bibfnamefont {P.~M.}\
  \bibnamefont {de~Freitas}}, \bibinfo {author} {\bibfnamefont
  {Y.}~\bibnamefont {Morita}}, \bibinfo {author} {\bibfnamefont
  {K.}~\bibnamefont {Murakami}}, \bibinfo {author} {\bibfnamefont
  {M.}~\bibnamefont {Nagamatu}}, \bibinfo {author} {\bibfnamefont
  {R.}~\bibnamefont {Nartallo}}, \bibinfo {author} {\bibfnamefont
  {P.}~\bibnamefont {Nieminen}}, \bibinfo {author} {\bibfnamefont
  {T.}~\bibnamefont {Nishimura}}, \bibinfo {author} {\bibfnamefont
  {K.}~\bibnamefont {Ohtsubo}}, \bibinfo {author} {\bibfnamefont
  {M.}~\bibnamefont {Okamura}}, \bibinfo {author} {\bibfnamefont
  {S.}~\bibnamefont {O'Neale}}, \bibinfo {author} {\bibfnamefont
  {Y.}~\bibnamefont {Oohata}}, \bibinfo {author} {\bibfnamefont
  {K.}~\bibnamefont {Paech}}, \bibinfo {author} {\bibfnamefont
  {J.}~\bibnamefont {Perl}}, \bibinfo {author} {\bibfnamefont {A.}~\bibnamefont
  {Pfeiffer}}, \bibinfo {author} {\bibfnamefont {M.~G.}\ \bibnamefont {Pia}},
  \bibinfo {author} {\bibfnamefont {F.}~\bibnamefont {Ranjard}}, \bibinfo
  {author} {\bibfnamefont {A.}~\bibnamefont {Rybin}}, \bibinfo {author}
  {\bibfnamefont {S.}~\bibnamefont {Sadilov}}, \bibinfo {author} {\bibfnamefont
  {E.}~\bibnamefont {di~Salvo}}, \bibinfo {author} {\bibfnamefont
  {G.}~\bibnamefont {Santin}}, \bibinfo {author} {\bibfnamefont
  {T.}~\bibnamefont {Sasaki}}, \bibinfo {author} {\bibfnamefont
  {N.}~\bibnamefont {Savvas}}, \bibinfo {author} {\bibfnamefont
  {Y.}~\bibnamefont {Sawada}}, \bibinfo {author} {\bibfnamefont
  {S.}~\bibnamefont {Scherer}}, \bibinfo {author} {\bibfnamefont
  {S.}~\bibnamefont {Sei}}, \bibinfo {author} {\bibfnamefont {V.}~\bibnamefont
  {Sirotenko}}, \bibinfo {author} {\bibfnamefont {D.}~\bibnamefont {Smith}},
  \bibinfo {author} {\bibfnamefont {N.}~\bibnamefont {Starkov}}, \bibinfo
  {author} {\bibfnamefont {H.}~\bibnamefont {Stoecker}}, \bibinfo {author}
  {\bibfnamefont {J.}~\bibnamefont {Sulkimo}}, \bibinfo {author} {\bibfnamefont
  {M.}~\bibnamefont {Takahata}}, \bibinfo {author} {\bibfnamefont
  {S.}~\bibnamefont {Tanaka}}, \bibinfo {author} {\bibfnamefont
  {E.}~\bibnamefont {Tcherniaev}}, \bibinfo {author} {\bibfnamefont {E.~S.}\
  \bibnamefont {Tehrani}}, \bibinfo {author} {\bibfnamefont {M.}~\bibnamefont
  {Tropeano}}, \bibinfo {author} {\bibfnamefont {P.}~\bibnamefont {Truscott}},
  \bibinfo {author} {\bibfnamefont {H.}~\bibnamefont {Uno}}, \bibinfo {author}
  {\bibfnamefont {L.}~\bibnamefont {Urban}}, \bibinfo {author} {\bibfnamefont
  {P.}~\bibnamefont {Urban}}, \bibinfo {author} {\bibfnamefont
  {M.}~\bibnamefont {Verderi}}, \bibinfo {author} {\bibfnamefont
  {A.}~\bibnamefont {Walkden}}, \bibinfo {author} {\bibfnamefont
  {W.}~\bibnamefont {Wander}}, \bibinfo {author} {\bibfnamefont
  {H.}~\bibnamefont {Weber}}, \bibinfo {author} {\bibfnamefont {J.~P.}\
  \bibnamefont {Wellisch}}, \bibinfo {author} {\bibfnamefont {T.}~\bibnamefont
  {Wenaus}}, \bibinfo {author} {\bibfnamefont {D.~C.}\ \bibnamefont
  {Williams}}, \bibinfo {author} {\bibfnamefont {D.}~\bibnamefont {Wright}},
  \bibinfo {author} {\bibfnamefont {T.}~\bibnamefont {Yamada}}, \bibinfo
  {author} {\bibfnamefont {H.}~\bibnamefont {Yoshida}}, \ and\ \bibinfo
  {author} {\bibfnamefont {D.}~\bibnamefont {Zschiesche}},\ }\bibfield  {title}
  {\enquote {\bibinfo {title} {Geant4 - a simulation toolkit},}\ }\href
  {\doibase 10.1016/S0168-9002(03)01368-8} {\bibfield  {journal} {\bibinfo
  {journal} {Nuclear Instruments and Methods in Physics Research, Section A:
  Accelerators, Spectrometers, Detectors and Associated Equipment}\ }\textbf
  {\bibinfo {volume} {506}},\ \bibinfo {pages} {250--303} (\bibinfo {year}
  {2003})}\BibitemShut {NoStop}%
\bibitem [{\citenamefont {Allison}\ \emph {et~al.}(2006)\citenamefont
  {Allison}, \citenamefont {Amako}, \citenamefont {Apostolakis}, \citenamefont
  {Araujo}, \citenamefont {Dubois}, \citenamefont {Asai}, \citenamefont
  {Barrand}, \citenamefont {Capra}, \citenamefont {Chauvie}, \citenamefont
  {Chytracek}, \citenamefont {Cirrone}, \citenamefont {Cooperman},
  \citenamefont {Cosmo}, \citenamefont {Cuttone}, \citenamefont {Daquino},
  \citenamefont {Donszelmann}, \citenamefont {Dressel}, \citenamefont {Folger},
  \citenamefont {Foppiano}, \citenamefont {Generowicz}, \citenamefont
  {Grichine}, \citenamefont {Guatelli}, \citenamefont {Gumplinger},
  \citenamefont {Heikkinen}, \citenamefont {Hrivnacova}, \citenamefont
  {Howard}, \citenamefont {Incerti}, \citenamefont {Ivanchenko}, \citenamefont
  {Johnson}, \citenamefont {Jones}, \citenamefont {Koi}, \citenamefont
  {Kokoulin}, \citenamefont {Kossov}, \citenamefont {Kurashige}, \citenamefont
  {Lara}, \citenamefont {Larsson}, \citenamefont {Lei}, \citenamefont {Longo},
  \citenamefont {Maire}, \citenamefont {Mantero}, \citenamefont {Mascialino},
  \citenamefont {McLaren}, \citenamefont {Lorenzo}, \citenamefont {Minamimoto},
  \citenamefont {Murakami}, \citenamefont {Nieminen}, \citenamefont {Pandola},
  \citenamefont {Parlati}, \citenamefont {Peralta}, \citenamefont {Perl},
  \citenamefont {Pfeiffer}, \citenamefont {Pia}, \citenamefont {Ribon},
  \citenamefont {Rodrigues}, \citenamefont {Russo}, \citenamefont {Sadilov},
  \citenamefont {Santin}, \citenamefont {Sasaki}, \citenamefont {Smith},
  \citenamefont {Starkov}, \citenamefont {Tanaka}, \citenamefont {Tcherniaev},
  \citenamefont {Tomé}, \citenamefont {Trindade}, \citenamefont {Truscott},
  \citenamefont {Urban}, \citenamefont {Verderi}, \citenamefont {Walkden},
  \citenamefont {Wellisch}, \citenamefont {Williams}, \citenamefont {Wright},
  \citenamefont {Yoshida},\ and\ \citenamefont {Peirgentili}}]{Allison2006}%
  \BibitemOpen
  \bibfield  {author} {\bibinfo {author} {\bibfnamefont {J.}~\bibnamefont
  {Allison}}, \bibinfo {author} {\bibfnamefont {K.}~\bibnamefont {Amako}},
  \bibinfo {author} {\bibfnamefont {J.}~\bibnamefont {Apostolakis}}, \bibinfo
  {author} {\bibfnamefont {H.}~\bibnamefont {Araujo}}, \bibinfo {author}
  {\bibfnamefont {P.~A.}\ \bibnamefont {Dubois}}, \bibinfo {author}
  {\bibfnamefont {M.}~\bibnamefont {Asai}}, \bibinfo {author} {\bibfnamefont
  {G.}~\bibnamefont {Barrand}}, \bibinfo {author} {\bibfnamefont
  {R.}~\bibnamefont {Capra}}, \bibinfo {author} {\bibfnamefont
  {S.}~\bibnamefont {Chauvie}}, \bibinfo {author} {\bibfnamefont
  {R.}~\bibnamefont {Chytracek}}, \bibinfo {author} {\bibfnamefont {G.~A.}\
  \bibnamefont {Cirrone}}, \bibinfo {author} {\bibfnamefont {G.}~\bibnamefont
  {Cooperman}}, \bibinfo {author} {\bibfnamefont {G.}~\bibnamefont {Cosmo}},
  \bibinfo {author} {\bibfnamefont {G.}~\bibnamefont {Cuttone}}, \bibinfo
  {author} {\bibfnamefont {G.~G.}\ \bibnamefont {Daquino}}, \bibinfo {author}
  {\bibfnamefont {M.}~\bibnamefont {Donszelmann}}, \bibinfo {author}
  {\bibfnamefont {M.}~\bibnamefont {Dressel}}, \bibinfo {author} {\bibfnamefont
  {G.}~\bibnamefont {Folger}}, \bibinfo {author} {\bibfnamefont
  {F.}~\bibnamefont {Foppiano}}, \bibinfo {author} {\bibfnamefont
  {J.}~\bibnamefont {Generowicz}}, \bibinfo {author} {\bibfnamefont
  {V.}~\bibnamefont {Grichine}}, \bibinfo {author} {\bibfnamefont
  {S.}~\bibnamefont {Guatelli}}, \bibinfo {author} {\bibfnamefont
  {P.}~\bibnamefont {Gumplinger}}, \bibinfo {author} {\bibfnamefont
  {A.}~\bibnamefont {Heikkinen}}, \bibinfo {author} {\bibfnamefont
  {I.}~\bibnamefont {Hrivnacova}}, \bibinfo {author} {\bibfnamefont
  {A.}~\bibnamefont {Howard}}, \bibinfo {author} {\bibfnamefont
  {S.}~\bibnamefont {Incerti}}, \bibinfo {author} {\bibfnamefont
  {V.}~\bibnamefont {Ivanchenko}}, \bibinfo {author} {\bibfnamefont
  {T.}~\bibnamefont {Johnson}}, \bibinfo {author} {\bibfnamefont
  {F.}~\bibnamefont {Jones}}, \bibinfo {author} {\bibfnamefont
  {T.}~\bibnamefont {Koi}}, \bibinfo {author} {\bibfnamefont {R.}~\bibnamefont
  {Kokoulin}}, \bibinfo {author} {\bibfnamefont {M.}~\bibnamefont {Kossov}},
  \bibinfo {author} {\bibfnamefont {H.}~\bibnamefont {Kurashige}}, \bibinfo
  {author} {\bibfnamefont {V.}~\bibnamefont {Lara}}, \bibinfo {author}
  {\bibfnamefont {S.}~\bibnamefont {Larsson}}, \bibinfo {author} {\bibfnamefont
  {F.}~\bibnamefont {Lei}}, \bibinfo {author} {\bibfnamefont {F.}~\bibnamefont
  {Longo}}, \bibinfo {author} {\bibfnamefont {M.}~\bibnamefont {Maire}},
  \bibinfo {author} {\bibfnamefont {A.}~\bibnamefont {Mantero}}, \bibinfo
  {author} {\bibfnamefont {B.}~\bibnamefont {Mascialino}}, \bibinfo {author}
  {\bibfnamefont {I.}~\bibnamefont {McLaren}}, \bibinfo {author} {\bibfnamefont
  {P.~M.}\ \bibnamefont {Lorenzo}}, \bibinfo {author} {\bibfnamefont
  {K.}~\bibnamefont {Minamimoto}}, \bibinfo {author} {\bibfnamefont
  {K.}~\bibnamefont {Murakami}}, \bibinfo {author} {\bibfnamefont
  {P.}~\bibnamefont {Nieminen}}, \bibinfo {author} {\bibfnamefont
  {L.}~\bibnamefont {Pandola}}, \bibinfo {author} {\bibfnamefont
  {S.}~\bibnamefont {Parlati}}, \bibinfo {author} {\bibfnamefont
  {L.}~\bibnamefont {Peralta}}, \bibinfo {author} {\bibfnamefont
  {J.}~\bibnamefont {Perl}}, \bibinfo {author} {\bibfnamefont {A.}~\bibnamefont
  {Pfeiffer}}, \bibinfo {author} {\bibfnamefont {M.~G.}\ \bibnamefont {Pia}},
  \bibinfo {author} {\bibfnamefont {A.}~\bibnamefont {Ribon}}, \bibinfo
  {author} {\bibfnamefont {P.}~\bibnamefont {Rodrigues}}, \bibinfo {author}
  {\bibfnamefont {G.}~\bibnamefont {Russo}}, \bibinfo {author} {\bibfnamefont
  {S.}~\bibnamefont {Sadilov}}, \bibinfo {author} {\bibfnamefont
  {G.}~\bibnamefont {Santin}}, \bibinfo {author} {\bibfnamefont
  {T.}~\bibnamefont {Sasaki}}, \bibinfo {author} {\bibfnamefont
  {D.}~\bibnamefont {Smith}}, \bibinfo {author} {\bibfnamefont
  {N.}~\bibnamefont {Starkov}}, \bibinfo {author} {\bibfnamefont
  {S.}~\bibnamefont {Tanaka}}, \bibinfo {author} {\bibfnamefont
  {E.}~\bibnamefont {Tcherniaev}}, \bibinfo {author} {\bibfnamefont
  {B.}~\bibnamefont {Tomé}}, \bibinfo {author} {\bibfnamefont
  {A.}~\bibnamefont {Trindade}}, \bibinfo {author} {\bibfnamefont
  {P.}~\bibnamefont {Truscott}}, \bibinfo {author} {\bibfnamefont
  {L.}~\bibnamefont {Urban}}, \bibinfo {author} {\bibfnamefont
  {M.}~\bibnamefont {Verderi}}, \bibinfo {author} {\bibfnamefont
  {A.}~\bibnamefont {Walkden}}, \bibinfo {author} {\bibfnamefont {J.~P.}\
  \bibnamefont {Wellisch}}, \bibinfo {author} {\bibfnamefont {D.~C.}\
  \bibnamefont {Williams}}, \bibinfo {author} {\bibfnamefont {D.}~\bibnamefont
  {Wright}}, \bibinfo {author} {\bibfnamefont {H.}~\bibnamefont {Yoshida}}, \
  and\ \bibinfo {author} {\bibfnamefont {M.}~\bibnamefont {Peirgentili}},\
  }\bibfield  {title} {\enquote {\bibinfo {title} {Geant4 developments and
  applications},}\ }\href {\doibase 10.1109/TNS.2006.869826} {\bibfield
  {journal} {\bibinfo  {journal} {IEEE Transactions on Nuclear Science}\
  }\textbf {\bibinfo {volume} {53}},\ \bibinfo {pages} {270--278} (\bibinfo
  {year} {2006})}\BibitemShut {NoStop}%
\bibitem [{\citenamefont {Allison}\ \emph {et~al.}(2016)\citenamefont
  {Allison}, \citenamefont {Amako}, \citenamefont {Apostolakis}, \citenamefont
  {Arce}, \citenamefont {Asai}, \citenamefont {Aso}, \citenamefont {Bagli},
  \citenamefont {Bagulya}, \citenamefont {Banerjee}, \citenamefont {Barrand},
  \citenamefont {Beck}, \citenamefont {Bogdanov}, \citenamefont {Brandt},
  \citenamefont {Brown}, \citenamefont {Burkhardt}, \citenamefont {Canal},
  \citenamefont {Cano-Ott}, \citenamefont {Chauvie}, \citenamefont {Cho},
  \citenamefont {Cirrone}, \citenamefont {Cooperman}, \citenamefont
  {Cortés-Giraldo}, \citenamefont {Cosmo}, \citenamefont {Cuttone},
  \citenamefont {Depaola}, \citenamefont {Desorgher}, \citenamefont {Dong},
  \citenamefont {Dotti}, \citenamefont {Elvira}, \citenamefont {Folger},
  \citenamefont {Francis}, \citenamefont {Galoyan}, \citenamefont {Garnier},
  \citenamefont {Gayer}, \citenamefont {Genser}, \citenamefont {Grichine},
  \citenamefont {Guatelli}, \citenamefont {Guèye}, \citenamefont {Gumplinger},
  \citenamefont {Howard}, \citenamefont {Hřivnáčová}, \citenamefont
  {Hwang}, \citenamefont {Incerti}, \citenamefont {Ivanchenko}, \citenamefont
  {Ivanchenko}, \citenamefont {Jones}, \citenamefont {Jun}, \citenamefont
  {Kaitaniemi}, \citenamefont {Karakatsanis}, \citenamefont {Karamitrosi},
  \citenamefont {Kelsey}, \citenamefont {Kimura}, \citenamefont {Koi},
  \citenamefont {Kurashige}, \citenamefont {Lechner}, \citenamefont {Lee},
  \citenamefont {Longo}, \citenamefont {Maire}, \citenamefont {Mancusi},
  \citenamefont {Mantero}, \citenamefont {Mendoza}, \citenamefont {Morgan},
  \citenamefont {Murakami}, \citenamefont {Nikitina}, \citenamefont {Pandola},
  \citenamefont {Paprocki}, \citenamefont {Perl}, \citenamefont {Petrović},
  \citenamefont {Pia}, \citenamefont {Pokorski}, \citenamefont {Quesada},
  \citenamefont {Raine}, \citenamefont {Reis}, \citenamefont {Ribon},
  \citenamefont {Fira}, \citenamefont {Romano}, \citenamefont {Russo},
  \citenamefont {Santin}, \citenamefont {Sasaki}, \citenamefont {Sawkey},
  \citenamefont {Shin}, \citenamefont {Strakovsky}, \citenamefont {Taborda},
  \citenamefont {Tanaka}, \citenamefont {Tomé}, \citenamefont {Toshito},
  \citenamefont {Tran}, \citenamefont {Truscott}, \citenamefont {Urban},
  \citenamefont {Uzhinsky}, \citenamefont {Verbeke}, \citenamefont {Verderi},
  \citenamefont {Wendt}, \citenamefont {Wenzel}, \citenamefont {Wright},
  \citenamefont {Wright}, \citenamefont {Yamashita}, \citenamefont {Yarba},\
  and\ \citenamefont {Yoshida}}]{Allison2016}%
  \BibitemOpen
  \bibfield  {author} {\bibinfo {author} {\bibfnamefont {J.}~\bibnamefont
  {Allison}}, \bibinfo {author} {\bibfnamefont {K.}~\bibnamefont {Amako}},
  \bibinfo {author} {\bibfnamefont {J.}~\bibnamefont {Apostolakis}}, \bibinfo
  {author} {\bibfnamefont {P.}~\bibnamefont {Arce}}, \bibinfo {author}
  {\bibfnamefont {M.}~\bibnamefont {Asai}}, \bibinfo {author} {\bibfnamefont
  {T.}~\bibnamefont {Aso}}, \bibinfo {author} {\bibfnamefont {E.}~\bibnamefont
  {Bagli}}, \bibinfo {author} {\bibfnamefont {A.}~\bibnamefont {Bagulya}},
  \bibinfo {author} {\bibfnamefont {S.}~\bibnamefont {Banerjee}}, \bibinfo
  {author} {\bibfnamefont {G.}~\bibnamefont {Barrand}}, \bibinfo {author}
  {\bibfnamefont {B.~R.}\ \bibnamefont {Beck}}, \bibinfo {author}
  {\bibfnamefont {A.~G.}\ \bibnamefont {Bogdanov}}, \bibinfo {author}
  {\bibfnamefont {D.}~\bibnamefont {Brandt}}, \bibinfo {author} {\bibfnamefont
  {J.~M.}\ \bibnamefont {Brown}}, \bibinfo {author} {\bibfnamefont
  {H.}~\bibnamefont {Burkhardt}}, \bibinfo {author} {\bibfnamefont
  {P.}~\bibnamefont {Canal}}, \bibinfo {author} {\bibfnamefont
  {D.}~\bibnamefont {Cano-Ott}}, \bibinfo {author} {\bibfnamefont
  {S.}~\bibnamefont {Chauvie}}, \bibinfo {author} {\bibfnamefont
  {K.}~\bibnamefont {Cho}}, \bibinfo {author} {\bibfnamefont {G.~A.}\
  \bibnamefont {Cirrone}}, \bibinfo {author} {\bibfnamefont {G.}~\bibnamefont
  {Cooperman}}, \bibinfo {author} {\bibfnamefont {M.~A.}\ \bibnamefont
  {Cortés-Giraldo}}, \bibinfo {author} {\bibfnamefont {G.}~\bibnamefont
  {Cosmo}}, \bibinfo {author} {\bibfnamefont {G.}~\bibnamefont {Cuttone}},
  \bibinfo {author} {\bibfnamefont {G.}~\bibnamefont {Depaola}}, \bibinfo
  {author} {\bibfnamefont {L.}~\bibnamefont {Desorgher}}, \bibinfo {author}
  {\bibfnamefont {X.}~\bibnamefont {Dong}}, \bibinfo {author} {\bibfnamefont
  {A.}~\bibnamefont {Dotti}}, \bibinfo {author} {\bibfnamefont {V.~D.}\
  \bibnamefont {Elvira}}, \bibinfo {author} {\bibfnamefont {G.}~\bibnamefont
  {Folger}}, \bibinfo {author} {\bibfnamefont {Z.}~\bibnamefont {Francis}},
  \bibinfo {author} {\bibfnamefont {A.}~\bibnamefont {Galoyan}}, \bibinfo
  {author} {\bibfnamefont {L.}~\bibnamefont {Garnier}}, \bibinfo {author}
  {\bibfnamefont {M.}~\bibnamefont {Gayer}}, \bibinfo {author} {\bibfnamefont
  {K.~L.}\ \bibnamefont {Genser}}, \bibinfo {author} {\bibfnamefont {V.~M.}\
  \bibnamefont {Grichine}}, \bibinfo {author} {\bibfnamefont {S.}~\bibnamefont
  {Guatelli}}, \bibinfo {author} {\bibfnamefont {P.}~\bibnamefont {Guèye}},
  \bibinfo {author} {\bibfnamefont {P.}~\bibnamefont {Gumplinger}}, \bibinfo
  {author} {\bibfnamefont {A.~S.}\ \bibnamefont {Howard}}, \bibinfo {author}
  {\bibfnamefont {I.}~\bibnamefont {Hřivnáčová}}, \bibinfo {author}
  {\bibfnamefont {S.}~\bibnamefont {Hwang}}, \bibinfo {author} {\bibfnamefont
  {S.}~\bibnamefont {Incerti}}, \bibinfo {author} {\bibfnamefont
  {A.}~\bibnamefont {Ivanchenko}}, \bibinfo {author} {\bibfnamefont {V.~N.}\
  \bibnamefont {Ivanchenko}}, \bibinfo {author} {\bibfnamefont {F.~W.}\
  \bibnamefont {Jones}}, \bibinfo {author} {\bibfnamefont {S.~Y.}\ \bibnamefont
  {Jun}}, \bibinfo {author} {\bibfnamefont {P.}~\bibnamefont {Kaitaniemi}},
  \bibinfo {author} {\bibfnamefont {N.}~\bibnamefont {Karakatsanis}}, \bibinfo
  {author} {\bibfnamefont {M.}~\bibnamefont {Karamitrosi}}, \bibinfo {author}
  {\bibfnamefont {M.}~\bibnamefont {Kelsey}}, \bibinfo {author} {\bibfnamefont
  {A.}~\bibnamefont {Kimura}}, \bibinfo {author} {\bibfnamefont
  {T.}~\bibnamefont {Koi}}, \bibinfo {author} {\bibfnamefont {H.}~\bibnamefont
  {Kurashige}}, \bibinfo {author} {\bibfnamefont {A.}~\bibnamefont {Lechner}},
  \bibinfo {author} {\bibfnamefont {S.~B.}\ \bibnamefont {Lee}}, \bibinfo
  {author} {\bibfnamefont {F.}~\bibnamefont {Longo}}, \bibinfo {author}
  {\bibfnamefont {M.}~\bibnamefont {Maire}}, \bibinfo {author} {\bibfnamefont
  {D.}~\bibnamefont {Mancusi}}, \bibinfo {author} {\bibfnamefont
  {A.}~\bibnamefont {Mantero}}, \bibinfo {author} {\bibfnamefont
  {E.}~\bibnamefont {Mendoza}}, \bibinfo {author} {\bibfnamefont
  {B.}~\bibnamefont {Morgan}}, \bibinfo {author} {\bibfnamefont
  {K.}~\bibnamefont {Murakami}}, \bibinfo {author} {\bibfnamefont
  {T.}~\bibnamefont {Nikitina}}, \bibinfo {author} {\bibfnamefont
  {L.}~\bibnamefont {Pandola}}, \bibinfo {author} {\bibfnamefont
  {P.}~\bibnamefont {Paprocki}}, \bibinfo {author} {\bibfnamefont
  {J.}~\bibnamefont {Perl}}, \bibinfo {author} {\bibfnamefont {I.}~\bibnamefont
  {Petrović}}, \bibinfo {author} {\bibfnamefont {M.~G.}\ \bibnamefont {Pia}},
  \bibinfo {author} {\bibfnamefont {W.}~\bibnamefont {Pokorski}}, \bibinfo
  {author} {\bibfnamefont {J.~M.}\ \bibnamefont {Quesada}}, \bibinfo {author}
  {\bibfnamefont {M.}~\bibnamefont {Raine}}, \bibinfo {author} {\bibfnamefont
  {M.~A.}\ \bibnamefont {Reis}}, \bibinfo {author} {\bibfnamefont
  {A.}~\bibnamefont {Ribon}}, \bibinfo {author} {\bibfnamefont {A.~R.}\
  \bibnamefont {Fira}}, \bibinfo {author} {\bibfnamefont {F.}~\bibnamefont
  {Romano}}, \bibinfo {author} {\bibfnamefont {G.}~\bibnamefont {Russo}},
  \bibinfo {author} {\bibfnamefont {G.}~\bibnamefont {Santin}}, \bibinfo
  {author} {\bibfnamefont {T.}~\bibnamefont {Sasaki}}, \bibinfo {author}
  {\bibfnamefont {D.}~\bibnamefont {Sawkey}}, \bibinfo {author} {\bibfnamefont
  {J.~I.}\ \bibnamefont {Shin}}, \bibinfo {author} {\bibfnamefont {I.~I.}\
  \bibnamefont {Strakovsky}}, \bibinfo {author} {\bibfnamefont
  {A.}~\bibnamefont {Taborda}}, \bibinfo {author} {\bibfnamefont
  {S.}~\bibnamefont {Tanaka}}, \bibinfo {author} {\bibfnamefont
  {B.}~\bibnamefont {Tomé}}, \bibinfo {author} {\bibfnamefont
  {T.}~\bibnamefont {Toshito}}, \bibinfo {author} {\bibfnamefont {H.~N.}\
  \bibnamefont {Tran}}, \bibinfo {author} {\bibfnamefont {P.~R.}\ \bibnamefont
  {Truscott}}, \bibinfo {author} {\bibfnamefont {L.}~\bibnamefont {Urban}},
  \bibinfo {author} {\bibfnamefont {V.}~\bibnamefont {Uzhinsky}}, \bibinfo
  {author} {\bibfnamefont {J.~M.}\ \bibnamefont {Verbeke}}, \bibinfo {author}
  {\bibfnamefont {M.}~\bibnamefont {Verderi}}, \bibinfo {author} {\bibfnamefont
  {B.~L.}\ \bibnamefont {Wendt}}, \bibinfo {author} {\bibfnamefont
  {H.}~\bibnamefont {Wenzel}}, \bibinfo {author} {\bibfnamefont {D.~H.}\
  \bibnamefont {Wright}}, \bibinfo {author} {\bibfnamefont {D.~M.}\
  \bibnamefont {Wright}}, \bibinfo {author} {\bibfnamefont {T.}~\bibnamefont
  {Yamashita}}, \bibinfo {author} {\bibfnamefont {J.}~\bibnamefont {Yarba}}, \
  and\ \bibinfo {author} {\bibfnamefont {H.}~\bibnamefont {Yoshida}},\
  }\bibfield  {title} {\enquote {\bibinfo {title} {Recent developments in
  geant4},}\ }\href {\doibase 10.1016/j.nima.2016.06.125} {\bibfield  {journal}
  {\bibinfo  {journal} {Nuclear Instruments and Methods in Physics Research,
  Section A: Accelerators, Spectrometers, Detectors and Associated Equipment}\
  }\textbf {\bibinfo {volume} {835}},\ \bibinfo {pages} {186--225} (\bibinfo
  {year} {2016})}\BibitemShut {NoStop}%
\bibitem [{\citenamefont {Shreter}\ and\ \citenamefont
  {Kalish}()}]{Shretter1979}%
  \BibitemOpen
  \bibfield  {author} {\bibinfo {author} {\bibfnamefont {U.}~\bibnamefont
  {Shreter}}\ and\ \bibinfo {author} {\bibfnamefont {R.}~\bibnamefont
  {Kalish}},\ }\href@noop {} {\enquote {\bibinfo {title} {A simple method for
  preparing thin 241am sources*},}\ }\BibitemShut {NoStop}%
\bibitem [{\citenamefont {Hokin}\ and\ \citenamefont {Stek}(1988)}]{Hokin1988}%
  \BibitemOpen
  \bibfield  {author} {\bibinfo {author} {\bibfnamefont {S.~A.}\ \bibnamefont
  {Hokin}}\ and\ \bibinfo {author} {\bibfnamefont {P.~C.}\ \bibnamefont
  {Stek}},\ }\bibfield  {title} {\enquote {\bibinfo {title} {Multichannel
  scintillator probe for energetic electron measurements},}\ }\href {\doibase
  10.1063/1.1139962} {\bibfield  {journal} {\bibinfo  {journal} {Review of
  Scientific Instruments}\ }\textbf {\bibinfo {volume} {59}},\ \bibinfo {pages}
  {2366--2369} (\bibinfo {year} {1988})}\BibitemShut {NoStop}%
\bibitem [{\citenamefont {Nefedov}\ and\ \citenamefont
  {Usenko}(2016)}]{Nefedov2016}%
  \BibitemOpen
  \bibfield  {author} {\bibinfo {author} {\bibfnamefont {Y.~Y.}\ \bibnamefont
  {Nefedov}}\ and\ \bibinfo {author} {\bibfnamefont {P.~L.}\ \bibnamefont
  {Usenko}},\ }\bibfield  {title} {\enquote {\bibinfo {title} {A scintillation
  detector of pulsed soft x rays},}\ }\href {\doibase
  10.1134/S0020441216010097} {\bibfield  {journal} {\bibinfo  {journal}
  {Instruments and Experimental Techniques}\ }\textbf {\bibinfo {volume}
  {59}},\ \bibinfo {pages} {115--119} (\bibinfo {year} {2016})}\BibitemShut
  {NoStop}%
\bibitem [{\citenamefont {Ashrafi}\ and\ \citenamefont
  {Gol}(2011)}]{Ashrafi2011}%
  \BibitemOpen
  \bibfield  {author} {\bibinfo {author} {\bibfnamefont {S.}~\bibnamefont
  {Ashrafi}}\ and\ \bibinfo {author} {\bibfnamefont {M.~G.}\ \bibnamefont
  {Gol}},\ }\bibfield  {title} {\enquote {\bibinfo {title} {Energy calibration
  of thin plastic scintillators using compton scattered gammarays},}\ }\href
  {\doibase 10.1016/j.nima.2011.04.003} {\bibfield  {journal} {\bibinfo
  {journal} {Nuclear Instruments and Methods in Physics Research, Section A:
  Accelerators, Spectrometers, Detectors and Associated Equipment}\ }\textbf
  {\bibinfo {volume} {642}},\ \bibinfo {pages} {70--74} (\bibinfo {year}
  {2011})}\BibitemShut {NoStop}%
\bibitem [{\citenamefont {Kelleter}\ and\ \citenamefont
  {Jolly}(2020)}]{Kelleter2020}%
  \BibitemOpen
  \bibfield  {author} {\bibinfo {author} {\bibfnamefont {L.}~\bibnamefont
  {Kelleter}}\ and\ \bibinfo {author} {\bibfnamefont {S.}~\bibnamefont
  {Jolly}},\ }\bibfield  {title} {\enquote {\bibinfo {title} {A mathematical
  expression for depth-light curves of therapeutic proton beams in a quenching
  scintillator},}\ }\href {\doibase 10.1002/mp.14099} {\bibfield  {journal}
  {\bibinfo  {journal} {Medical Physics}\ }\textbf {\bibinfo {volume} {47}},\
  \bibinfo {pages} {2300--2308} (\bibinfo {year} {2020})}\BibitemShut {NoStop}%
\bibitem [{\citenamefont {Tretyak}(2010)}]{Tretyak2010}%
  \BibitemOpen
  \bibfield  {author} {\bibinfo {author} {\bibfnamefont {V.~I.}\ \bibnamefont
  {Tretyak}},\ }\bibfield  {title} {\enquote {\bibinfo {title} {Semi-empirical
  calculation of quenching factors for ions in scintillators},}\ }\href
  {\doibase 10.1016/j.astropartphys.2009.11.002} {\bibfield  {journal}
  {\bibinfo  {journal} {Astroparticle Physics}\ }\textbf {\bibinfo {volume}
  {33}},\ \bibinfo {pages} {40--53} (\bibinfo {year} {2010})}\BibitemShut
  {NoStop}%
\bibitem [{\citenamefont {Ziegler}, \citenamefont {Ziegler},\ and\
  \citenamefont {Biersack}(2010)}]{Ziegler2010}%
  \BibitemOpen
  \bibfield  {author} {\bibinfo {author} {\bibfnamefont {J.~F.}\ \bibnamefont
  {Ziegler}}, \bibinfo {author} {\bibfnamefont {M.~D.}\ \bibnamefont
  {Ziegler}}, \ and\ \bibinfo {author} {\bibfnamefont {J.~P.}\ \bibnamefont
  {Biersack}},\ }\bibfield  {title} {\enquote {\bibinfo {title} {Srim - the
  stopping and range of ions in matter (2010)},}\ }\href {\doibase
  10.1016/j.nimb.2010.02.091} {\bibfield  {journal} {\bibinfo  {journal}
  {Nuclear Instruments and Methods in Physics Research, Section B: Beam
  Interactions with Materials and Atoms}\ }\textbf {\bibinfo {volume} {268}},\
  \bibinfo {pages} {1818--1823} (\bibinfo {year} {2010})}\BibitemShut {NoStop}%
\bibitem [{\citenamefont {Severijns}, \citenamefont {Beck},\ and\ \citenamefont
  {Naviliat-Cuncic}(2006)}]{Severijns2006}%
  \BibitemOpen
  \bibfield  {author} {\bibinfo {author} {\bibfnamefont {N.}~\bibnamefont
  {Severijns}}, \bibinfo {author} {\bibfnamefont {M.}~\bibnamefont {Beck}}, \
  and\ \bibinfo {author} {\bibfnamefont {O.}~\bibnamefont {Naviliat-Cuncic}},\
  }\bibfield  {title} {\enquote {\bibinfo {title} {Tests of the standard
  electroweak model in nuclear beta decay},}\ }\href {\doibase
  10.1103/RevModPhys.78.991} {\bibfield  {journal} {\bibinfo  {journal}
  {Reviews of Modern Physics}\ }\textbf {\bibinfo {volume} {78}},\ \bibinfo
  {pages} {991--1040} (\bibinfo {year} {2006})}\BibitemShut {NoStop}%
\bibitem [{\citenamefont {Saul}\ \emph {et~al.}(2020)\citenamefont {Saul},
  \citenamefont {Roick}, \citenamefont {Abele}, \citenamefont {Mest},
  \citenamefont {Klopf}, \citenamefont {Petukhov}, \citenamefont {Soldner},
  \citenamefont {Wang}, \citenamefont {Werder},\ and\ \citenamefont
  {Märkisch}}]{Saul2020}%
  \BibitemOpen
  \bibfield  {author} {\bibinfo {author} {\bibfnamefont {H.}~\bibnamefont
  {Saul}}, \bibinfo {author} {\bibfnamefont {C.}~\bibnamefont {Roick}},
  \bibinfo {author} {\bibfnamefont {H.}~\bibnamefont {Abele}}, \bibinfo
  {author} {\bibfnamefont {H.}~\bibnamefont {Mest}}, \bibinfo {author}
  {\bibfnamefont {M.}~\bibnamefont {Klopf}}, \bibinfo {author} {\bibfnamefont
  {A.~K.}\ \bibnamefont {Petukhov}}, \bibinfo {author} {\bibfnamefont
  {T.}~\bibnamefont {Soldner}}, \bibinfo {author} {\bibfnamefont
  {X.}~\bibnamefont {Wang}}, \bibinfo {author} {\bibfnamefont {D.}~\bibnamefont
  {Werder}}, \ and\ \bibinfo {author} {\bibfnamefont {B.}~\bibnamefont
  {Märkisch}},\ }\bibfield  {title} {\enquote {\bibinfo {title} {Limit on the
  fierz interference term b from a measurement of the beta asymmetry in neutron
  decay},}\ }\href {\doibase 10.1103/PHYSREVLETT.125.112501} {\bibfield
  {journal} {\bibinfo  {journal} {Physical Review Letters}\ }\textbf {\bibinfo
  {volume} {125}} (\bibinfo {year} {2020}),\
  10.1103/PHYSREVLETT.125.112501}\BibitemShut {NoStop}%
\bibitem [{\citenamefont {Plaster}\ \emph {et~al.}(2012)\citenamefont
  {Plaster}, \citenamefont {Rios}, \citenamefont {Back}, \citenamefont
  {Bowles}, \citenamefont {Broussard}, \citenamefont {Carr}, \citenamefont
  {Clayton}, \citenamefont {Currie}, \citenamefont {Filippone}, \citenamefont
  {García}, \citenamefont {Geltenbort}, \citenamefont {Hickerson},
  \citenamefont {Hoagland}, \citenamefont {Hogan}, \citenamefont {Hona},
  \citenamefont {Holley}, \citenamefont {Ito}, \citenamefont {Liu},
  \citenamefont {Liu}, \citenamefont {Makela}, \citenamefont {Mammei},
  \citenamefont {Martin}, \citenamefont {Melconian}, \citenamefont
  {Mendenhall}, \citenamefont {Morris}, \citenamefont {Mortensen},
  \citenamefont {Pattie}, \citenamefont {Galván}, \citenamefont {Pitt},
  \citenamefont {Ramsey}, \citenamefont {Russell}, \citenamefont {Saunders},
  \citenamefont {Schmid}, \citenamefont {Seestrom}, \citenamefont {Sjue},
  \citenamefont {Sondheim}, \citenamefont {Tatar}, \citenamefont {Tipton},
  \citenamefont {Vogelaar}, \citenamefont {Vorndick}, \citenamefont {Wrede},
  \citenamefont {Xu}, \citenamefont {Yan}, \citenamefont {Young},\ and\
  \citenamefont {Yuan}}]{Plaster2012}%
  \BibitemOpen
  \bibfield  {author} {\bibinfo {author} {\bibfnamefont {B.}~\bibnamefont
  {Plaster}}, \bibinfo {author} {\bibfnamefont {R.}~\bibnamefont {Rios}},
  \bibinfo {author} {\bibfnamefont {H.~O.}\ \bibnamefont {Back}}, \bibinfo
  {author} {\bibfnamefont {T.~J.}\ \bibnamefont {Bowles}}, \bibinfo {author}
  {\bibfnamefont {L.~J.}\ \bibnamefont {Broussard}}, \bibinfo {author}
  {\bibfnamefont {R.}~\bibnamefont {Carr}}, \bibinfo {author} {\bibfnamefont
  {S.}~\bibnamefont {Clayton}}, \bibinfo {author} {\bibfnamefont
  {S.}~\bibnamefont {Currie}}, \bibinfo {author} {\bibfnamefont {B.~W.}\
  \bibnamefont {Filippone}}, \bibinfo {author} {\bibfnamefont {A.}~\bibnamefont
  {García}}, \bibinfo {author} {\bibfnamefont {P.}~\bibnamefont {Geltenbort}},
  \bibinfo {author} {\bibfnamefont {K.~P.}\ \bibnamefont {Hickerson}}, \bibinfo
  {author} {\bibfnamefont {J.}~\bibnamefont {Hoagland}}, \bibinfo {author}
  {\bibfnamefont {G.~E.}\ \bibnamefont {Hogan}}, \bibinfo {author}
  {\bibfnamefont {B.}~\bibnamefont {Hona}}, \bibinfo {author} {\bibfnamefont
  {A.~T.}\ \bibnamefont {Holley}}, \bibinfo {author} {\bibfnamefont {T.~M.}\
  \bibnamefont {Ito}}, \bibinfo {author} {\bibfnamefont {C.~Y.}\ \bibnamefont
  {Liu}}, \bibinfo {author} {\bibfnamefont {J.}~\bibnamefont {Liu}}, \bibinfo
  {author} {\bibfnamefont {M.}~\bibnamefont {Makela}}, \bibinfo {author}
  {\bibfnamefont {R.~R.}\ \bibnamefont {Mammei}}, \bibinfo {author}
  {\bibfnamefont {J.~W.}\ \bibnamefont {Martin}}, \bibinfo {author}
  {\bibfnamefont {D.}~\bibnamefont {Melconian}}, \bibinfo {author}
  {\bibfnamefont {M.~P.}\ \bibnamefont {Mendenhall}}, \bibinfo {author}
  {\bibfnamefont {C.~L.}\ \bibnamefont {Morris}}, \bibinfo {author}
  {\bibfnamefont {R.}~\bibnamefont {Mortensen}}, \bibinfo {author}
  {\bibfnamefont {R.~W.}\ \bibnamefont {Pattie}}, \bibinfo {author}
  {\bibfnamefont {A.~P.}\ \bibnamefont {Galván}}, \bibinfo {author}
  {\bibfnamefont {M.~L.}\ \bibnamefont {Pitt}}, \bibinfo {author}
  {\bibfnamefont {J.~C.}\ \bibnamefont {Ramsey}}, \bibinfo {author}
  {\bibfnamefont {R.}~\bibnamefont {Russell}}, \bibinfo {author} {\bibfnamefont
  {A.}~\bibnamefont {Saunders}}, \bibinfo {author} {\bibfnamefont
  {R.}~\bibnamefont {Schmid}}, \bibinfo {author} {\bibfnamefont {S.~J.}\
  \bibnamefont {Seestrom}}, \bibinfo {author} {\bibfnamefont {S.}~\bibnamefont
  {Sjue}}, \bibinfo {author} {\bibfnamefont {W.~E.}\ \bibnamefont {Sondheim}},
  \bibinfo {author} {\bibfnamefont {E.}~\bibnamefont {Tatar}}, \bibinfo
  {author} {\bibfnamefont {B.}~\bibnamefont {Tipton}}, \bibinfo {author}
  {\bibfnamefont {R.~B.}\ \bibnamefont {Vogelaar}}, \bibinfo {author}
  {\bibfnamefont {B.}~\bibnamefont {Vorndick}}, \bibinfo {author}
  {\bibfnamefont {C.}~\bibnamefont {Wrede}}, \bibinfo {author} {\bibfnamefont
  {Y.~P.}\ \bibnamefont {Xu}}, \bibinfo {author} {\bibfnamefont
  {H.}~\bibnamefont {Yan}}, \bibinfo {author} {\bibfnamefont {A.~R.}\
  \bibnamefont {Young}}, \ and\ \bibinfo {author} {\bibfnamefont
  {J.}~\bibnamefont {Yuan}},\ }\bibfield  {title} {\enquote {\bibinfo {title}
  {Measurement of the neutron beta-asymmetry parameter a0 with ultracold
  neutrons},}\ }\href {\doibase 10.1103/PhysRevC.86.055501} {\bibfield
  {journal} {\bibinfo  {journal} {Physical Review C - Nuclear Physics}\
  }\textbf {\bibinfo {volume} {86}} (\bibinfo {year} {2012}),\
  10.1103/PhysRevC.86.055501}\BibitemShut {NoStop}%
\bibitem [{\citenamefont {Konrad}\ \emph {et~al.}(2012)\citenamefont {Konrad},
  \citenamefont {Abele}, \citenamefont {Beck}, \citenamefont {Drescher},
  \citenamefont {Dubbers}, \citenamefont {Erhart}, \citenamefont {Fillunger},
  \citenamefont {Gösselsberger}, \citenamefont {Heil}, \citenamefont
  {Horvath}, \citenamefont {Jericha}, \citenamefont {Klauser}, \citenamefont
  {Klenke}, \citenamefont {Märkisch}, \citenamefont {Maix}, \citenamefont
  {Mest}, \citenamefont {Nowak}, \citenamefont {Rebrova}, \citenamefont
  {Roick}, \citenamefont {Sauerzopf}, \citenamefont {Schmidt}, \citenamefont
  {Soldner}, \citenamefont {Wang},\ and\ \citenamefont {Zimmer}}]{Konrad2012}%
  \BibitemOpen
  \bibfield  {author} {\bibinfo {author} {\bibfnamefont {G.}~\bibnamefont
  {Konrad}}, \bibinfo {author} {\bibfnamefont {H.}~\bibnamefont {Abele}},
  \bibinfo {author} {\bibfnamefont {M.}~\bibnamefont {Beck}}, \bibinfo {author}
  {\bibfnamefont {C.}~\bibnamefont {Drescher}}, \bibinfo {author}
  {\bibfnamefont {D.}~\bibnamefont {Dubbers}}, \bibinfo {author} {\bibfnamefont
  {J.}~\bibnamefont {Erhart}}, \bibinfo {author} {\bibfnamefont
  {H.}~\bibnamefont {Fillunger}}, \bibinfo {author} {\bibfnamefont
  {C.}~\bibnamefont {Gösselsberger}}, \bibinfo {author} {\bibfnamefont
  {W.}~\bibnamefont {Heil}}, \bibinfo {author} {\bibfnamefont {M.}~\bibnamefont
  {Horvath}}, \bibinfo {author} {\bibfnamefont {E.}~\bibnamefont {Jericha}},
  \bibinfo {author} {\bibfnamefont {C.}~\bibnamefont {Klauser}}, \bibinfo
  {author} {\bibfnamefont {J.}~\bibnamefont {Klenke}}, \bibinfo {author}
  {\bibfnamefont {B.}~\bibnamefont {Märkisch}}, \bibinfo {author}
  {\bibfnamefont {R.~K.}\ \bibnamefont {Maix}}, \bibinfo {author}
  {\bibfnamefont {H.}~\bibnamefont {Mest}}, \bibinfo {author} {\bibfnamefont
  {S.}~\bibnamefont {Nowak}}, \bibinfo {author} {\bibfnamefont
  {N.}~\bibnamefont {Rebrova}}, \bibinfo {author} {\bibfnamefont
  {C.}~\bibnamefont {Roick}}, \bibinfo {author} {\bibfnamefont
  {C.}~\bibnamefont {Sauerzopf}}, \bibinfo {author} {\bibfnamefont
  {U.}~\bibnamefont {Schmidt}}, \bibinfo {author} {\bibfnamefont
  {T.}~\bibnamefont {Soldner}}, \bibinfo {author} {\bibfnamefont
  {X.}~\bibnamefont {Wang}}, \ and\ \bibinfo {author} {\bibfnamefont
  {O.}~\bibnamefont {Zimmer}},\ }\bibfield  {title} {\enquote {\bibinfo {title}
  {Neutron decay with perc: A progress report},}\ \ }(\bibinfo  {publisher}
  {Institute of Physics Publishing},\ \bibinfo {year} {2012})\BibitemShut
  {NoStop}%
\bibitem [{\citenamefont {Roick}\ \emph {et~al.}(2019)\citenamefont {Roick},
  \citenamefont {Saul}, \citenamefont {Abele},\ and\ \citenamefont
  {Märkisch}}]{Roick2019}%
  \BibitemOpen
  \bibfield  {author} {\bibinfo {author} {\bibfnamefont {C.}~\bibnamefont
  {Roick}}, \bibinfo {author} {\bibfnamefont {H.}~\bibnamefont {Saul}},
  \bibinfo {author} {\bibfnamefont {H.}~\bibnamefont {Abele}}, \ and\ \bibinfo
  {author} {\bibfnamefont {B.}~\bibnamefont {Märkisch}},\ }\bibfield  {title}
  {\enquote {\bibinfo {title} {Undetected electron backscattering in perkeo
  iii},}\ }\href {\doibase 10.1051/epjconf/201921904005} {\bibfield  {journal}
  {\bibinfo  {journal} {EPJ Web of Conferences}\ }\textbf {\bibinfo {volume}
  {219}},\ \bibinfo {pages} {04005} (\bibinfo {year} {2019})}\BibitemShut
  {NoStop}%
\end{thebibliography}%

\end{document}